\documentclass[aps,prd,twocolumn,nofootinbib,groupedaddress,preprintnumbers,floatfix]{revtex4-2}

\usepackage{amsmath,amssymb,bm}
\usepackage[dvipdfmx]{graphicx}
\graphicspath{{./}}
\usepackage{float}
\usepackage{color}
\usepackage{hyperref}
\usepackage{physics}

\newcommand{\Mpl}{M_{\rm Pl}}
\renewcommand{\dd}{{\rm d}}

\newcommand{\be}{\begin{equation}}
\newcommand{\ee}{\end{equation}}
\newcommand{\ba}{\begin{eqnarray}}
\newcommand{\ea}{\end{eqnarray}}
\newcommand{\taK}{\tilde{\alpha}_K}

\makeatletter
\@addtoreset{equation}{section}
\makeatother

\begin{document}

\preprint{WUCG-26-07}

\title{Phantom-divide crossing and suppressed structure growth in \\ 
kinetically braided dark energy with momentum exchange}

\author{Masroor C.~Pookkillath}
\email{masroorcp@gmail.com}
\affiliation{Department of Physics, Waseda University, 3-4-1 Okubo, Shinjuku, Tokyo 169-8555, Japan}

\author{Shinji Tsujikawa}
\email{tsujikawa@waseda.jp}
\affiliation{Department of Physics, Waseda University, 3-4-1 Okubo, Shinjuku, Tokyo 169-8555, Japan}

\date{\today}

\begin{abstract}
We construct a linearly stable scalar-field model that realizes both an
upward crossing of the dark-energy equation of state, from
$w_{\rm DE}<-1$ to $w_{\rm DE}>-1$, and weakened gravitational
clustering in the cold dark matter (CDM) sector.  An exponential potential
breaks shift symmetry and drives the background from a stable phantom phase
toward the nonphantom regime, while a pure momentum-transfer interaction
increases the dynamical inertia of CDM without altering its background
dilution law.  We derive the background and linear perturbation equations
and establish the no-ghost and Laplacian-stability conditions.  For
perturbations deep inside the Hubble radius, where the quasi-static
approximation applies, the effective gravitational coupling for CDM can
fall below Newton's constant, suppressing late-time growth and small-scale
matter power, while the baryonic coupling remains enhanced by Galileon
braiding.  A modified CLASS calculation, including the scalar-field
perturbation and the full Boltzmann hierarchies, reveals signatures of
transient braiding around radiation--matter equality.  
For the representative stable solutions studied here, these signatures
include an enhancement of matter power toward the lowest wavenumbers
probed numerically and a reduction of CMB temperature power over the
multipole range $2\leq\ell\leq30$.
We also find small shifts in the acoustic scale and
the position of the first temperature peak.
These results motivate a full likelihood analysis of the model.
\end{abstract}

\maketitle

\section{Introduction}

A broad range of observations indicates that the cosmic energy budget is dominated by two dark components: dark matter (DM) and dark energy (DE). DM is required by the dynamics of galaxies and clusters, gravitational lensing, the cosmic microwave background (CMB), and the formation of large-scale structure. DE, by contrast, drives the present cosmic acceleration, first established with Type Ia supernovae and subsequently corroborated by CMB and baryon-acoustic-oscillation (BAO) measurements~\cite{SupernovaSearchTeam:1998fmf,SupernovaCosmologyProject:1998vns,WMAP:2003elm,SDSS:2005xqv,Planck:2018vyg}. 
Despite this compelling gravitational evidence, the microscopic nature of both components remains unknown~\cite{Copeland:2006wr,Silvestri:2009hh,DeFelice:2010aj,Clifton:2011jh,Tsujikawa:2013fta,Joyce:2014kja,Koyama:2015vza,Amendola:2016saw,Kase:2018aps}.

The minimal cosmological description is the $\Lambda$ cold dark matter
($\Lambda$CDM) model~\cite{Peebles:1982ff,Peebles:1984ge,Turner:1984nf,Efstathiou:1990xe,Ostriker:1995su}, in which $\Lambda$ is a cosmological constant with
$w_{\rm DE}\equiv P_{\rm DE}/\rho_{\rm DE}=-1$, while DM is cold and
collisionless; we refer to the latter component as CDM below. Here
$\rho_{\rm DE}$ and $P_{\rm DE}$ denote the DE energy density and pressure,
respectively. Recent BAO measurements by the Dark Energy Spectroscopic
Instrument (DESI), when combined with CMB and supernova data, have
nevertheless strengthened the preference for an evolving DE sector over a
strict cosmological constant. 
Within the Chevallier--Polarski--Linder (CPL)
parametrization~\cite{Chevallier:2000qy,Linder:2002et}, the best-fit DE histories cross the phantom divide
from $w_{\rm DE}<-1$ at intermediate redshifts to $w_{\rm DE}>-1$ at lower
redshifts, typically around
$z\simeq0.5$~\cite{DESI:2024mwx,DESI:2024aqx,DESI:2025zgx,DESI:2025fii}.

Meanwhile, weak-lensing, cluster, and redshift-space-distortion measurements
have often inferred a late-time clustering amplitude below that obtained by
extrapolating the Planck-normalized $\Lambda$CDM
model~\cite{KiDS:2016kqt,DES:2017myr,Asgari:2020wuj,Heymans:2020gsg,DES:2021bvc,Li:2023tui}.
This amplitude is commonly characterized 
by $\sigma_8$, the present-day
root-mean-square linear matter density contrast 
smoothed over spheres of
comoving radius $8h_{100}^{-1}\,{\rm Mpc}$, or by
$S_8\equiv\sigma_8\sqrt{\Omega_{m0}/0.3}$. 
Here $h_{100}\equiv H_0/(100\,{\rm km}\,
{\rm s}^{-1}\,{\rm Mpc}^{-1})$,
$H_0$ is the present Hubble rate, 
and $\Omega_{m0}$ is the
present matter density parameter. 
Taken together, these observational indications motivate a theoretically consistent framework that realizes the CPL-favored phantom-divide crossing
while weakening the late-time growth of matter perturbations.

Producing such a crossing with a single scalar degree of freedom is nontrivial. Positivity of the canonical kinetic energy enforces $w_{\rm DE}\geq-1$ for quintessence~\cite{Fujii:1982ms,Ratra:1987rm,Wetterich:1987fm,Chiba:1997ej,Ferreira:1997au,Caldwell:1997ii,Copeland:1997et}. A stable k-essence field~\cite{Armendariz-Picon:1999hyi,Chiba:1999ka,Armendariz-Picon:2000nqq} obeys the same restriction, because the no-ghost and Laplacian-stability conditions preclude a regular passage through $w_{\rm DE}=-1$. These constraints are also central to recent DESI-oriented studies of canonical one- and two-field quintessence~\cite{Shlivko:2025fgv,Akrami:2025zlb,Bayat:2025xfr,Cline:2025sbt,Gialamas:2025pwv,Alestas:2025syk,Shlivko:2025krk}. Reversing the sign of the canonical kinetic term permits $w_{\rm DE}<-1$~\cite{Caldwell:1999ew,Caldwell:2003vq,Singh:2003vx}, but introduces a ghost and a severe vacuum instability~\cite{Carroll:2003st,Cline:2003gs}.
Two-field ``quintom'' constructions do not remove this fundamental difficulty unless an additional degeneracy or constraint arises in the ultraviolet completion~\cite{Feng:2004ad,Guo:2004fq}.

Derivative self-interactions offer a qualitatively different route. Galileon interactions~\cite{Nicolis:2008in,Deffayet:2009wt} can support
phantom-like background evolution without necessarily 
introducing a ghost degree of freedom~\cite{DeFelice:2010pv,DeFelice:2010nf,Nesseris:2010pc}.
These interactions are contained within Horndeski theories, the most general four-dimensional scalar--tensor 
theories yielding second-order field equations~\cite{Horndeski:1974wa,Deffayet:2011gz,Kobayashi:2011nu,Charmousis:2011bf}. This second-order structure avoids the Ostrogradsky instability that generally arises in nondegenerate higher-derivative theories. 
The multimessenger observation of the binary 
neutron-star merger GW170817 and its electromagnetic counterpart constrains the propagation
speed $c_T$ of tensor perturbations to agree 
with the speed of light $c$ to extremely high precision~\cite{LIGOScientific:2017vwq,
LIGOScientific:2017zic,Goldstein:2017mmi}. 
Within Horndeski theories,
imposing $c_T=c$ without relying on 
cancellations among independent
operators restricts the viable Lagrangian, 
up to boundary terms, to
the form~\cite{Creminelli:2017sry,Ezquiaga:2017ekz,
Sakstein:2017xjx,Baker:2017hug}
\begin{equation}
L=G_2(\phi,X)+G_3(\phi,X)\Box\phi
+G_4(\phi)R\,.
\label{HornLag}
\end{equation}
Here, $G_2$ and $G_3$ are functions of the DE 
scalar field $\phi$ and the kinetic scalar $X\equiv-(1/2)\nabla_\mu\phi\nabla^\mu\phi$, whereas $G_4$ depends only on $\phi$; $R$ is the Ricci scalar, $\Box\phi\equiv\nabla_\mu\nabla^\mu\phi$ is the covariant d'Alembertian of $\phi$, and $\nabla_\mu$ denotes the covariant derivative associated with the spacetime metric $g_{\mu\nu}$. 
Importantly, this reduced class still accommodates the cubic
Galileon interaction employed in the present model, for which
$G_3(X)\propto X$.

Shift symmetry, however, imposes a decisive restriction even within this surviving class. Reference~\cite{Tsujikawa:2025wca} showed that, if $G_2$, $G_3$, and $G_4$ are all shift symmetric, a healthy late-time solution cannot evolve from $w_{\rm DE}<-1$ to $w_{\rm DE}>-1$. The covariant cubic Galileon and Galileon ghost-condensate models are representative examples~\cite{DeFelice:2010pv,Peirone:2019aua}. The desired crossing therefore requires explicit $\phi$ dependence in at least one Horndeski function.

One possibility is a $\phi$-dependent $G_4$, as in the scalar-tensor representation of metric $f(R)$ gravity and, more generally, in nonminimally coupled scalar-tensor theories. Such theories can realize phantom-divide crossing~\cite{Amendola:2006we,Hu:2007nk,Starobinsky:2007hu,Appleby:2007vb,Tsujikawa:2007xu,Amendola:2007nt,Motohashi:2010tb,Boisseau:2000pr,Perivolaropoulos:2005yv}. The nonminimal coupling, however, mediates a fifth force and makes the effective Planck mass field dependent. Viable $f(R)$ models must screen this force locally, typically through the chameleon mechanism~\cite{Khoury:2003aq,Faulkner:2006ub,Capozziello:2007eu}. Solar-System and cosmic-structure-growth constraints then require $w_{\rm DE}$ to remain close to $-1$~\cite{Hu:2007nk,Brax:2008hh}. In more general scalar-tensor theories, nonlinear derivative interactions can instead provide Vainshtein screening~\cite{Vainshtein:1972sx}. 
Even with such screening, cosmological scalar evolution can induce a local time variation of the effective gravitational coupling~\cite{Babichev:2011iz,Kimura:2011dc}, and this variation is tightly constrained by lunar-laser-ranging measurements~\cite{Hofmann:2018myc,Tsujikawa:2019pih}.
Although several nonminimally coupled DE models have recently been proposed to realize phantom-divide crossing~\cite{Ye:2024ywg,Wolf:2024stt,Pan:2025psn,Wang:2025znm,Adam:2025kve,SanchezLopez:2025uzw,Chakraborty:2025syu}, it remains to be established whether an observationally appreciable crossing survives all local-gravity and cosmological structure-growth constraints.

A cleaner alternative is to introduce a scalar potential $V(\phi)$ in $G_2$ while keeping $G_4=\Mpl^2/2$ constant, where $\Mpl\equiv(8\pi G)^{-1/2}$ is the reduced Planck mass and $G$ is Newton's gravitational constant. Reference~\cite{Tsujikawa:2025wca} demonstrated that a cubic Galileon supplemented by such a potential can evolve from a stable phase with $w_{\rm DE}<-1$ to one with $w_{\rm DE}>-1$ at low redshifts (see also 
Refs.~\cite{Aoki:2024ktc,Wolf:2025acj,Tsujikawa:2026xqm,Calderon:2026hbr,Naidoo:2026umv,Garcia-Garcia:2026nzy,Hallam:2026qsk} for related works). Because the Planck mass remains constant and ordinary matter is minimally coupled, this construction induces neither the unscreened fifth force nor the local variation of the gravitational coupling characteristic of a nonminimal $G_4(\phi)R$ interaction. The cubic Galileon provides the kinetic braiding required for the phantom phase, while the potential ultimately redirects the background trajectory toward the nonphantom side.

This background-level success is accompanied by an important limitation. The same braiding that generates $w_{\rm DE}<-1$ mediates an attractive scalar interaction, causing the effective gravitational coupling $G_{\rm eff}$ governing matter clustering to exceed Newton's constant, $G_{\rm eff}>G$~\cite{DeFelice:2010as,DeFelice:2011hq,Peirone:2019aua,Tsujikawa:2025wca}. 
The inclusion of the scalar potential can suppress the growth of matter perturbations relative to that in the potential-free cubic Galileon, but the potential alone cannot realize weak gravity relative to 
$\Lambda$CDM, namely, $G_{\rm eff}<G$.
More generally, in Horndeski theories without a direct DE--DM interaction and with matter minimally coupled to the metric, 
a stable scalar perturbation mode generically adds a non-negative attractive contribution to the quasi-static gravitational coupling~\cite{DeFelice:2011hq,Kase:2018aps}. Alleviating the $\sigma_8$ tension therefore calls for an additional mechanism that suppresses DM clustering without disrupting the successful background evolution.

Momentum exchange between the dark sectors provides precisely such a mechanism. If DE and DM exchange momentum but not background energy, the resulting drag can suppress clustering while preserving the standard homogeneous dilution law for DM. Phenomenological elastic-scattering models, forecasts, and nonlinear studies have explored this possibility extensively~\cite{Simpson:2010vh,Asghari:2019qld,BeltranJimenez:2020tme,BeltranJimenez:2020qdu,BeltranJimenez:2021wbq,Cardona:2022lcz,BeltranJimenez:2022ixm,BeltranJimenez:2024nbz,BeltranJimenez:2024rkg,Poulin:2022sgp,Cruickshank:2025wjy,Cruickshank:2025rqa,BeltranJimenez:2025nls}. A covariant Lagrangian realization was introduced in Ref.~\cite{Pourtsidou:2013nha} and further developed in Refs.~\cite{Boehmer:2015sha,Skordis:2015yra,Koivisto:2015qua,Pourtsidou:2016ico,Dutta:2017fjw,Linton:2017cep,Kase:2019veo,Kase:2019mox,Chamings:2019qak,Kase:2020hst,Amendola:2020lnd,DeFelice:2020cpt,Linton:2021gqk,Liu:2023yaw,Aoki:2025bmj,BeltranJimenez:2026ymd}. In particular, the interaction Lagrangian density $L_{\rm int}=\beta Z^2$, where $\beta$ is a constant coupling, $Z\equiv u_c^\mu\nabla_\mu\phi$, and $u_c^\mu$ is the CDM four-velocity, modifies the CDM Euler equation and can reduce the effective gravitational coupling felt by CDM below $G$. Most previous scalar-field applications, however, assumed quintessence and therefore remained 
in the range $w_{\rm DE}\geq-1$.

In this paper, we combine these two mechanisms by augmenting the
potential-extended cubic-Galileon model of Ref.~\cite{Tsujikawa:2025wca}
with the pure momentum-transfer interaction $L_{\rm int}=\beta Z^2$.
We show analytically and numerically that the resulting theory can realize an upward phantom-divide crossing from $w_{\rm DE}<-1$ to $w_{\rm DE}>-1$ at low redshifts, while yielding a weak effective gravitational coupling for CDM, $G_c<G$.
We impose the no-ghost and Laplacian-stability conditions and
derive the complete Newtonian-gauge scalar perturbation system.  We then use
a modified version of the Cosmic Linear Anisotropy Solving System
(CLASS)~\cite{Lesgourgues:2011re,Blas:2011rf} to evolve the scalar-field,
metric, matter, and radiation perturbations without applying the
quasi-static approximation on large scales.  
On sub-Hubble scales, the representative solutions exhibit a suppressed
present-day growth rate of matter perturbations and reduced small-scale
matter power. By contrast, transient braiding produces an enhancement of
the matter power spectrum toward the lowest wavenumbers covered by the
numerical calculation, while the large-angle CMB temperature power is
reduced. These results provide the theoretical basis for a systematic
likelihood analysis.

The remainder of this paper is organized as follows.  In
Sec.~\ref{sec:model}, we introduce the action and background equations.  In
Sec.~\ref{subsec:stability_background}, we analyze the background dynamics,
stability conditions, and representative numerical solutions.  In
Sec.~\ref{persec}, we derive the linear perturbation equations and their
quasi-static limit.  Section~\ref{sec:superHubble} studies the
super-Hubble response to the transient braiding peak, and
Sec.~\ref{sec:powerCMB} presents the linear matter and CMB temperature power
spectra.  Section~\ref{sec:conclude} summarizes our results and future
directions.
Throughout this paper, we use natural units with $c=\hbar=1$.

\section{The model and background equations}
\label{sec:model}

We consider a scalar-tensor theory governed by the action
\ba
{\cal S}&=&\int \dd^4x\sqrt{-g}\biggl[
\frac{\Mpl^2}{2}R+a_1X+a_2X^2+3a_3X\Box\phi
\nonumber\\
&&\qquad\qquad\quad\,
-V(\phi)+\beta Z^2\biggr]+{\cal S}_m\,,
\label{action}
\ea
where $a_1$, $a_2$, $a_3$, 
and $\beta$ are constant model parameters,
$V(\phi)$ is the scalar-field potential, and
$g\equiv\det(g_{\mu\nu})$.
The two scalar contractions entering the action are
\begin{equation}
X=-\frac12 \nabla_\mu\phi\nabla^\mu\phi,\qquad
Z=u_c^\mu \nabla_\mu \phi\,,
\end{equation}
where $u_c^\mu$ is the four-velocity of CDM.  
The term $\beta Z^2$ describes a pure momentum transfer between CDM and the scalar field. 
The matter action ${\cal S}_m$ contains contributions from CDM, baryons, and radiation. We assume that baryons and radiation are minimally coupled to the metric, so their continuity and Euler equations retain their standard forms.

The gravity--scalar part of the action~\eqref{action} belongs to the
luminal Horndeski class, while $\beta Z^2$ is an additional covariant
CDM--scalar interaction.  In the notation of Eq.~\eqref{HornLag}, the
Horndeski sector is specified by
\ba 
G_2&=&a_1X+a_2X^2-V(\phi)\,,
\qquad
G_3=3a_3X,\nonumber\\
G_4&=&\frac{\Mpl^2}{2}\,.
\ea
In the interacting DE--CDM framework of Ref.~\cite{Kase:2020hst}, the interaction term in Eq.~\eqref{action} corresponds to
\begin{equation}
f_2(Z)=\beta Z^2\,.
\end{equation}
Since the dark-sector interaction $f_2$ is independent of the CDM number density $n_c$, the CDM particle number is conserved, and no energy is transferred between CDM and the scalar field at the background level. The interaction nevertheless affects linear perturbations because $Z$ depends on the CDM four-velocity. For the scalar-field potential, we adopt the exponential form
\begin{equation}
V(\phi)=V_0 e^{-\lambda\phi/\Mpl}\,,
\label{Vexp}
\end{equation}
where $V_0>0$ sets the energy scale of the potential and the dimensionless constant $\lambda$ determines its slope.

We consider a spatially flat Friedmann--Lema\^itre--Robertson--Walker
(FLRW) background with the line element
\begin{equation}
\dd s^2=-\dd t^2+a^2(t)\delta_{ij}\dd x^i\dd x^j\,,
\end{equation}
where $t$ is cosmic time, $a(t)$ is the scale factor, $x^i$ are comoving
spatial coordinates, and $\delta_{ij}$ is the Kronecker delta. We normalize
the scale factor to unity at the present epoch, $a_0=1$, and define the
redshift as $z\equiv 1/a-1$. Hereafter, a subscript $0$ denotes a quantity evaluated at $z=0$. For a homogeneous scalar field $\phi=\phi(t)$, the CDM four-velocity is $u_c^\mu=(1,0,0,0)$, giving
\begin{equation}
X=\frac12\dot\phi^2,\qquad Z=\dot\phi\,,
\end{equation}
where a dot denotes differentiation 
with respect to $t$. At the background
level, the momentum-transfer term therefore reduces to
$\beta Z^2=2\beta X$, shifting the coefficient of $X$ from $a_1$ to
$a_1+2\beta$. It is thus convenient to define
\begin{equation}
A\equiv a_1+2\beta\,.
\label{Adef}
\end{equation}
This definition simplifies the background equations, 
whereas $\beta$
remains explicit in the perturbation sector.

The energy densities of CDM, baryons, and radiation are denoted by
$\rho_c$, $\rho_b$, and $\rho_r$, respectively.  Their continuity
equations are
\be
\dot{\rho}_{I}+3H (1+w_I) \rho_I=0\,,
\label{coneq}
\ee
where $I=c,b,r$, $w_c=w_b=0$, $w_r=1/3$, and
$w_I\equiv P_I/\rho_I$ is the equation-of-state parameter of species $I$.
Here $P_I$ denotes the pressure of the fluid component $I$, and $H=\dot a/a$ is the Hubble parameter.  
For baryons, CDM, and radiation,
we set $P_b=P_c=0$ and $P_r=\rho_r/3$. 

The Friedmann equations can be written as
\begin{align}
 3\Mpl^2H^2 &=\rho_c+\rho_b+\rho_r+\rho_{\rm DE},
\label{fried1}\\
 2\Mpl^2\dot H&=-\left( \rho_c+\rho_b
 +\frac{4}{3}\rho_r+\rho_{\rm DE}+P_{\rm DE} \right),
\label{fried2}
\end{align}
where the effective DE density and pressure are
\begin{align}
\rho_{\rm DE}&=A X+3a_2X^2-9a_3H\dot\phi^3+V,
\label{rhoDE}\\
P_{\rm DE}&=A X+a_2X^2+3a_3\dot\phi^2\ddot\phi
-V\,.
\label{pDE}
\end{align}
Taking the time derivative of Eq.~\eqref{fried1} and using
Eqs.~\eqref{coneq} and \eqref{fried2}, we obtain
\be
\dot{\rho}_{\rm DE}+3H \left( \rho_{\rm DE}
+P_{\rm DE} \right)=0\,.
\label{DEcon}
\ee
This equation is equivalent to the scalar-field 
equation of motion on the FLRW background.

To study the background dynamics, we introduce the dimensionless variables
\begin{align}
x_1&=\frac{\dot\phi}{\sqrt6\Mpl H},\qquad
x_2=\frac{a_2\dot\phi^4}{4\Mpl^2H^2},\qquad
x_3=-\frac{3a_3\dot\phi^3}{\Mpl^2H},
\nonumber\\
x_4&=\frac{V}{3\Mpl^2H^2},\qquad
\Omega_I=\frac{\rho_I}{3\Mpl^2H^2}\,,\qquad
(I=c,b,r)\,.
\label{xdefs}
\end{align}
The DE density parameter and equation of state are then
\begin{align}
\Omega_{\rm DE}&=A x_1^2+x_2+x_3+x_4\,,
\label{OmDE}\\
w_{\rm DE}&=\frac{3A x_1^2+x_2-x_3\epsilon_\phi-3x_4}
{3(Ax_1^2+x_2+x_3+x_4)}\,,
\label{wDE}
\end{align}
where $\epsilon_\phi=\ddot\phi/(H\dot\phi)$.  Equation~\eqref{fried1}
gives
\begin{equation}
\Omega_c=1-\Omega_{\rm DE}-\Omega_b-\Omega_r\,.
\label{Omm}
\end{equation}

The background equations can be recast 
as the autonomous system
\begin{align}
 x_1'&=x_1(\epsilon_\phi-h),
&x_2'&=2x_2(2\epsilon_\phi-h),
\nonumber\\
 x_3'&=x_3(3\epsilon_\phi-h),
&x_4'&=-x_4(\sqrt6\lambda x_1+2h),
\nonumber\\
\Omega_b'&=-\Omega_b(3+2h),
&\Omega_r'&=-2\Omega_r(2+h),
\label{autonomous}
\end{align}
where $h=\dot H/H^2$, and a prime denotes differentiation with respect to $N=\ln a$. 
Solving Eqs.~\eqref{fried2} and \eqref{DEcon} 
for $\epsilon_\phi$ and $h$, we obtain
\begin{align}
\epsilon_\phi
&=\frac{1}{{\cal D}}
\biggl[
 x_3\left(3Ax_1^2+x_2+\Omega_r-3x_4-3\right)
 \nonumber\\
&\hspace{1.4cm}{}-4\left(3Ax_1^2+2x_2\right)
 +2\sqrt6\lambda x_1x_4\biggr],
\label{epssol}\\
 h&=-\frac{1}{{\cal D}}\biggl[
 2Ax_1^2\left(3Ax_1^2+\Omega_r+7x_2+6x_3-3x_4+3\right)
 \nonumber\\
&\hspace{1.4cm}{}+2x_2\left(2\Omega_r+2x_2+3x_3-6x_4+6\right)
 \nonumber\\
&\hspace{1.4cm}{}+x_3\left(2\Omega_r-\sqrt6\lambda x_1x_4
+3x_3-6x_4+6\right)\biggr]\,,
\label{hsol}
\end{align}
where
\begin{equation}
{\cal D}=4Ax_1^2+8x_2+4x_3+x_3^2\,.
\label{Ddef}
\end{equation}
\section{Background dynamics}
\label{subsec:stability_background}

In this section, we analyze the cosmological background dynamics of the model defined by the action~\eqref{action}, with particular emphasis on the stability of linear perturbations. We first summarize the no-ghost and Laplacian-stability conditions and then derive analytic estimates for the evolution of the background variables and the quantities entering these conditions at high, intermediate, and low redshifts. Finally, we solve the autonomous system numerically to verify the analytic estimates and elucidate how the evolution of the background variables drives the transition from $w_{\rm DE}<-1$ to $w_{\rm DE}>-1$ without inducing ghost or Laplacian instabilities.

\subsection{Linear stability conditions}

For the subclass of Horndeski theories defined by the Lagrangian~\eqref{HornLag}, the stability conditions in the presence of a more general form of momentum-transfer coupling between DE and CDM were derived in Ref.~\cite{Kase:2020hst} from the quadratic actions for tensor and scalar perturbations.
Specializing those results to the present model yields the following conditions, written in terms of the background variables introduced above.

Since $G_4=\Mpl^2/2$ is constant, tensor perturbations propagate at the speed of light and are free from ghost instabilities.  In the scalar sector, the coefficient associated with the no-ghost condition for the field perturbation $\delta\phi$ is given by\footnote{The quantity denoted by $q_s$ in
Ref.~\cite{Kase:2020hst} is dimensionful and equals $\Mpl^2 q_s$, where
$q_s$ is defined in Eq.~\eqref{qsX}. In terms of the notation
$Q_s$ used in Ref.~\cite{Tsujikawa:2025wca}, our $q_s$ corresponds to
$q_s=Q_s/(3x_1^2)$.}
\be
q_s=\frac{{\cal D}}{2x_1^2}
=2A+\frac{4x_2}{x_1^2}
+\frac{2x_3}{x_1^2}
+\frac{x_3^2}{2x_1^2}\,,
\label{qsX}
\ee
where ${\cal D}$ is defined in Eq.~\eqref{Ddef}.  
We restrict 
our analysis to solutions for which $\dot{\phi}$ retains a 
fixed nonzero sign throughout the cosmological evolution, so 
that $x_1^2>0$. The absence of scalar ghosts then requires
\begin{equation}
q_s>0\,,
\label{qpos}
\end{equation}
which is equivalent to
${\cal D}>0$. Under this condition, the denominators in Eqs.~\eqref{epssol} and \eqref{hsol} remain nonzero, thereby preventing the corresponding background quantities from diverging.

The no-ghost condition for CDM is given by 
\begin{equation}
q_c=1+\frac{4\beta x_1^2}{\Omega_c}>0,
\label{qcX}
\end{equation}
where $\Omega_c$ is defined in Eq.~\eqref{xdefs}.  
For $\beta>0$ and $\Omega_c>0$, this inequality 
is automatically satisfied.
Negative $\beta$ can lead to $q_c<0$ and 
should generally be avoided unless a restricted 
parameter region is explicitly verified to be stable.
Therefore, in what follows, we focus on the parameter range
\be
\beta>0\,,
\ee
for which $q_c>1$.
The squared sound speed of the scalar-field 
perturbation $\delta \phi$ is expressed 
in the form
\begin{equation}
c_s^2=\hat c_s^2+\Delta c_s^2\,,
\label{csX}
\end{equation}
where $\hat c_s^2$ denotes the contribution 
without the explicit CDM--scalar mixing correction, 
while $\Delta c_s^2$ represents 
the correction induced by this mixing:
\begin{align}
\hat c_s^2
&=-\frac{2}{3x_1^2 q_s}\biggl[
h-\frac{x_3}{2}\left(1+3\epsilon_\phi
-\frac{x_3}{2}\right)
\nonumber\\
&\hspace{1.7cm}{}+
\frac32\left(q_c\Omega_c+\Omega_b
+\frac43\Omega_r\right)\biggr]\,,
\label{hatcsX}\\
\Delta c_s^2&= \frac{\Omega_c(1-q_c)^2}
{x_1^2 q_s q_c}\,.
\end{align}
The absence of Laplacian instabilities 
requires 
\begin{equation}
c_s^2>0\,.
\label{lapmodel}
\end{equation}
Since $\Delta c_s^2>0$ under the no-ghost 
conditions $q_s>0$ and $q_c>0$, the inequality
in Eq.~\eqref{lapmodel} is always satisfied for
$\hat{c}_s^2>0$.

Equations~\eqref{qpos}, \eqref{qcX}, and \eqref{lapmodel} are the stability conditions imposed on the numerical background solutions throughout the expansion history, from the early radiation era through the present epoch and into the asymptotic future.

\subsection{High-redshift behavior}
\label{highsec}

We first examine the high-redshift behavior of the background 
solutions and the associated stability conditions. Since the 
potential contribution is negligible in this regime, we set 
$x_4=0$ when deriving the analytic estimates. In the absence 
of the momentum-transfer interaction ($\beta=0$), the 
corresponding asymptotic behavior was analyzed in 
Ref.~\cite{Tsujikawa:2025wca}. This analysis carries over to 
the present model upon making the replacement 
$a_1\to A=a_1+2\beta$.

In the earliest high-redshift regime, the cubic Galileon variable $x_3$ dominates over the other DE variables.  More explicitly, we impose
\ba
& &
\{ |Ax_1^2|,\quad |x_2|,\quad |\beta x_1^2| \}
\ll |x_3| \ll 1,\nonumber \\
& &
|\beta x_1^2| \ll \Omega_c\,.
\label{earlyhierarchy}
\ea
By introducing the variables
\begin{equation}
r\equiv \frac{A x_1^2}{x_3},
\qquad
s\equiv \frac{x_2}{x_3}\,,
\label{rsdef}
\end{equation}
the first line of Eq.~\eqref{earlyhierarchy} implies
$|r|\ll1$, $|s|\ll1$, and 
$|\beta x_1^2/x_3|\ll1$, with $|x_3|\ll1$.
The second line of Eq.~\eqref{earlyhierarchy} gives 
$q_c\simeq1$ and suppresses the explicit CDM-coupling 
contribution to $c_s^2$, namely 
$\Delta c_s^2$, 
in the deep radiation and matter eras.
Expanding Eqs.~\eqref{epssol} and \eqref{hsol} 
under the hierarchy \eqref{earlyhierarchy}, 
we obtain
\ba
\epsilon_\phi 
&=& 
\frac{\Omega_r-3}{4}
+\frac{3-\Omega_r}{16}x_3 
+\cdots\,,\label{earlyepsh0}\\
h &=& -\frac{3+\Omega_r}{2}
-\frac{3-\Omega_r}{8}x_3+\cdots\,.
\label{earlyepsh}
\ea
Then, from Eqs.~\eqref{wDE}, \eqref{qsX}, and \eqref{csX}, 
the leading-order contributions to $w_{\rm DE}$, $q_s$, and 
$c_s^2$ are given, respectively, by
\begin{equation}
w_{\rm DE}\simeq \frac{3-\Omega_r}{12},
\qquad
q_s\simeq \frac{2x_3}{x_1^2},
\qquad
c_s^2\simeq \frac{5+\Omega_r}{12}\,.
\label{earlywcs}
\end{equation}
In the radiation era, where $\Omega_r\simeq1$, Eq.~\eqref{earlywcs} gives
$w_{\rm DE}\simeq1/6$ and $c_s^2\simeq1/2$, whereas in the 
matter era, where $\Omega_r\simeq0$, one obtains 
$w_{\rm DE}\simeq1/4$ and $c_s^2\simeq5/12$.  The no-ghost 
condition is satisfied in this earliest regime for 
\be
x_3>0\,, 
\label{x3con}
\ee
which is imposed in the following.

The above expansion also determines the leading power-law behavior of the background variables $x_1$, $x_2$, $x_3$, 
and $x_4$. 
Substituting the leading-order expressions in 
Eqs.~\eqref{earlyepsh0} and \eqref{earlyepsh} into the autonomous system 
\eqref{autonomous}, treating $\Omega_r$ as approximately 
constant, and neglecting $\sqrt{6}\lambda x_1$ relative to 
$2h$ in the $x_4$ equation, we obtain
\begin{equation}
\begin{aligned}
x_1&\propto a^{3(1+\Omega_r)/4},\qquad
x_2\propto a^{2\Omega_r},\\
x_3&\propto a^{(5\Omega_r-3)/4},\qquad
x_4\propto a^{3+\Omega_r}.
\end{aligned}
\label{earlyscaling}
\end{equation}
During radiation domination, this gives
$x_1\propto a^{3/2}$, $x_2\propto a^2$, 
$x_3\propto a^{1/2}$, and
$x_4\propto a^4$. Hence, $q_s\simeq 2x_3/x_1^2\propto a^{-5/2}$,
so that $q_s$ rapidly grows toward the asymptotic past. During
the early matter era, Eq.~\eqref{earlyscaling} gives
$x_1\propto a^{3/4}$, $x_2\simeq {\rm const.}$, $x_3\propto a^{-3/4}$,
and $x_4\propto a^3$, which leads to $q_s\propto a^{-9/4}$ as long as
the hierarchy \eqref{earlyhierarchy} remains valid. 
The decrease of $x_3$ relative to $x_1^2$ after the onset of matter domination 
also signals the eventual breakdown of the earliest hierarchy and the transition 
to the intermediate regime discussed below.

\subsection{Intermediate-redshift behavior}
\label{intersec}

The hierarchy \eqref{earlyhierarchy} applies only to the earliest stage, where $|r|\ll1$ and $|s|\ll1$.  Once $r$ and $s$ become finite, this hierarchy should be replaced by a more general intermediate approximation.  In particular, the condition $|r|\ll1$ is incompatible with the tracker value $r=-1/2$ discussed below.  To describe the regime in which the solution approaches the covariant-Galileon tracker originally found in Ref.~\cite{DeFelice:2010pv}, we keep $r$ and $s$ finite while retaining
\begin{equation}
x_3\ll1,\qquad 
|\beta x_1^2|\ll\Omega_c\,.
\label{interhierarchy}
\end{equation}
Thus, in this intermediate regime, the inequalities
$|A x_1^2|\ll x_3$ and $|x_2|\ll x_3$ in Eq.~\eqref{earlyhierarchy}
are relaxed and replaced by finite $r=A x_1^2/x_3$ and
$s=x_2/x_3$.  We also do not impose $|\beta x_1^2|\ll x_3$, since this would amount to assuming $|\beta/A|\ll1$ near the tracker
$r\simeq -1/2$.  Instead, we keep the ratio
$\beta x_1^2/x_3=\beta r/A$ finite.  Meanwhile, the condition
$|\beta x_1^2|\ll\Omega_c$ ensures $q_c\simeq1$ and suppresses the explicit term proportional to $(1-q_c)^2$ in Eq.~\eqref{csX}.
Expanding Eqs.~\eqref{epssol} and \eqref{hsol} under
Eq.~\eqref{interhierarchy}, we obtain
\begin{align}
\epsilon_\phi
&\simeq
\frac{\Omega_r-3-12r-8s}
{4(1+r+2s)}+{\cal O}(x_3),
\nonumber\\
h
&\simeq -\frac{3+\Omega_r}{2}\nonumber\\
&\quad +\frac{x_3}{8(1+r+2s)}
\left[\Omega_r-3-12r(r+2)\right.\nonumber\\
&\quad\left. -4s(7r+3)-8s^2\right]
+{\cal O}(x_3^2).
\label{interepsh}
\end{align}
The ${\cal O}(x_3)$ correction to $h$ 
has been kept explicitly
because it is required for the leading-order evaluation of
$c_s^2$, due to the cancellation of the zeroth-order terms in
the square bracket of Eq.~\eqref{hatcsX}. 
The covariant-Galileon result is recovered from 
Eq.~\eqref{interepsh} by setting $s=0$, while all the 
$s$-dependent terms represent corrections 
induced by a nonzero $x_2$, with $s=x_2/x_3$.
The DE equation of state is 
then given by
\begin{equation}
w_{\rm DE}
\simeq
\frac{3-\Omega_r+12r(r+2)
+4s(7r+3)+8s^2}
{12(1+r+s)(1+r+2s)}.
\label{wDEintermediate}
\end{equation}
The first three terms in the numerator reproduce the result for
$s=0$, while the terms proportional to $s$ and $s^2$ represent the
corrections arising from $x_2$. The corresponding leading-order
expressions for the stability quantities are
\begin{equation}
q_s
\simeq
\frac{2x_3}{x_1^2}\left(1+r+2s\right)\,,
\label{qsintermediate}
\end{equation}
and
\ba
c_s^2 &\simeq&
\frac{5+\Omega_r+12r^2+8r
+16s(2r+1)+16s^2}
{12(1+r+2s)^2} \nonumber \\
& &-\frac{2\beta r}{A(1+r+2s)}\,.
\label{csintermediateB}
\ea
The last term in Eq.~\eqref{csintermediateB} arises from $q_c\Omega_c=\Omega_c+4\beta x_1^2$ in 
Eq.~\eqref{hatcsX}.  
By contrast, the explicit contribution 
$\Delta c_s^2$ in Eq.~\eqref{csX} is of higher 
order under the condition 
$|\beta x_1^2|\ll\Omega_c$ in 
Eq.~\eqref{interhierarchy}.

The evolution equations for $r$ and 
$s$ follow from
$r'/r=2x_1'/x_1-x_3'/x_3=-\epsilon_\phi-h$ and 
$s'/s=x_2'/x_2-x_3'/x_3=\epsilon_\phi-h$.  Substituting Eq.~\eqref{interepsh} gives, at leading order,
\begin{align}
r'
&\simeq
r\frac{(9+\Omega_r)(1+2r)+4s(5+\Omega_r)}
{4(1+r+2s)}\,,
\label{rprime}\\
s'
&\simeq
s \frac{3(1-2r)+\Omega_r(3+2r)+4s(1+\Omega_r)}
{4(1+r+2s)}\,.
\label{sprime}
\end{align}
For $s=0$, Eq.~\eqref{rprime} reduces to
\begin{equation}
r'
\simeq
\frac{9+\Omega_r}{4}\,
\frac{r(1+2r)}{1+r}\,.
\label{rprimes0}
\end{equation}
Thus, the covariant-Galileon tracker
appears as the fixed point
\begin{equation}
r=-\frac12\,,
\label{trackerR}
\end{equation}
which is equivalent to $x_3=-2Ax_1^2$.  Since $x_3>0$, 
the existence of the tracker requires $A<0$.  
For approximately constant $\Omega_r$, 
Eq.~\eqref{rprimes0} can be integrated as
\begin{equation}
\frac{r^2}{1+2r}=c_{\rm tr} a^{(9+\Omega_r)/2},
\label{rimplicit}
\end{equation}
where $c_{\rm tr}$ is a positive integration constant.  Equivalently, 
choosing the branch that starts from $r\simeq0^-$ and approaches the tracker, we obtain
\begin{equation}
r=-\left[1+\sqrt{1+c_0 a^{-(9+\Omega_r)/2}}
\right]^{-1}\,,
\label{rsolution}
\end{equation}
where $c_0=1/c_{\rm tr}$.
During radiation and matter domination, Eq.~\eqref{rsolution} 
reduces, respectively, to
$r=-[1+\sqrt{1+c_0a^{-5}}]^{-1}$ and
$r=-[1+\sqrt{1+c_0a^{-9/2}}]^{-1}$. In both epochs, $r$ 
evolves from $0^-$ toward the tracker value $-1/2$ as the 
scale factor increases. This evolution is consistent with the 
early-time hierarchy~\eqref{earlyhierarchy}, 
which applies only in the regime $|r|\ll1$.

Away from an exact fixed point, the $x_i$ do not obey 
universal power laws. When $r$ and $s$ vary slowly, however, 
the ratios $x_1^2/x_3=r/A$ and $x_2/x_3=s$ are approximately 
constant, implying that $x_1^2$, $x_2$, and $x_3$ approximately 
share the same scaling. In particular, near the 
covariant-Galileon tracker characterized by $s\simeq0$ and 
$r\simeq-1/2$, one has $x_3\simeq-2A x_1^2$. For approximately 
constant $\Omega_r$, this gives
$x_1\propto a^{3+\Omega_r}$ and
$x_3\propto a^{2(3+\Omega_r)}$. 
In this limit, the no-ghost coefficient approaches a finite 
value rather than retaining the power-law behavior found in 
the earlier high-redshift regime. Indeed,
$q_s\simeq2A(1+r+2s)/r$ reduces to $q_s\simeq-2A$ on the 
tracker.

Along the tracker, the leading-order contributions 
to $w_{\rm DE}$, $q_s$, and $c_s^2$ are obtained from Eqs.~\eqref{wDEintermediate}, 
\eqref{qsintermediate}, and \eqref{csintermediateB} 
by setting $s=0$ and $r=-1/2$.  
They are given by 
\begin{equation}
w_{\rm DE} \simeq
-2-\frac{\Omega_r}{3},\qquad
q_s\simeq -2A\,,\qquad
c_s^2 \simeq
\frac{4+\Omega_r}{3}
+\frac{2\beta}{A},
\label{trackerwqs}
\end{equation}
where we used $x_3=-2A x_1^2$ on the tracker.
During the radiation and matter eras, the tracker gives
$w_{\rm DE}=-7/3$ and $w_{\rm DE}=-2$, respectively.  
For $A<0$, the no-ghost condition $q_s>0$ 
is satisfied on the tracker.  
The scalar perturbation is free from 
Laplacian instabilities on the tracker provided that 
$(4+\Omega_r)/3+2\beta/A>0$.  
For $\beta>0$, this condition is satisfied 
if $2\beta/|A| \ll 1$.

The intermediate formula \eqref{wDEintermediate} 
also explains why the solution does not necessarily reach the exact tracker value 
$w_{\rm DE}=-2$ before approaching the late-time de Sitter attractor.  
In the matter era, $\Omega_r\simeq0$, and for $s \ll 1$, one has
\begin{equation}
w_{\rm DE}
\simeq
\frac{1+8r+4r^2}{4(1+r)^2}
+\frac{3-32r-8r^2}{12(1+r)^3}s
+{\cal O}(s^2)\,.
\label{wDEmatterrs}
\end{equation}
For $s=0$, the first phantom-divide crossing during matter 
domination, from $w_{\rm DE}>-1$ to $w_{\rm DE}<-1$, 
occurs at
\begin{equation}
r=-1+\frac{\sqrt6}{4}\,.
\label{phantomcrossr}
\end{equation}
By contrast, the covariant-Galileon tracker value 
$w_{\rm DE}=-2$ is attained only at $r=-1/2$. 
Hence, the DE equation of state in the range 
\begin{equation}
-2<w_{\rm DE}<-1
\end{equation}
can be realized during the matter era for
\begin{equation}
-\frac12<r<-1+\frac{\sqrt6}{4}\,.
\label{phantomrangeR}
\end{equation}
The initial value of $r$ determines how closely the solution 
approaches the tracker before the late-time dynamics becomes 
important. The solutions of interest start from a negative 
value of $r$ close to zero, after which $r$ decreases without 
changing sign. Since $x_1^2>0$ and $x_3>0$, the relation 
$r=Ax_1^2/x_3<0$ requires
\be
A<0\,.
\label{Asign}
\ee
A nonzero value of $s=x_2/x_3$ describes an additional 
departure from the pure covariant-Galileon trajectory. 
Expanding Eq.~\eqref{wDEmatterrs} to first order in $s$, the 
coefficient of the linear correction is
\be
w_s \equiv \frac{3-32r-8r^2}{12(1+r)^3}\,,
\ee
which is positive throughout the interval~\eqref{phantomrangeR}. 
For example, at $r=-0.45$, the DE equation of state is 
approximately
$w_{\rm DE}=-1.48+7.90s+{\cal O}(s^2)$. Thus, a positive $s$ 
increases $w_{\rm DE}$ relative to the pure covariant-Galileon 
trajectory with $s=0$. This upward shift can prevent the 
solution from reaching the exact tracker value $w_{\rm DE}=-2$ 
and favors a minimum in the shallow-phantom range
$-2<w_{\rm DE}<-1$. By contrast, a negative $s$ decreases 
$w_{\rm DE}$ and drives it further below $-1$.

For $x_3>0$, the condition $s>0$ is equivalent to
\be
x_2>0\,.
\label{x2con}
\ee
This parameter region is therefore preferred for realizing a 
shallow phantom regime before the solution approaches the 
late-time attractor. The choice $s>0$ is also favored by the 
no-ghost condition. Indeed, in the intermediate regime,
$q_s\simeq2A(1+r+2s)/r$, and a positive $s$ increases the 
factor $1+r+2s$, thereby helping to maintain $q_s>0$ for 
$A<0$ and $r<0$. In terms of the original model coefficient, 
$x_2>0$ corresponds to $a_2>0$ for $\dot\phi\neq 0$.

A positive value of $s$ is also compatible with the 
Laplacian stability condition, as can be seen from 
Eq.~\eqref{csintermediateB}. In the interval
$-1/2<r<-1+\sqrt6/4$, which corresponds to
$-2<w_{\rm DE}<-1$ for $s=0$ during matter domination, one 
has $2r+1>0$. The $s$-independent part of the numerator of 
the first term in Eq.~\eqref{csintermediateB} is positive 
throughout this interval, while the terms proportional to 
$s$ and $s^2$ provide additional positive contributions for 
$s>0$. Moreover, $1+r+2s>0$ automatically holds in this 
interval for $s>0$. The first term in 
Eq.~\eqref{csintermediateB} is therefore positive.
The last term in Eq.~\eqref{csintermediateB} represents the 
correction induced by the CDM coupling. Since $r/A>0$ for 
$A<0$ and $r<0$, this term decreases $c_s^2$ when $\beta>0$. 
The Laplacian stability condition $c_s^2>0$ can thus be 
satisfied for $s>0$, provided that $\beta/|A|\ll1$. 
For $\beta\leq0$, the coupling term is nonnegative and 
hence does not induce a Laplacian instability, although the 
CDM no-ghost condition $q_c>0$ must still be imposed.

\subsection{Low-redshift behavior}
\label{lowsec}

At lower redshifts, $x_4$ generally becomes non-negligible, 
and the approximation $x_4=0$ employed in the high- and 
intermediate-redshift analytic regimes breaks down. The 
phantom-divide crossing must therefore be analyzed using the 
full background equations, without assuming either $x_3\ll1$ 
or $x_4=0$. Provided that
$\Omega_{\rm DE}=Ax_1^2+x_2+x_3+x_4>0$, Eq.~\eqref{wDE} 
shows that the condition $w_{\rm DE}>-1$ 
is equivalent to
\begin{equation}
6Ax_1^2+4x_2+x_3(3-\epsilon_\phi)>0\,.
\label{wgtminusonecond}
\end{equation}
The explicit $x_4$ terms in Eq.~\eqref{wDE} cancel in this inequality. Nevertheless, $x_4$ affects the 
phantom-divide crossing through its contribution to
$\epsilon_\phi$.  Substituting Eq.~\eqref{epssol} into
Eq.~\eqref{wgtminusonecond}, we obtain the exact condition
\begin{equation}
{\cal C} \equiv {\cal C}_0+{\cal B} x_4>0\,,
\label{Cdef}
\end{equation}
where
\begin{align}
{\cal C}_0
&={\cal D}(6Ax_1^2+4x_2+3x_3)
\nonumber\\
&\quad
-x_3\left[
 x_3(3Ax_1^2+x_2+\Omega_r-3)
 -4(3Ax_1^2+2x_2)
 \right] \nonumber \\
&=x_3^2 \bigl[15-\Omega_r+
24r^2+64rs+48r+32s^2+48s
\nonumber\\
&\quad+3x_3(1+r+s) \bigr]\,,
\label{C0def} \\
{\cal B}&=x_3(3x_3-2\sqrt6\lambda x_1)\,.
\label{Bdef}
\end{align}
Here $r$ and $s$ are defined in Eq.~\eqref{rsdef}.  In what follows, we
focus on the parameter region
\be
x_4>0\,,
\ee
which ensures the positivity of the scalar potential.  We also assume
\be
\lambda x_1>0\,.
\label{lamcon}
\ee
As shown below, this condition guarantees the dynamical stability of the
future de Sitter point.

The phantom-divide crossing occurs at ${\cal C}=0$, with
${\cal C}<0$ corresponding to $w_{\rm DE}<-1$ and
${\cal C}>0$ to $w_{\rm DE}>-1$. 
For fixed instantaneous values of $(x_1,x_2,x_3,\Omega_r)$ and 
${\cal B} \neq 0$, the critical value of $x_4$ at the crossing is
\begin{equation}
x_{4,c}=-\frac{{\cal C}_0}{{\cal B}}
=-\frac{{\cal C}_0}{x_3(3x_3-2\sqrt6\lambda x_1)}\,.
\label{x4critical}
\end{equation}
A positive value of $x_{4,c}$ requires ${\cal C}_0$ and 
${\cal B}$ to have opposite signs. If ${\cal C}_0<0$ and 
${\cal B}>0$, one has $w_{\rm DE}<-1$ for 
$0<x_4<x_{4,c}$ and $w_{\rm DE}>-1$ for $x_4>x_{4,c}$. 
In this case, the positive potential contribution increases 
${\cal C}$ and drives a solution that would otherwise lie in 
the phantom regime across the divide into the region 
$w_{\rm DE}>-1$. If instead ${\cal C}_0>0$ and 
${\cal B}<0$, the two regions are interchanged: 
$0<x_4<x_{4,c}$ corresponds to $w_{\rm DE}>-1$, whereas 
$x_4>x_{4,c}$ corresponds to $w_{\rm DE}<-1$. Therefore, the 
condition $3x_3-2\sqrt6\lambda x_1>0$ is not generally 
required for a phantom-divide crossing. The relevant criterion 
is the sign of the full quantity ${\cal C}$ in 
Eq.~\eqref{Cdef}, which must be evaluated along the dynamical 
trajectory because ${\cal C}_0$, ${\cal B}$, and $x_4$ all 
evolve in time.

In the deep high-redshift regime, where $|r|\ll1$, 
$|s|\ll1$, and $x_3\ll1$, Eq.~\eqref{C0def} reduces to
\begin{equation}
{\cal C}_0\simeq(15-\Omega_r)x_3^2>0\,.
\label{C0early}
\end{equation}
The numerical solutions presented later in 
Sec.~\ref{numesec} also 
satisfy $3x_3\gg2\sqrt6\lambda x_1$ during the initial stage 
of their evolution. Equation~\eqref{Bdef} then gives
\begin{equation}
{\cal B}\simeq3x_3^2>0\,.
\label{Bearlypositive}
\end{equation}
Since the potential contribution $x_4$ is extremely small in 
this regime, $|{\cal B}x_4|\ll{\cal C}_0$, and hence
${\cal C}\simeq{\cal C}_0>0$. The solutions therefore 
initially lie on the $w_{\rm DE}>-1$ side of the divide, in 
agreement with the high-redshift estimate derived in 
Sec.~\ref{highsec}.

As the universe evolves toward lower redshifts, the terms 
involving $r$ and $s$ in Eq.~\eqref{C0def} can no longer be 
neglected. Moreover, the ratio 
$2\sqrt6\lambda x_1/(3x_3)$ increases, causing ${\cal B}$ to 
change from positive to negative before the first 
phantom-divide crossing. Around this crossing, at $z=z_1$, 
the numerical solutions typically satisfy
$2\sqrt6\lambda x_1\gg3x_3$, so that
\begin{equation}
{\cal B}\simeq-2\sqrt6\lambda x_1x_3<0\,.
\label{Bnearz1}
\end{equation}
The high-redshift estimate~\eqref{C0early} is no longer 
sufficient to evaluate ${\cal C}_0$ around $z=z_1$, because 
the $r$- and $s$-dependent terms in Eq.~\eqref{C0def} can 
become important. For the numerical solutions presented later 
in Sec.~\ref{numesec}, however, ${\cal C}_0$ 
evaluated from the 
full expression~\eqref{C0def} remains positive around $z=z_1$. 
The first crossing from $w_{\rm DE}>-1$ to $w_{\rm DE}<-1$ 
then occurs when the negative contribution ${\cal B}x_4$ 
increases in magnitude until
$|{\cal B}x_4|={\cal C}_0$. Immediately after the crossing,
$|{\cal B}x_4|>{\cal C}_0$, giving ${\cal C}<0$ and driving 
the system into the phantom regime.

The second phantom-divide crossing at $z=z_c$ can occur when ${\cal C}_0$ grows again at low redshifts and overtakes 
$|{\cal B}x_4|$.  Since ${\cal B}x_4<0$ around this crossing, the transition from $w_{\rm DE}<-1$ to $w_{\rm DE}>-1$ is realized when ${\cal C}={\cal C}_0+{\cal B}x_4$ changes from negative to positive.  This increase of ${\cal C}_0$ is not a direct contribution of $x_4$ to ${\cal C}$, since ${\cal B}x_4$ remains negative near $z=z_c$. 
Rather, it is induced indirectly by the change of the background trajectory caused by the scalar potential, through the evolution of $x_1$, $x_2$, and $x_3$.  If $x_4=0$, corresponding to the absence of the potential contribution, numerical integration shows that ${\cal C}_0$ remains negative after the first phantom-divide crossing and that the solution stays in the region $w_{\rm DE}<-1$. The restoration of ${\cal C}_0$ to positive values for $x_4\neq0$ is therefore essential for the second crossing from $w_{\rm DE}<-1$ to $w_{\rm DE}>-1$, as illustrated by the numerical solutions in Sec.~\ref{numesec}.

In the asymptotic future, two fixed points can in principle be relevant to cosmic acceleration. 
The first is the ordinary quintessence point,
given by
\begin{equation}
\begin{aligned}
x_{1,Q}&=\frac{\lambda}{\sqrt{6}A},
\qquad
x_{4,Q}=1-\frac{\lambda^2}{6A},\\
x_{2,Q}&=x_{3,Q}=0,\\
\Omega_{c,Q}&=\Omega_{b,Q}=\Omega_{r,Q}=0\,.
\end{aligned}
\label{Qpoint}
\end{equation}
For $A=1$, this reduces to the standard quintessence point
$x_{1,Q}=\lambda/\sqrt6$ and $x_{4,Q}=1-\lambda^2/6$
~\cite{Copeland:1997et}.
At this fixed point,
\begin{equation}
w_{\rm DE}=w_{\rm eff}
=-1+\frac{\lambda^2}{3A}\,,
\label{wQpoint}
\end{equation}
where
$w_{\rm eff}\equiv-1-2\dot H/(3H^2)=-1-2h/3$ is the effective
equation-of-state parameter governing the total cosmological 
expansion. For $A>0$, this fixed point gives 
rise to accelerated expansion 
when $\lambda^2<2A$. Linearizing the 
autonomous system~\eqref{autonomous}
around this point, we 
obtain the eigenvalues
\begin{equation}
\mu_Q=\left\{
\begin{array}{c}
-3+\lambda^2/A,\quad
-3+\lambda^2/(2A),\\[1mm]
-\lambda^2/A,\quad
-\lambda^2/A,\\[1mm]
-3+\lambda^2/A,\quad
-4+\lambda^2/A
\end{array}
\right\}\,.
\label{eigQpoint}
\end{equation}
For $A>0$, all the eigenvalues are negative when 
$\lambda^2<3A$. Thus, on the ordinary quintessence branch, 
this fixed point is an accelerated attractor for 
$\lambda^2<2A$. By contrast, the Galileon branch relevant to 
the present solutions has $A<0$, as required by 
Eq.~\eqref{Asign}. At the fixed point~\eqref{Qpoint}, the 
scalar no-ghost coefficient is then $q_s=2A<0$, while the two 
eigenvalues $-\lambda^2/A$ are positive for $\lambda\neq0$. 
Therefore, on the $A<0$ Galileon branch, the fixed 
point~\eqref{Qpoint} suffers from a scalar ghost and is 
dynamically unstable.

The second accelerated fixed point corresponds to a de Sitter 
solution on the Galileon branch. It satisfies 
$\epsilon_\phi=h=0$, while the background variables obey
\begin{equation}
\begin{aligned}
A x_{1,{\rm dS}}^2
&=\frac{1}{2}x_{3,{\rm dS}}-2,\\
x_{2,{\rm dS}}
&=3-\frac{3}{2}x_{3,{\rm dS}},\\
x_{4,{\rm dS}}&=0,\\
\Omega_{c,{\rm dS}}
&=\Omega_{b,{\rm dS}}=\Omega_{r,{\rm dS}}=0\,.
\end{aligned}
\label{dSpointA}
\end{equation}
This branch satisfies $\Omega_{\rm DE}=1$ 
and $w_{\rm DE}=-1$. Although $x_4$ can be 
important for the low-redshift phantom-divide
crossing, the asymptotic Galileon de Sitter branch itself has
$x_{4,{\rm dS}}=0$.  For an exponential potential, this corresponds to the asymptotic limit in which the potential contribution becomes
negligible along the de Sitter branch.  
The eigenvalues of homogeneous
perturbations around the fixed point \eqref{dSpointA} are
\begin{equation}
\mu_{\rm dS}=\left\{
\begin{array}{c}
0,\quad -3,\quad -3,\\[1mm]
-3,\quad -4,\quad
-\sqrt6\lambda x_{1,{\rm dS}}
\end{array}
\right\}\,.
\label{eigdS}
\end{equation}
The zero eigenvalue reflects the fact that Eq.~\eqref{dSpointA}
represents a continuous line of de Sitter points.  
The remaining eigenvalues are negative for
\begin{equation}
\lambda x_{1,{\rm dS}}>0\,.
\label{dSdynstab}
\end{equation}
Thus, the de Sitter branch is dynamically stable in the directions
transverse to the fixed-point line.  For $A<0$, the first of
Eqs.~\eqref{dSpointA} admits real solutions for
$x_{3,{\rm dS}}<4$.  The scalar no-ghost coefficient at this de Sitter
point is
\begin{equation}
q_s |_{\rm dS}=
\frac{A [(x_{3,{\rm dS}}-3)^2+7]}
{x_{3,{\rm dS}}-4}\,,
\label{qsdS}
\end{equation}
which is positive for $A<0$ and 
$x_{3,{\rm dS}}<4$.  Although $q_c$
itself becomes singular in the limit $\Omega_c\to0$, this does not lead
to a singularity in $c_s^2$.  Indeed, after substituting the de Sitter
fixed-point values into Eq.~\eqref{csX}, all $\beta$-dependent
contributions cancel, and the scalar sound speed remains finite.  The
resulting squared sound speed is
\begin{equation}
c_s^2 |_{\rm dS}=
\frac{x_{3,{\rm dS}}(2-x_{3,{\rm dS}})}
{3[(x_{3,{\rm dS}}-3)^2+7]}\,.
\label{csdS}
\end{equation}
The absence of Laplacian instabilities therefore requires
\be
0<x_{3,{\rm dS}}<2\,.
\ee
Under this condition, $x_{2,{\rm dS}}>0$.
Combining the dynamical condition \eqref{dSdynstab} with the no-ghost and Laplacian stability conditions, the stable late-time attractor of
the $A<0$ Galileon branch is selected as the de Sitter branch
\eqref{dSpointA}, rather than the ordinary quintessence point
\eqref{Qpoint}.  This is consistent with Eq.~\eqref{x2con}, since the solutions considered below keep $x_2$ positive throughout the cosmological evolution.

\subsection{Numerical solutions}
\label{numesec}

We now present three representative background solutions that 
interpolate between the analytic regimes discussed in 
Secs.~\ref{highsec}--\ref{lowsec}. The preceding analysis 
identifies the following viable branch:
\ba
& &
x_2>0,\qquad x_3>0,\qquad 
\lambda x_1>0,\qquad x_4>0,\nonumber \\
& &
A<0\,.
\ea

\begin{figure*}[t]
\centering
\includegraphics[width=0.98\textwidth]{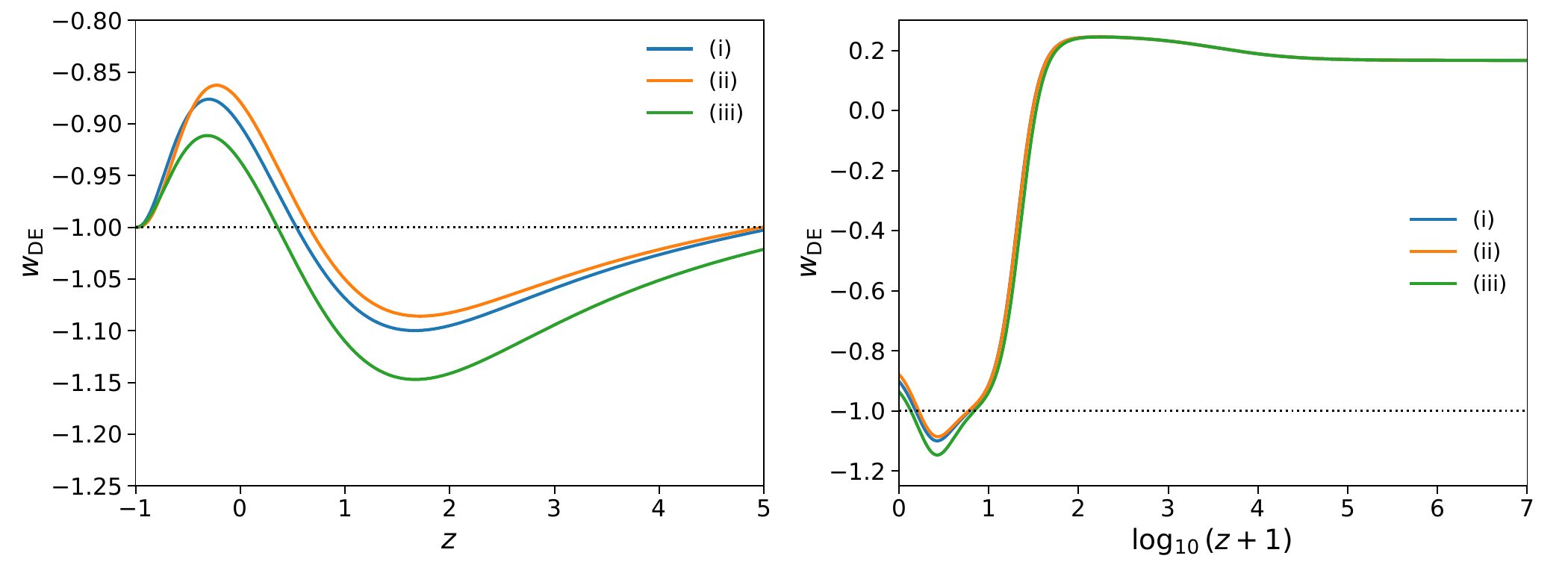}
\caption{Evolution of $w_{\rm DE}$, defined in Eq.~\eqref{wDE}, 
for the three representative cases (i), (ii), and
(iii) obtained by numerically integrating the 
background equations. The left panel shows 
$w_{\rm DE}$ as a function of $z=1/a-1$ over
$-1\leq z\leq 5$. It displays the low-redshift upward crossing from $w_{\rm DE}<-1$ to $w_{\rm DE}>-1$, 
followed by the approach to the de Sitter value 
$w_{\rm DE}=-1$ from above as $z\to -1$. 
The endpoint $z=-1$ corresponds to the asymptotic 
future limit. The right panel shows the same quantity over the range
$0\leq\log_{10}(z+1)\leq7$, highlighting its behavior at high 
and intermediate redshifts.
The common parameters are $a_1=-1$, 
$\Omega_{c0}=0.27$, $\Omega_{b0}=0.05$, and
$\Omega_{r0}=9.0\times10^{-5}$, while the three 
cases are defined in Eq.~\eqref{stableCasesTableBackground}.  The low-redshift upward crossings occur at $z_c\simeq0.5347$, $0.6572$, and $0.3556$, and the
minimum values are $w_{{\rm DE},{\rm min}}\simeq-1.100$, $-1.086$, and $-1.147$ at redshifts $z=1.6647$, $z=1.7186$, and $z=1.6714$ for cases (i), (ii), and (iii), respectively.}
 \label{fig:fig1}
\end{figure*}

We choose representative solutions 
with the present-day 
density parameters fixed to
\be
\Omega_{c0}=0.27,\qquad \Omega_{b0}=0.05,\qquad 
\Omega_{r0}=9.0\times10^{-5}\,,
\label{LCDMparams}
\ee
together with
\be
a_1=-1\,.
\ee
We then verify numerically that these solutions undergo two 
phantom-divide crossings while remaining free from ghost and 
Laplacian instabilities throughout their evolution, including 
the transition between the analytic regimes. The 
high-redshift evolution is obtained by integrating the 
background equations backward in time from the present epoch, 
$z=0$, to $z=10^7$. For the left panel of 
Fig.~\ref{fig:fig1}, we also integrate the equations forward 
in time to $N=12$ ($z\simeq-0.999994$), by which time the 
solutions have numerically approached the de Sitter fixed 
line. The plotted endpoint $z=-1$ represents the asymptotic 
future limit.

\begin{figure}[t]
\centering
\includegraphics[width=\columnwidth]{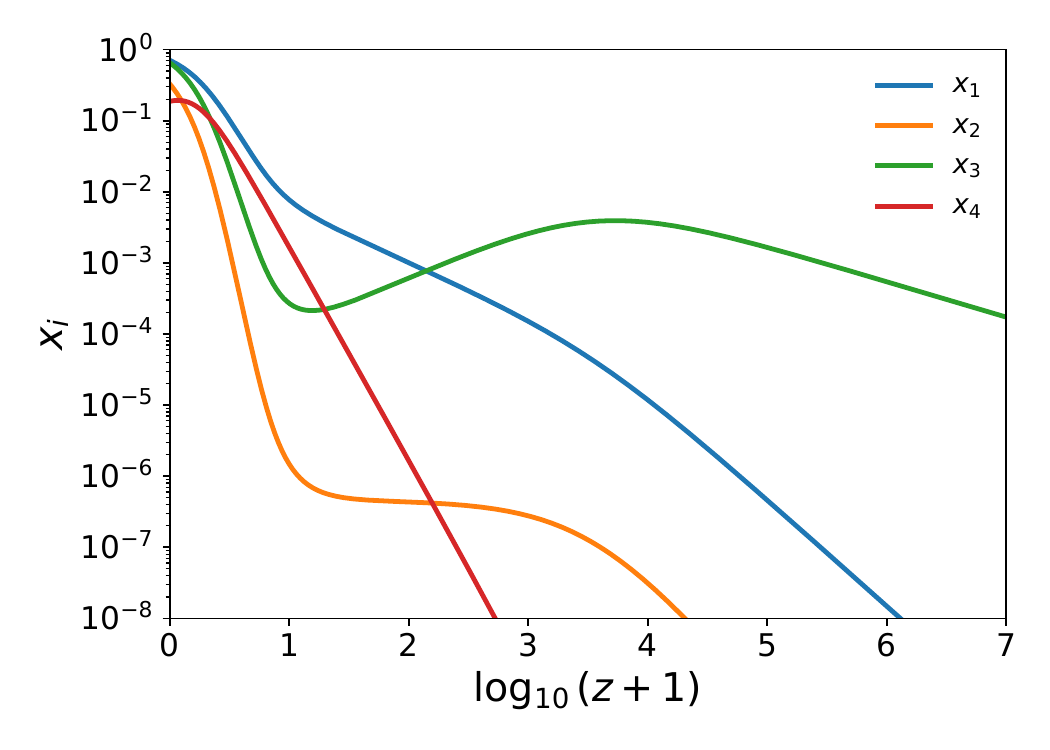}
\caption{Evolution of the background variables $x_1$, $x_2$, $x_3$, and $x_4$, defined in Eq.~\eqref{xdefs}, for case (i) of Fig.~\ref{fig:fig1}.
The horizontal axis is $\log_{10}(z+1)$ over
$0\leq \log_{10}(z+1)\leq 7$, while the vertical axis is logarithmic. 
The solution shown corresponds to
$\beta=1.1077541\times10^{-2}$ and
$x_3^{\rm peak}\simeq3.873008\times10^{-3}$, where
$x_3^{\rm peak}$ denotes the maximum value attained by $x_3$
during the transient braiding epoch. 
At early times, the high-redshift hierarchy $\{|A|x_1^2,x_2,x_4\}\ll x_3\ll1$ is realized.
Toward low redshifts, the growth of $x_4$ drives the departure from the shift-symmetric Galileon trajectory and enables the upward phantom-divide crossing.}
 \label{fig:fig2}
\end{figure}

\begin{figure}[t]
\centering
\includegraphics[width=\columnwidth]{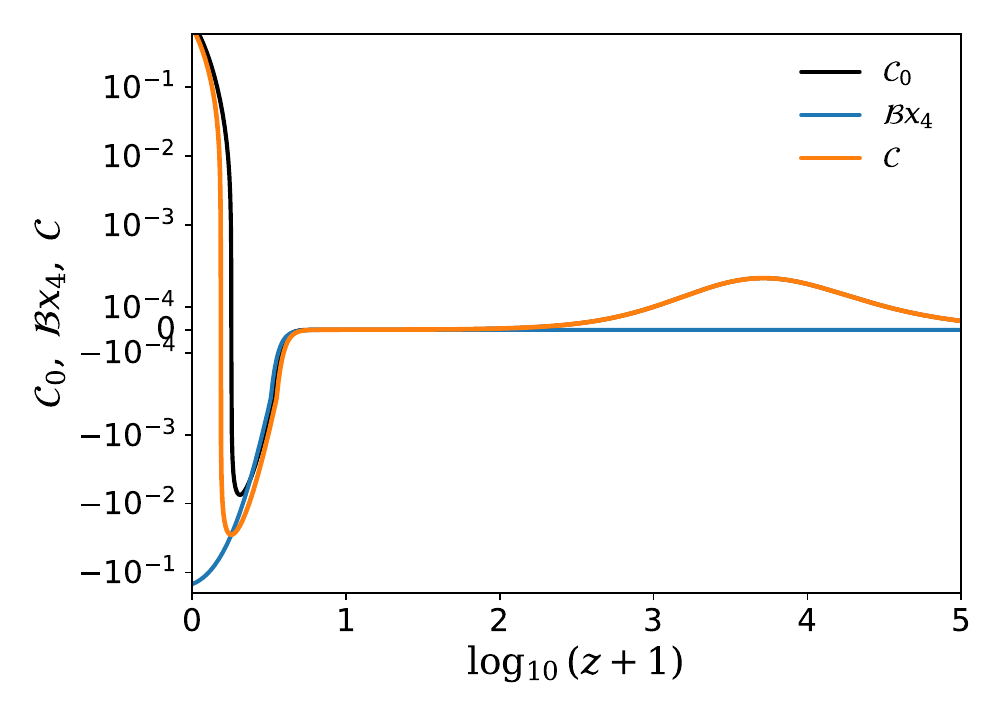}
\caption{Evolution of the diagnostic quantities
${\cal C}_0$, ${\cal B}x_4$, and
${\cal C}={\cal C}_0+{\cal B}x_4$, defined in
Eqs.~\eqref{Cdef}--\eqref{Bdef}, for case (i) of
Fig.~\ref{fig:fig1}, over the range
$0\leq\log_{10}(z+1)\leq5$.
The zeros of ${\cal C}$ determine where
$w_{\rm DE}$ crosses the phantom divide.
For the solution shown here, the low-redshift upward crossing occurs
at $z_c\simeq0.5347$, while the earlier crossing takes place at a
higher redshift. The figure shows that the transition is controlled by the full
combination ${\cal C}$ rather than by the sign of ${\cal B}$ alone.}
\label{fig:fig3}
\end{figure}

We consider the following three representative cases:
\be
\begin{aligned}
&\begin{array}{c|cc}
 & \beta & \lambda \\
\hline
{\rm (i)} & 1.1077541\times10^{-2} & 0.9148 \\
{\rm (ii)} & 1.9617328\times10^{-2} & 1.1140892 \\
{\rm (iii)} & 3.6001860\times10^{-2} & 0.8149903
\end{array}
\\[1.5mm]
&\begin{array}{c|cccc}
 & x_{1,0} & x_{2,0} & x_{3,0} & x_{4,0} \\
\hline
{\rm (i)} & 0.708068 & 0.328828 & 0.654177 & 0.187158 \\
{\rm (ii)} & 0.720630 & 0.324659 & 0.694523 & 0.159660 \\
{\rm (iii)} & 0.778347 & 0.377507 & 0.711101 & 0.153504\,,
\end{array}
\end{aligned}
\label{stableCasesTableBackground}
\ee
where $x_{i,0}$ denotes the present-day value of $x_i$, with
$i=1,2,3,4$.
In all three cases, $A=a_1+2\beta<0$.

Figure~\ref{fig:fig1} displays the full temporal sequence anticipated
from the analytic discussion in Secs.~\ref{highsec}--\ref{lowsec}, while
Fig.~\ref{fig:fig2} shows the corresponding evolution of the background
variables for case (i).  In the deep high-redshift regime, the hierarchy
$\{|A|x_1^2,x_2,x_4\}\ll x_3\ll1$ is realized.  In this regime, the
potential contribution $x_4$ is strongly suppressed, whereas the cubic
Galileon variable $x_3$ gives the leading contribution to the small DE
density.  The solutions therefore approach the radiation-era value
$w_{\rm DE}\simeq1/6$, in agreement with Eq.~\eqref{earlywcs}.  In the
same regime, ${\cal C}_0\simeq(15-\Omega_r)x_3^2>0$,
${\cal B}\simeq3x_3^2>0$, and $x_4$ is negligible, so that
${\cal C}\simeq{\cal C}_0>0$.

The scaling laws in Eq.~\eqref{earlyscaling} are visible in
Fig.~\ref{fig:fig2}.  Evolving forward from the radiation era, one has
$x_1\propto a^{3/2}$, $x_2\propto a^2$, $x_3\propto a^{1/2}$, and
$x_4\propto a^4$.  Since $x_4$ decreases much faster than $x_3$ toward
the past, the potential contribution becomes negligible at high redshift,
while the cubic Galileon contribution remains the dominant DE component.
After the universe enters the early matter era, the scaling of $x_3$
changes to $x_3\propto a^{-3/4}$.  Consequently, $x_3$ grows during
radiation domination, reaches a localized maximum around the
radiation--matter transition, and then decreases during the early matter
era.  For case (i), this maximum is
$x_3^{\rm peak}\simeq 3.873008\times10^{-3} $.  
We refer to this localized maximum as 
the transient braiding peak, whose impact on
large-scale perturbations will be analyzed in Sec.~\ref{sec:superHubble}.

Figure~\ref{fig:fig3} shows how the diagnostic quantities introduced in
Sec.~\ref{lowsec} behave along the case-(i) trajectory.  In the intermediate
regime, ${\cal B}=x_3(3x_3-2\sqrt6\lambda x_1)$ becomes negative, so the
potential contribution ${\cal B}x_4$ competes with the positive contribution
${\cal C}_0$.  The sign of the full combination
${\cal C}={\cal C}_0+{\cal B}x_4$ determines on which side of the
phantom divide the solution lies.  The first zero of ${\cal C}$ drives the
transition from $w_{\rm DE}>-1$ to $w_{\rm DE}<-1$, while the later zero
produces the low-redshift upward crossing at $z_c\simeq0.5347$.  Thus, the
potential is not merely a spectator: 
its growth at low redshifts changes the background trajectory and allows the system to leave the phantom regime.
Among the three examples, case (iii) reaches 
the most negative value of $w_{\rm DE}$, 
whereas case (ii) undergoes the upward crossing 
at the largest redshift.

\begin{figure}[t]
 \centering
 \includegraphics[width=\columnwidth]{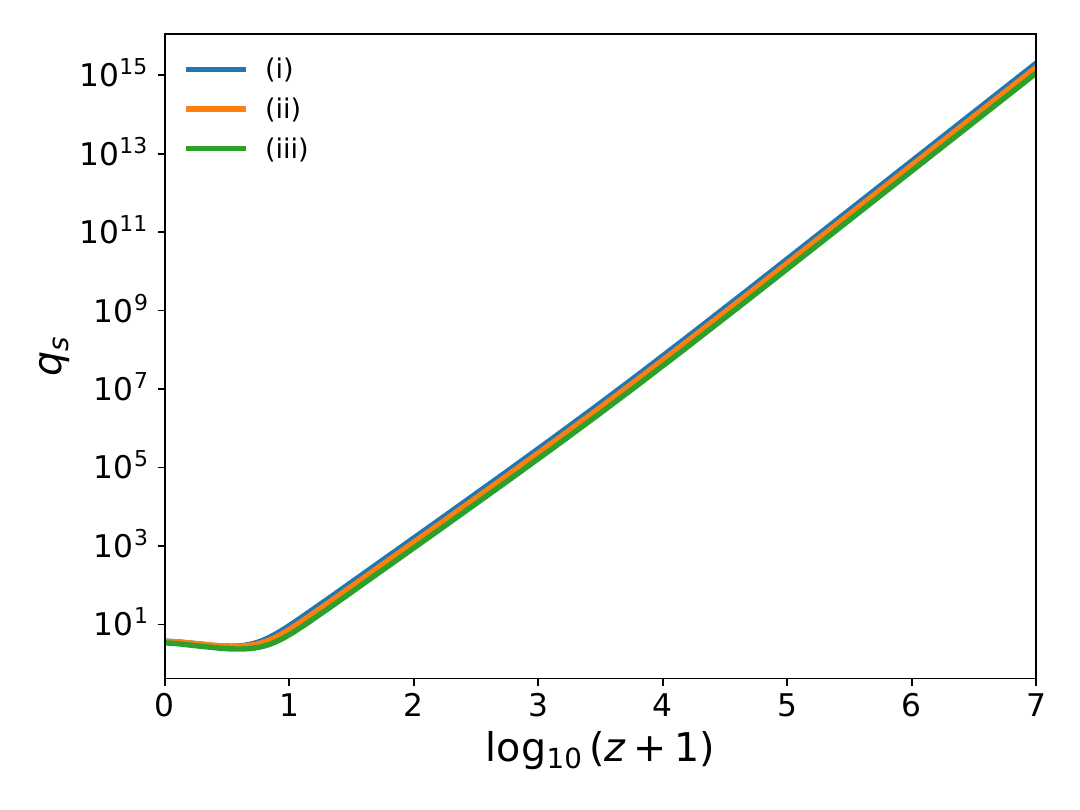}
 \caption{Evolution of the scalar-field no-ghost coefficient $q_s$, defined in
Eq.~\eqref{qsX}, for the three cases of Fig.~\ref{fig:fig1}, over the
range $0\leq\log_{10}(z+1)\leq7$. The vertical axis is logarithmic.
The coefficient remains positive throughout the plotted range for all
three trajectories, with minimum values
$q_{s,{\rm min}}\simeq2.680$, $2.778$, and $2.366$ for cases (i),
(ii), and (iii), respectively.}
 \label{fig:fig4}
\end{figure}

\begin{figure}[t]
\centering
\includegraphics[width=\columnwidth]{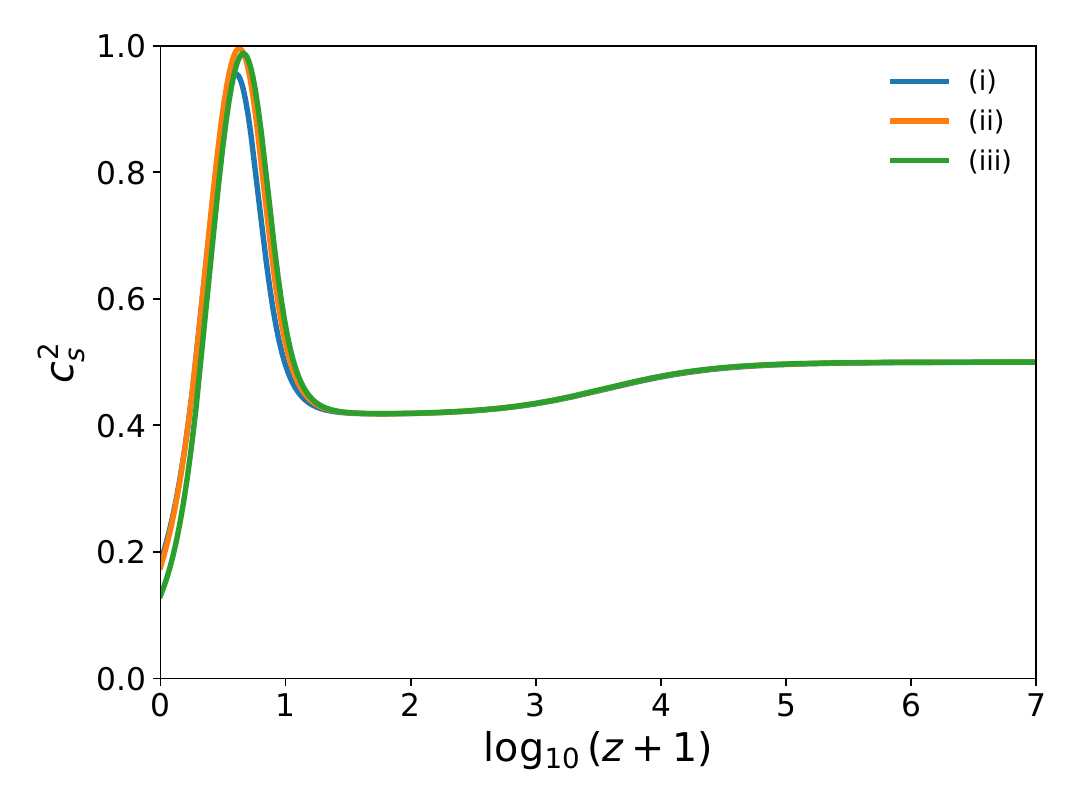}
\caption{Evolution of the squared scalar propagation speed $c_s^2$, defined in
Eq.~\eqref{csX}, for the three cases of Fig.~\ref{fig:fig1}, over the
range $0\leq\log_{10}(z+1)\leq7$. All three solutions satisfy
$0<c_s^2<1$ throughout the plotted range. The minimum values are
$c_{s,{\rm min}}^2\simeq0.176$, $0.173$, and $0.121$ for cases (i),
(ii), and (iii), respectively.}
 \label{fig:fig5}
\end{figure}

The future extension in the left panel of Fig.~\ref{fig:fig1}
confirms the attractor analysis of Sec.~\ref{lowsec}.  
After the upward crossing, all three solutions stay on the nonphantom side and
approach $w_{\rm DE}=-1$ from above as $z\to -1$.  Numerically, one
finds $x_4\to 0$, together with $h\to 0$ and $\epsilon_\phi\to 0$.
The asymptotic values of $x_3$ are approximately
$1.284$, $1.324$, and $1.291$ for cases (i), (ii), and (iii),
respectively, all of which lie in the stable de Sitter interval
$0<x_{3,{\rm dS}}<2$.  Thus, the potential controls the low-redshift exit
from the phantom regime, even though its dimensionless energy contribution
vanishes on the asymptotic Galileon de Sitter branch.

The no-ghost coefficient $q_s$ is shown in Fig.~\ref{fig:fig4}.
In the high-redshift regime the numerical curves follow the analytic
estimate $q_s\simeq 2x_3/x_1^2$ in Eq.~\eqref{earlywcs}, together with
the scaling laws in Eq.~\eqref{earlyscaling}.  During radiation
domination this gives $q_s\propto a^{-5/2}$, while in the early matter
era one has $q_s\propto a^{-9/4}$ as long as the hierarchy
\eqref{earlyhierarchy} remains valid.  
Hence, toward the asymptotic past $(z\to\infty)$, $q_s$ grows rather
than approaching zero.  Along the entire numerical evolution shown in
Fig.~\ref{fig:fig4}, it remains finite and positive and never approaches the
potentially strongly coupled limit $q_s\to0$.  Thus, all three examples are
free from scalar ghosts and avoid this limit.

Figure~\ref{fig:fig5} displays the squared scalar propagation speed
$c_s^2$. In the deep radiation era, the solutions approach
$c_s^2\simeq1/2$, in agreement with Eq.~\eqref{earlywcs}, while in the
early matter era they pass close to $c_s^2\simeq5/12$ before
intermediate-regime corrections become important. 
Numerical evaluation of Eq.~\eqref{csX} on the backgrounds shows that
$c_s^2$ remains positive in all three cases. Extending the integration
into the asymptotic future confirms that $c_s^2$ converges to the
de Sitter value given in Eq.~\eqref{csdS}, with
$c_s^2|_{\rm dS}\simeq0.0308$, $0.0304$, and $0.0308$ for cases (i),
(ii), and (iii), respectively.
Since $c_s^2$ remains positive and bounded away from the gradient-degenerate limit
$c_s^2\to0$, while $q_s$ remains positive, the backgrounds are free
from scalar ghost and Laplacian instabilities.

\section{Perturbation equations of motion}
\label{persec}

In this section, we derive the linear scalar perturbation equations for
the model~\eqref{action} in Newtonian gauge by specializing the gauge-ready
formulation of Ref.~\cite{Kase:2020hst} to the present theory. We then
consider the sub-Hubble limit and obtain the effective gravitational
couplings governing the growth of matter perturbations.

In the Newtonian gauge, the line element 
is written as
\be
{\rm d}s^2=-(1+2\Psi){\rm d}t^2+a^2(t) (1-2\Phi) \delta_{ij} 
{\rm d}x^i {\rm d}x^j\,,
\label{Newtonianmetric}
\ee
where $\Psi(t,x^i)$ is the Newtonian lapse potential and $\Phi(t,x^i)$ is the scalar curvature potential. 
With this metric convention, 
the curvature perturbation used here differs
by an overall minus sign from the quantity denoted by $\Phi$ in Ref.~\cite{Kase:2020hst}. 
This sign conversion has been applied in writing the equations below. 
We decompose the scalar field and the fluid energy densities into
background and first-order perturbations as
\be
\phi=\bar\phi(t)+\delta\phi(t,x^i)\,,
\qquad
\rho_I=\bar\rho_I(t)+\delta\rho_I(t,x^i)\,,
\ee
where $I=c,b,r$ labels CDM, baryons, and radiation, 
respectively. In what
follows, we drop the overbar from background 
quantities and define the density contrasts
\be
\delta_I\equiv \frac{\delta\rho_I}{\rho_I}\,,
\qquad I=c,b,r\,.
\label{deltadefN}
\ee
For each fluid, we write the components of 
the perturbed covariant four-velocity 
in terms of the scalar velocity potential $v_I$ as
\be
u_{I0}=-1-\Psi\,,\qquad u_{Ii}=-\partial_i v_I\,,
\ee
to first order in perturbations. 
Here $\partial_i\equiv\partial/\partial x^i$.

Following the notation of Ref.~\cite{Aoki:2025bmj}, 
we introduce the dimensionless effective-field-theory (EFT) functions
\ba
& &
\taK=6A x_1^2+12x_2+6x_3\,,\qquad
\alpha_B=-\frac{x_3}{2}\,,\nonumber \\
& &
\alpha_{m_2}=12\beta x_1^2\,.
\label{alphabetaModel}
\ea
Here, $\taK$, $\alpha_B$, and $\alpha_{m_2}$ characterize the scalar
kinetic sector, kinetic braiding, and the CDM--scalar momentum-transfer
operator, respectively.  The functions
$\alpha_K=6a_1 x_1^2+12x_2+6x_3$ and
$\beta_K=12\beta x_1^2$ used in Ref.~\cite{Kase:2020hst} are related to
the above quantities as
$\taK=\alpha_K+\beta_K$ and $\alpha_{m_2}=\beta_K$.  With these
definitions, the no-ghost coefficients are
\begin{equation}
q_s=\frac{\taK+6\alpha_B^2}{3x_1^2}\,,\qquad
q_c=1+\frac{\alpha_{m_2}}{3\Omega_c}\,.
\label{qsAlphaRelation}
\end{equation}
Thus, the scalar no-ghost condition $q_s>0$ is equivalent to
$\taK+6\alpha_B^2>0$, in agreement with Ref.~\cite{Aoki:2025bmj}.  
In the CDM sector, $\alpha_{m_2}>0$, or equivalently $\beta>0$, ensures
$q_c>1$ provided that $\Omega_c>0$.
We also use the shorthand 
\begin{equation}
\epsilon_Y\equiv \frac{\dot Y}{HY},
\qquad
Y\in\{\taK,\alpha_B,q_c,\Delta\}\,,
\label{epsYdef}
\end{equation}
where $\Delta$ will be defined below in Eq.~\eqref{Delta}.  
The intrinsic squared sound speeds are $c_b^2=c_c^2=0$ for baryons
and CDM and $c_r^2=1/3$ for radiation.

\subsection{Newtonian-gauge linear perturbation equations}

Using $\delta\rho_I=\rho_I\delta_I$ and
$\rho_I=3H^2\Mpl^2\Omega_I$, we write the linear perturbation
equations directly in terms of $\delta_I$ and $\Omega_I$.  In Fourier
space, $k\equiv |\bm{k}|$ denotes the comoving wavenumber of a
Fourier mode.  We also define
\be
\kappa_k\equiv \frac{k}{aH}\,.
\ee
The Hamiltonian constraint, momentum constraint,
scalar-field perturbation equation, and anisotropic-stress equation are
\begin{align}
&6(1+\alpha_B)\frac{\dot\Phi}{H}
 -(6\alpha_B-\taK)\frac{\dot{\delta\phi}}{\dot\phi}
 +2\kappa_k^2\Phi \nonumber\\
&+(6+12\alpha_B-\taK)\Psi
 +3(\Omega_c\delta_c+\Omega_b\delta_b+\Omega_r\delta_r)
 \nonumber\\
&
 -\biggl[
 2\alpha_B \kappa_k^2
 -6(1+\alpha_B)h
 -(6\alpha_B-\taK)\epsilon_\phi \nonumber\\
&-3(3\Omega_c+3\Omega_b+4\Omega_r)
\biggr]\frac{H\delta\phi}{\dot\phi}=0\,,
\label{HamNmodel}
\end{align}
\begin{align}
&\frac{\dot\Phi}{H}
 +(1+\alpha_B)\Psi
 -\alpha_B\frac{\dot{\delta\phi}}{\dot\phi}
 -\frac{3H}{2}q_c\Omega_c
 \left(v_c-\frac{\delta\phi}{\dot\phi}\right)
 \nonumber\\
&
 -\frac{3H}{2}\Omega_b
 \left(v_b-\frac{\delta\phi}{\dot\phi}\right)
 -2H\Omega_r
 \left(v_r-\frac{\delta\phi}{\dot\phi}\right)
 \nonumber\\
&+(h+\epsilon_\phi\alpha_B)
 \frac{H\delta\phi}{\dot\phi}=0\,,
\label{MomNmodel}
\end{align}
\begin{widetext}
\begin{align}
&\taK \frac{\ddot{\delta\phi}}{H\dot\phi}
 +\left[
 \epsilon_{\taK} \taK
 +(3+2h-2\epsilon_\phi)\taK
 \right]\frac{\dot{\delta\phi}}{\dot\phi}
  +\left[
 \kappa_k^2
 \left(2\alpha_B^2+3x_1^2q_s\hat c_s^2\right)
 +18\lambda^2x_1^2x_4
 \right]\frac{H\delta\phi}{\dot\phi}
  +3(1-q_c)\Omega_c\frac{\dot\delta_c}{H}
\nonumber\\
&+6\alpha_B\frac{\ddot\Phi}{H^2}
 +(6\alpha_B-\taK)\frac{\dot\Psi}{H}+3\left[
 2h+2(3+h+\epsilon_{\alpha_B})\alpha_B
 +3q_c\Omega_c+3\Omega_b+4\Omega_r
 \right]\frac{\dot\Phi}{H}
\nonumber\\
&
-\biggl[
 2\alpha_B\kappa_k^2
 + \epsilon_{\taK} \taK
 -6\epsilon_{\alpha_B}\alpha_B
 +(3+2h)(\taK-6\alpha_B)
 -6h(1+\alpha_B)
 -3(3\Omega_c+3\Omega_b+4\Omega_r)
 \biggr]\Psi=0\,,
\label{scalarNmodel}\\[1mm]
&\Psi=\Phi\,.
\label{anisomodel}
\end{align}
\end{widetext}
The fluid continuity equations are 
\begin{align}
&\dot\delta_c-3\dot\Phi+\kappa_k^2 H^2 v_c=0\,,
\label{contcNmodel}\\
&\dot\delta_b-3\dot\Phi+\kappa_k^2 H^2 v_b=0\,,
\label{contbNmodel}\\
&\dot\delta_r-4\dot\Phi+\frac{4}{3}\kappa_k^2 H^2 v_r=0\,,
\label{contrNmodel}
\end{align}
whereas the Euler equations are  
\begin{align}
&\dot v_c
 +H\epsilon_{q_c}v_c
 -\frac{\Psi}{q_c}
 +\frac{1-q_c}{q_c\dot\phi}
 \left(\dot{\delta\phi}-H\epsilon_\phi\delta\phi\right)
 \nonumber\\
&-\epsilon_{q_c}\frac{H\delta\phi}{\dot\phi}=0\,,
\label{EulerCDMNmodel}\\
&\dot v_b-\Psi=0\,,
\label{EulerbNmodel}\\
&\dot v_r-Hv_r-\frac{1}{4}\delta_r-\Psi=0\,.
\label{EulerrNmodel}
\end{align}
Equations~\eqref{HamNmodel}--\eqref{EulerrNmodel}, together with the
background equations in Sec.~\ref{sec:model}, close the Newtonian-gauge
scalar perturbation system in the perfect-fluid approximation.  In the full
CLASS calculation, the radiation-fluid equations are replaced by the photon
and neutrino Boltzmann hierarchies.  The traceless part of the spatial
Einstein equations then contains the photon and neutrino anisotropic
stresses, so Eq.~\eqref{anisomodel} no longer reduces to
$\Psi=\Phi$.

\subsection{Quasi-static effective gravitational couplings}

For large-scale-structure and redshift-space-distortion measurements, we
focus on modes satisfying $c_s^2\kappa_k^2\gg1$. As shown in
Fig.~\ref{fig:fig5}, the squared scalar propagation speed remains finite
and is not parametrically small for the representative backgrounds.
The scalar sound horizon is therefore only moderately smaller than the
Hubble horizon, so this condition effectively selects modes deep inside
the Hubble radius. For the exponential potential, the scalar mass
squared, $V_{,\phi\phi}=\lambda^2V/\Mpl^2$, is at most of order $H_0^2$
on the late-time backgrounds and is negligible compared with
$H^2\kappa_k^2=k^2/a^2$ on these scales.

We apply the quasi-static approximation
~\cite{Boisseau:2000pr,Tsujikawa:2007gd,DeFelice:2011hq} by retaining the
CDM and baryon density perturbations as sources and the leading
spatial-gradient terms of the metric and induced scalar-field
perturbations.  The free oscillating scalar mode and radiation
perturbations are neglected.  Equations~\eqref{HamNmodel} and
\eqref{scalarNmodel}, together with the anisotropic-stress relation
\eqref{anisomodel}, then form algebraic equations for $\Psi$ and
$\delta\phi$.  Solving these equations yields
\ba
& &
\Psi=\Phi \simeq 
-\frac{1}{2\Mpl^2 H^2 \kappa_k^2 \Delta}
\biggl[ (\alpha_B^2+\Delta)(\rho_c \delta_c
+\rho_b \delta_b) \nonumber \\
& &\qquad \qquad +(1-q_c) \alpha_B \rho_c 
\frac{\dot{\delta}_c}{H}
\biggr]\,,\label{PsiPhi}\\
& &
\delta \phi \simeq 
-\frac{\dot{\phi}}
{2\Mpl^2 H^3 \kappa_k^2 \Delta}
\biggl[ \alpha_B (\rho_c \delta_c
+\rho_b \delta_b) \nonumber \\
& &\qquad~+(1-q_c)  \rho_c 
\frac{\dot{\delta}_c}{H}
\biggr]\,,\label{deltaphi}
\ea
where 
\be
\Delta \equiv \frac{\dot{\phi}^2 q_s 
\hat{c}_s^2}{4H^2 \Mpl^2}
=\frac{3}{2}x_1^2 q_s \hat{c}_s^2\,.
\label{Delta}
\ee
Taking time derivatives of Eqs.~\eqref{contcNmodel} and \eqref{contbNmodel}, and then using Eqs.~\eqref{EulerCDMNmodel}, \eqref{EulerbNmodel}, \eqref{PsiPhi}, and \eqref{deltaphi}, 
we obtain the quasi-static growth equations 
for CDM and baryons,
\ba
& &
\ddot\delta_{c}+c_1 H\dot
\delta_{c}+c_2 H \dot\delta_{b}
-4\pi G_c \rho_m \delta_m=0\,,
\label{deltacQSmodel}\\
& &
\ddot\delta_{b}+2H\dot\delta_{b}
+c_3 \dot\delta_{c}
-4\pi G_b \rho_m \delta_m=0\,,
\label{deltabQSmodel}
\ea
where 
\be
\rho_m=\rho_c+\rho_b\,,\qquad
\delta_m=\frac{\rho_c}{\rho_m} \delta_c
+\frac{\rho_b}{\rho_m} \delta_b\,.
\label{delM}
\ee
The friction coefficients are given by 
\begin{align}
c_1={}&(2+\epsilon_{q_c})\frac{\hat c_s^2}{c_s^2}
+\biggl[
\frac{2\alpha_B-2q_c(\alpha_B+\epsilon_{q_c})}
{1-q_c} \nonumber \\
&-1-2\alpha_B-\epsilon_{\Delta}-2h
\biggr]\frac{\Delta c_s^2}{c_s^2},
\label{c1model}\\
c_2={}&\frac{3(1-q_c)\Omega_b\alpha_B}{2q_c\Delta}
\frac{\hat c_s^2}{c_s^2},
\label{c2model}\\
c_3={}&-\frac{3\alpha_B(1-q_c)\Omega_c}{2\Delta}H\,.
\label{c3model}
\end{align}
Here $\Delta c_s^2=c_s^2-\hat c_s^2$, and $\epsilon_Y=\dot Y/(HY)$ for $Y=\alpha_B,q_c,\Delta$.  
The effective gravitational couplings appearing in Eqs.~\eqref{deltacQSmodel} and \eqref{deltabQSmodel} are
\begin{align}
G_{c}
={}&\frac{G}{q_c}
\frac{\hat c_s^2}{c_s^2}
\left(
1+\frac{q_c\alpha_B^2}{\Delta}
+{\cal G}_{\rm mix} \right),
\label{Gccmodel}\\
G_{b}
={}& G\left( 1+\frac{\alpha_B^2}
{\Delta} \right) \,,
\label{Gbbmodel}
\end{align}
where 
\be
{\cal G}_{\rm mix}
\equiv
\frac{\alpha_B}{\Delta}
\left[ q_c\epsilon_{q_c}+(1-q_c)
(1+\alpha_B+h+\epsilon_\Delta-\epsilon_{\alpha_B})
\right],
\label{calG}
\ee
and $G=(8\pi \Mpl^2)^{-1}$ is the Newtonian gravitational constant.

In the absence of momentum transfer, $\beta=0$, 
one has $q_c=1$ and $c_s^2=\hat c_s^2$, so that
$G_c=G_b=G(1+\alpha_B^2/\Delta)$. 
Since $\Delta>0$ under the stability conditions 
$q_s>0$ and $\hat c_s^2>0$, the Galileon braiding term
$\alpha_B=-x_3/2$ enhances the gravitational couplings for both CDM and baryons relative to the $\Lambda$CDM value.

For $\beta>0$, the quantity $q_c=1+4\beta x_1^2/\Omega_c$ departs from unity 
and grows toward low redshifts, acting as an effective inertia for CDM velocity perturbations.
At the same time, the ratio $\hat c_s^2/c_s^2$ becomes smaller than unity because 
$\Delta c_s^2=c_s^2-\hat c_s^2$ is positive. 
These two effects reduce the prefactor $q_c^{-1}\hat c_s^2/c_s^2$ in $G_c$ and 
therefore tend to suppress the growth of the CDM density contrast
$\delta_c$.  Whether $G_c$ falls below $G$ 
is determined by the competition between this momentum-transfer suppression and the Galileon
braiding enhancement in the parentheses of Eq.~\eqref{Gccmodel}.

\begin{figure}[t]
 \centering
 \includegraphics[width=\columnwidth]{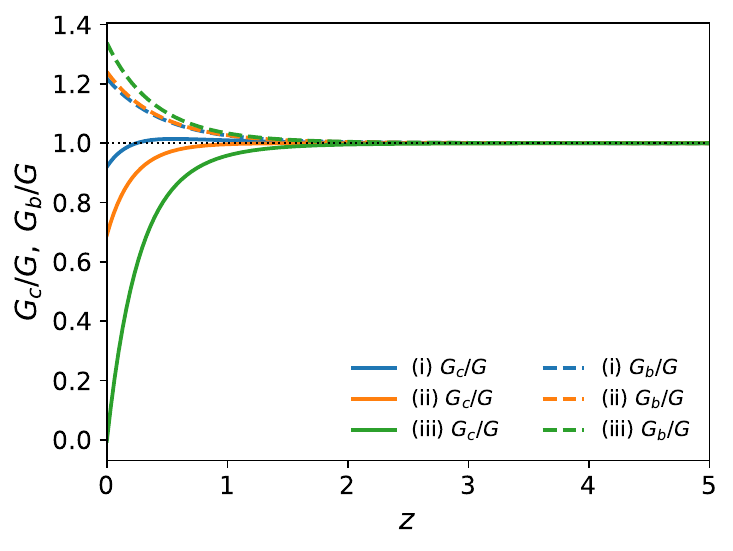}
 \caption{Evolution of the effective gravitational couplings $G_c/G$ and $G_b/G$,
defined in Eqs.~\eqref{Gccmodel} and \eqref{Gbbmodel}, for the three
background solutions of Fig.~\ref{fig:fig1} over $0\leq z\leq5$.
The solid and dashed curves represent $G_c/G$ and $G_b/G$, respectively.
At high redshifts, both couplings approach the general-relativistic value
of unity. Toward lower redshifts, $G_b/G$ is enhanced in all three cases,
with the enhancement increasing from case (i) to case (iii). By contrast,
$G_c/G$ is increasingly suppressed as the momentum-transfer coupling
becomes stronger. In case (i), $G_c/G$ undergoes a mild enhancement at
intermediate redshifts before decreasing below unity. The suppression is
stronger in cases (ii) and (iii), with $G_c/G$ approaching zero at the
present epoch in case (iii).}
\label{fig:fig6}
\end{figure}

Figure~\ref{fig:fig6} shows the two effective gravitational couplings
computed from Eqs.~\eqref{Gccmodel} and \eqref{Gbbmodel}.  At
$z\gtrsim2$, both couplings are close to $G$, since the braiding and
momentum-transfer corrections are small.  Evolving toward the present,
the baryonic coupling is enhanced in all three cases and departs
monotonically from unity, with the largest enhancement occurring in case
(iii).  The CDM coupling exhibits a qualitatively different behavior.  In
case (i), $G_c/G$ first develops a mild enhancement above unity at
intermediate redshift, but subsequently crosses below unity. 
In case (ii), $G_c/G$ is suppressed below unity at low redshifts, 
while in case (iii) it decreases rapidly toward the present epoch and becomes close to zero.  
Hence, the suppression of the effective CDM coupling
becomes progressively stronger as $\beta$ increases, while Galileon braiding continues to enhance the
baryonic coupling.

This behavior originates from the momentum-transfer terms that affect CDM
but not baryons.  In the CDM coupling, the inertia factor $q_c>1$, the ratio $\hat c_s^2/c_s^2<1$, and the negative
momentum-transfer mixing contribution ${\cal G}_{\rm mix}$ all work against the braiding enhancement.
These effects suppress the small-scale
growth of CDM perturbations as $\beta$ increases.  The impact of the transient $x_3$ peak on large-scale 
perturbations around radiation--matter
equality will be discussed separately in Sec.~\ref{sec:superHubble}.

\begin{figure}[t]
 \centering
 \includegraphics[width=\columnwidth]{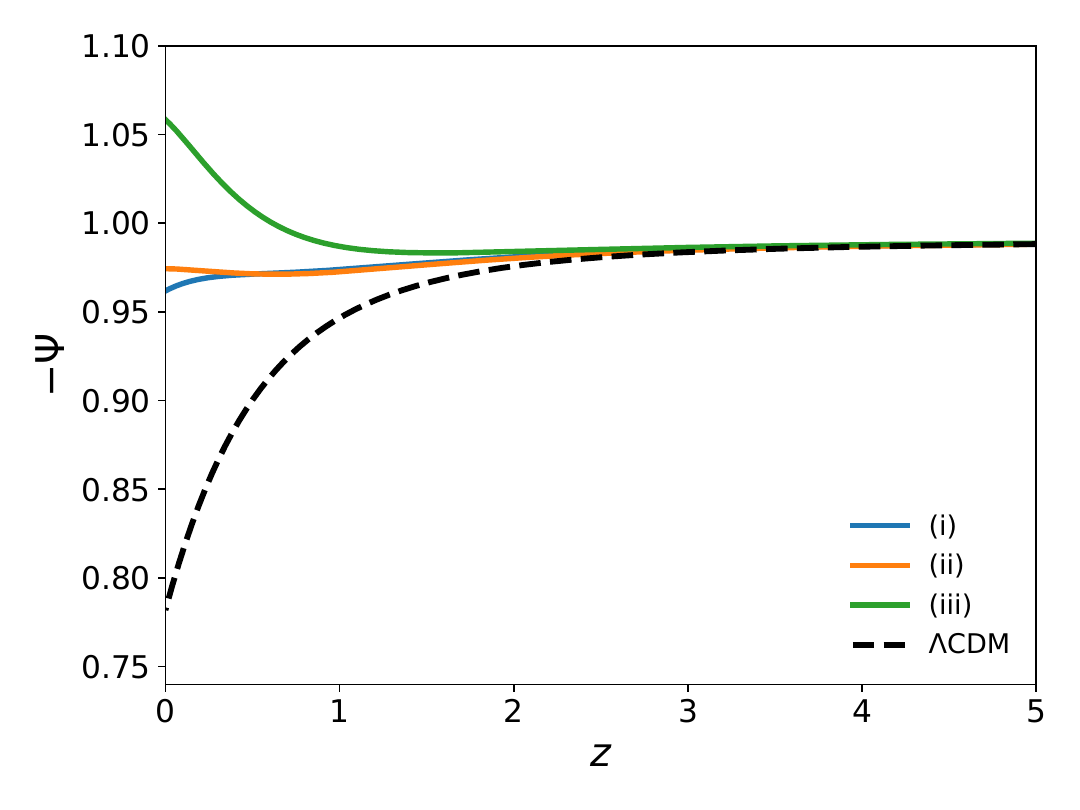}
 \caption{Evolution of the gravitational potential $-\Psi$ over $0\le z\le 5$
for the three cases of Fig.~\ref{fig:fig1} and $\Lambda$CDM.  The potential is obtained
from the quasi-static expression \eqref{PsiPhi} after solving
Eqs.~\eqref{deltacQSmodel} and \eqref{deltabQSmodel}, including the
velocity-mixing source term proportional to
$(1-q_c)\alpha_B\rho_c\dot\delta_c/H$.  It is normalized such that $-\Psi=1$ at $z=100$.  
The black dashed curve shows the corresponding
$\Lambda$CDM result with $\Omega_{c0}=0.27$ and $\Omega_{b0}=0.05$.}
 \label{fig:fig7}
\end{figure}

We next integrate Eqs.~\eqref{deltacQSmodel} and \eqref{deltabQSmodel} from the deep matter era to $z=0$, imposing the common growing-mode
initial conditions ${\rm d}\delta_c/{\rm d}N={\rm d}\delta_b/{\rm d}N=\delta_c=\delta_b$
at $z=100$. Figure~\ref{fig:fig7} shows the resulting evolution of the gravitational
potential. The normalized potential magnitude $-\Psi$ in $\Lambda$CDM decays toward lower
redshifts, whereas the corresponding curves for the three interacting cases remain above
the $\Lambda$CDM curve over most of the range shown. 
This behavior cannot be inferred from $G_c$
alone.  Even when the quasi-static CDM coupling is strongly suppressed, the Poisson equation \eqref{PsiPhi} contains the velocity-mixing
contribution proportional to $(1-q_c)\alpha_B \rho_c \dot\delta_c/H$, which is positive for the growing mode of the present solutions.  As a result,
case (iii), which has the largest $\beta$ 
and the smallest present value
of $G_c/G$, exhibits the largest late-time departure from the
$\Lambda$CDM reference potential.

To characterize the growth signal measured through 
redshift-space distortions, 
we use the product $f\sigma_8$. 
The logarithmic growth rate $f$ and the fluctuation amplitude $\sigma_8$ are defined by
\be
f(z)\equiv\frac{{\rm d}\ln\delta_m}{{\rm d}\ln a}
=\frac{\dot\delta_m}{H\delta_m}\,,
\qquad
\sigma_8(z)=\sigma_8(0)\,\frac{\delta_m(z)}{\delta_m(0)}\,.
\label{fsigma8def}
\ee
Here $\sigma_8(z)$ is the root-mean-square linear total-matter density
contrast at redshift $z$, smoothed with a spherical top-hat of comoving
radius $8h_{100}^{-1}\,{\rm Mpc}$.  In the quasi-static equations
\eqref{deltacQSmodel} and \eqref{deltabQSmodel}, the effective
gravitational couplings \eqref{Gccmodel} and \eqref{Gbbmodel} do not
depend on $k$.  The resulting linear growth is therefore scale independent, 
so the second relation in Eq.~\eqref{fsigma8def} applies to sub-Hubble
perturbations.

\begin{figure}[t]
 \centering
 \includegraphics[width=\columnwidth]{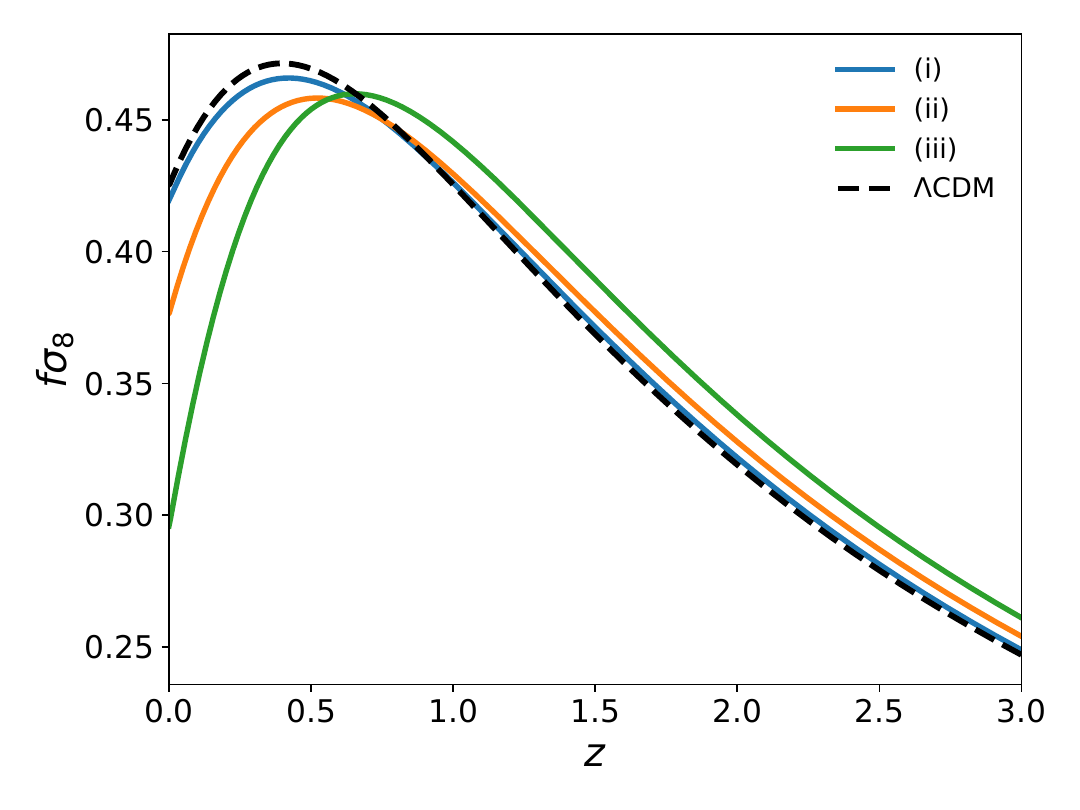}
 \caption{Evolution of the redshift-space-distortion growth observable
$f\sigma_8$ over $0\le z\le 3$ for the three cases of
Fig.~\ref{fig:fig1}, obtained by solving the quasi-static growth
equations \eqref{deltacQSmodel} and \eqref{deltabQSmodel}.  The total
matter density contrast is defined in Eq.~\eqref{delM}, and the common
present-day normalization $\sigma_8(0)=0.811$ is imposed.  The black
dashed curve shows the corresponding $\Lambda$CDM result.  At $z=0$,
the suppression of $f\sigma_8$ increases from case (i) to case (iii).}
 \label{fig:fig8}
\end{figure}

Figure~\ref{fig:fig8} shows that the momentum-transfer interaction suppresses the present-day growth observable $f\sigma_8$ relative to the $\Lambda$CDM prediction. 
Near $z=0$, the ordering of the curves follows
the hierarchy of the momentum-transfer parameter $\beta$ and of the effective CDM coupling: case (i) remains closest to the standard prediction, whereas case (iii) gives the largest reduction. 
This suppression is not uniform over $0\leq z\leq3$.  Since all curves are normalized to the same value of $\sigma_8(0)$, the modified growth
histories cross the $\Lambda$CDM result and can yield larger
$f\sigma_8$ at intermediate redshifts.  All three backgrounds remain in the stable region $c_s^2>0$.
Thus, a coupling of order $\beta={\cal O}(10^{-2})$ is sufficient to
produce a sizable present-day reduction of $f\sigma_8$. 
This differs from the $a_2=a_3=0$ model of
Ref.~\cite{Kase:2019mox}, where $\alpha_B=0$ and hence the
momentum-transfer mixing term ${\cal G}_{\rm mix}$ is absent.

\section{Super-Hubble perturbations around 
the transient braiding peak}
\label{sec:superHubble}

In this section, we focus on the evolution of perturbations on
super-Hubble scales around the transient braiding epoch.  The relevant
modes satisfy $\kappa_k=k/(aH)\ll 1$ when the cubic Galileon variable
$x_3$, or equivalently the braiding parameter $\alpha_B=-x_3/2$, develops
a localized peak close to radiation--matter equality.  Since such modes
lie outside the regime of validity of the quasi-static approximation
adopted in Sec.~\ref{persec}, we use the full Newtonian-gauge equations
\eqref{HamNmodel}--\eqref{EulerrNmodel}.  For the analytic estimates, we
take the super-Hubble limit of these equations. For the numerical analysis, we modify the CLASS code
~\cite{Lesgourgues:2011re,Blas:2011rf} to implement the present theory,
and use it to evolve the full linear system, including $\delta\phi$,
together with the standard Boltzmann hierarchies and recombination
physics.

We identify how the transient braiding peak affects the metric
potential, the CDM density contrast, and the CDM velocity variable.  We
first summarize the background evolution around the peak and derive useful
large-scale estimates for $\Phi$, $\delta_c$, and $V_c$.  We then compare
these estimates with the full numerical evolution for the representative
cases (i), (ii), and (iii).  The implications for the matter and CMB
spectra are discussed separately in Sec.~\ref{sec:powerCMB}.

\subsection{Background around the transient $x_3$ peak}

We first summarize the background relations needed to estimate the large-scale perturbations. 
As shown in Fig.~\ref{fig:fig2}, the variable
$x_3$ develops a localized peak (transient braiding peak) around radiation--matter equality. 
We focus on the early epoch in which the $x_3$ contribution gives the dominant part of 
the DE density, namely
\be
x_3\gg |A|x_1^2\,,\qquad x_3\gg x_2\,,\qquad x_3\gg x_4\,.
\label{x3peakHierarchy}
\ee
Throughout this subsection, we do not assume $x_3\ll 1$.
The Friedmann constraint
then gives
\be
\Omega_m+\Omega_r+x_3\simeq 1\,,
\qquad \Omega_m\equiv\Omega_c+\Omega_b\,.
\label{FriedPeak}
\ee
Expanding Eqs.~\eqref{epssol} and \eqref{hsol} 
under the hierarchy \eqref{x3peakHierarchy} and keeping the leading-order terms, we obtain
\ba
\epsilon_\phi &\simeq& \frac{\Omega_r-3}{4+x_3}\,,
\label{epsPeak}\\
h&\simeq&-\frac{6+3x_3+2\Omega_r}{4+x_3}\,.
\label{hPeak}
\ea
The evolution equation $x_3'=x_3(3\epsilon_\phi-h)$ then yields
\be
\frac{x_3'}{x_3}\simeq
\frac{5\Omega_r+3x_3-3}{4+x_3}\,.
\label{x3primePeak}
\ee
At the transient maximum of $x_3$, we have $x_3'=0$.
This condition gives
\be
\Omega_r^{\rm peak}\simeq \frac{3}{5}(1-x_3^{\rm peak})\,,
\qquad
\Omega_m^{\rm peak}\simeq \frac{2}{5}(1-x_3^{\rm peak})\,.
\label{OmegaPeak}
\ee
Thus, the peak occurs slightly before exact radiation--matter equality,
with 
\be
\frac{\Omega_r^{\rm peak}}{\Omega_m^{\rm peak}} \simeq \frac{3}{2}\,. 
\label{OmegarOmeagmPeakRatio}
\ee
At the same
point one has
\be
\epsilon_\phi^{\rm peak}\simeq -\frac{3}{5}\,,
\qquad
h^{\rm peak}\simeq -\frac{9}{5}\,.
\label{epshPeak}
\ee
These relations are independent of the peak amplitude $x_3^{\rm peak}$.

\subsection{Early-time super-Hubble evolution 
of $\Phi$, $\delta_c$, and $V_c$}

\begin{figure}[t]
 \centering
 \includegraphics[width=\columnwidth]{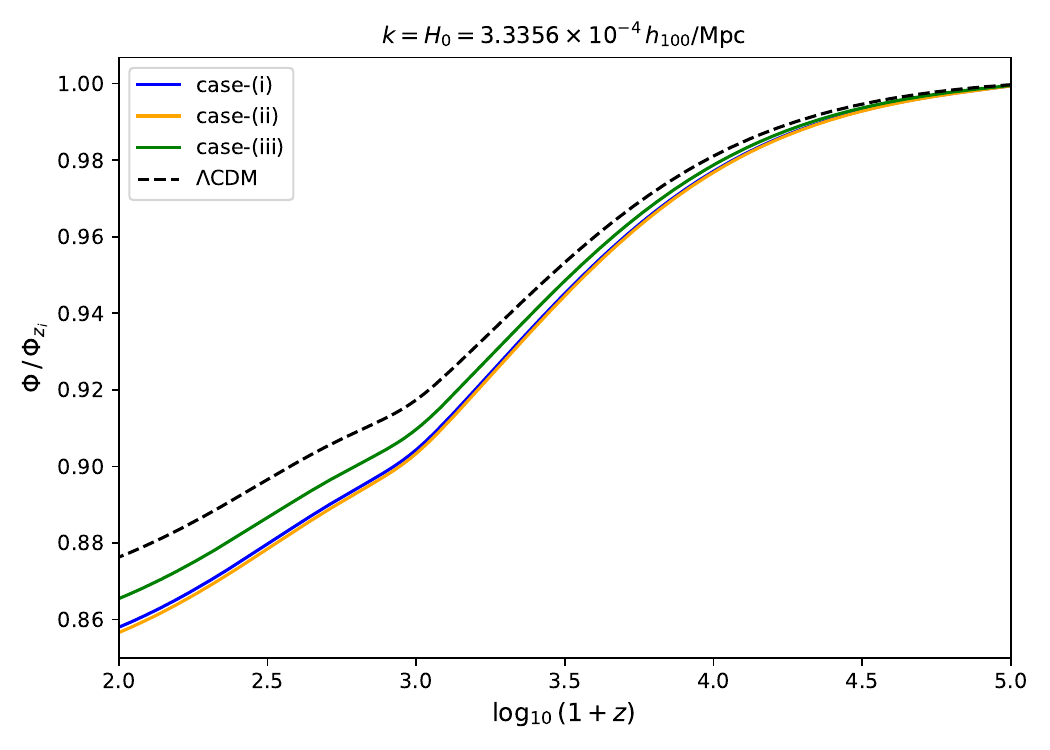}
 \caption{Full numerical CLASS evolution of the normalized curvature potential
$\Phi/\Phi_{z_i}$ for cases (i), (ii), and (iii), together with the
$\Lambda$CDM solution.  The plot covers
$2\leq\log_{10}(1+z)\leq5$ for $k=H_0$,
or equivalently $\kappa_{k0}=1$, with all curves normalized at
$\log_{10}(1+z_i)=5$.  In the units shown above the panel,
$k=3.3356\times10^{-4}\,h_{100}\,{\rm Mpc}^{-1}$.  The scalar-field
perturbation $\delta\phi$ is evolved as part of the full linear system.}
 \label{fig:fig9}
\end{figure}

\begin{figure}[!t]
 \centering
 \includegraphics[width=1.0\columnwidth]{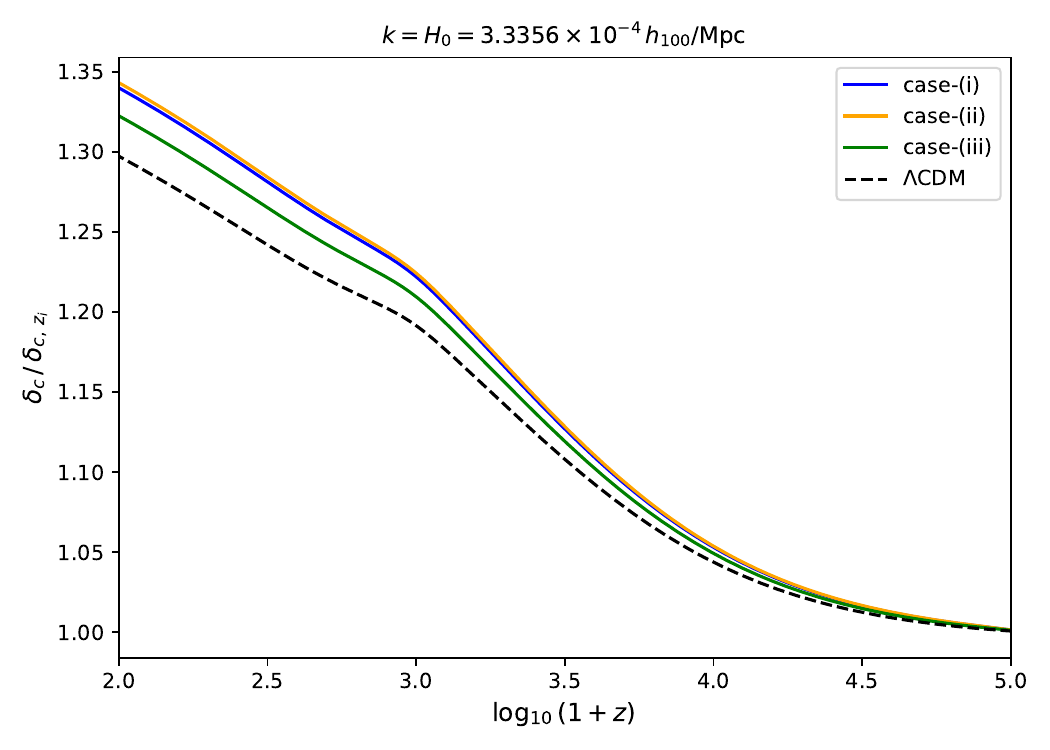}
 \vspace{-2mm}
 \includegraphics[width=1.0\columnwidth]{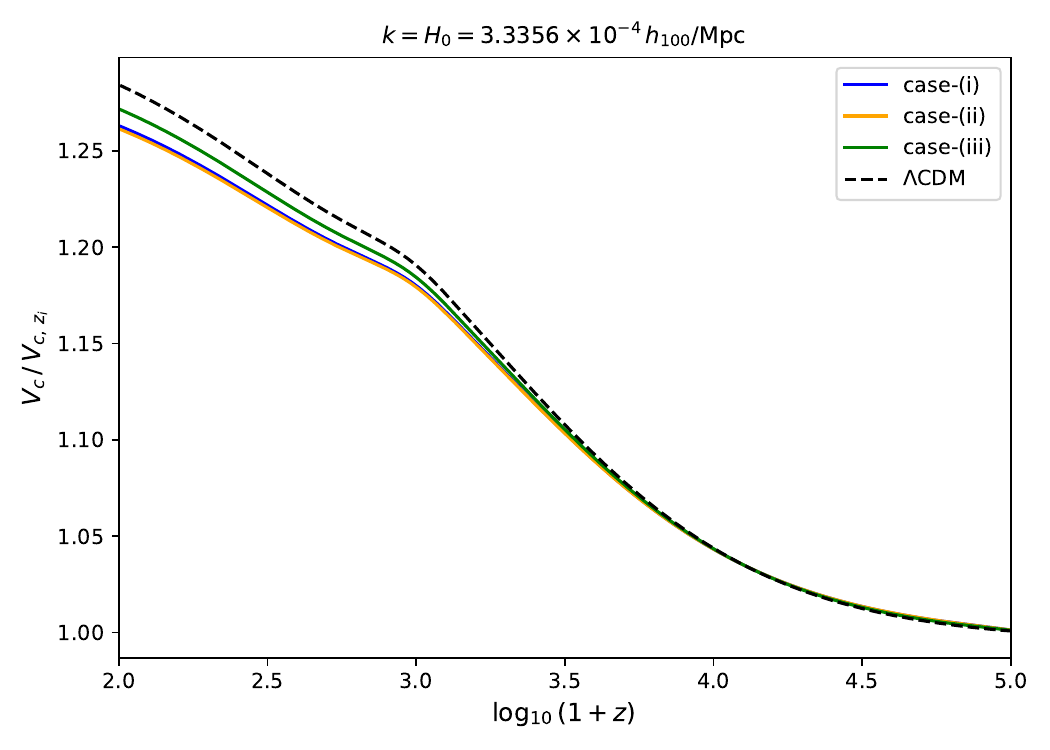}
 \caption{Full numerical CLASS evolution of the normalized CDM perturbations for
cases (i), (ii), and (iii), together with $\Lambda$CDM.  The upper panel
shows $\delta_c/\delta_{c,z_i}$, while the lower panel shows
$V_c/V_{c,z_i}$, where $V_c= H v_c$ is the dimensionless velocity
potential used in the text.  The scalar-field perturbation $\delta\phi$
is included in the complete evolution.  Both panels cover
$2\leq\log_{10}(1+z)\leq5$ for
$k=H_0=3.3356\times10^{-4}\,h_{100}\,{\rm Mpc}^{-1}$, and all curves are
normalized at $\log_{10}(1+z_i)=5$.  Toward lower redshift, the CDM
density contrast is enhanced relative to $\Lambda$CDM in all three cases,
whereas the normalized velocity response is slightly smaller.  Case (iii)
is closest to $\Lambda$CDM in both panels.}
 \label{fig:fig10}
\end{figure}

We now study the regular matter--radiation adiabatic growing mode on
super-Hubble scales during the transient braiding epoch.  
The scalar-field perturbation is initially set to zero and subsequently 
evolved as part of the coupled perturbation system.
To distinguish the different analytic approximations used below, we proceed in two steps.  
We first introduce an auxiliary analytic 
approximation in which $\delta\phi$ is set to zero, thereby isolating the evolution of the metric
potential across the radiation--matter transition. 
We then restore $\delta\phi$ and the
CDM momentum-transfer terms to decompose the response generated by the
localized $x_3$ peak.  The numerical results are obtained throughout from
the full CLASS evolution.  In Figs.~\ref{fig:fig9} and
\ref{fig:fig10}, we choose the comoving wavenumber
$k=H_0$, or equivalently $\kappa_{k0}=1$ at $z=0$. In the units shown in
the figures, this corresponds to
$k=3.3356\times10^{-4}\,h_{100}\,{\rm Mpc}^{-1}$.  Here $h_{100}$ is the
reduced Hubble constant defined in the Introduction and should not be
confused with the background variable $h=\dot H/H^2$.  This mode remains
outside the Hubble radius throughout the displayed interval and satisfies
$\kappa_k\ll1$ during the transient braiding peak.

We impose regular adiabatic growing-mode initial conditions for the matter
and radiation perturbations in the deep radiation era.  Neglecting neutrino
anisotropic stress, the conformal-Newtonian-gauge initial conditions are
given by~\cite{Ma:1995ey}
\be
\delta_{c,i}=\delta_{b,i}=\frac{3}{4}\delta_{r,i},
\qquad
\delta_{r,i}=-2\Psi_i\,,
\label{AdiabaticICNewtonian}
\ee
where a subscript $i$ denotes evaluation at the initial epoch.
The second
relation follows directly from the Hamiltonian constraint
\eqref{HamNmodel} in the present notation.  
At this early epoch, the
solution is in the radiation-era GR limit, for which
$\dot\Phi_i\simeq0$, $|\alpha_{B,i}|\ll1$,
$|\tilde{\alpha}_{K,i}|\ll1$, and $\Omega_{r,i}\simeq1$.  For the
scalar-field perturbation, we choose
\be
\delta\phi_i=0,\qquad
(\delta\phi_{,N})_i=0\,,
\ee
so that the $\delta\phi$-dependent terms in Eq.~\eqref{HamNmodel} do not
contribute at the initial epoch.  Thus, the matter and radiation
perturbations are initialized in their adiabatic growing mode, while the
scalar-field perturbation is independently chosen to vanish initially.
Neglecting neutrino anisotropic stress, Eq.~\eqref{anisomodel} gives
$\Psi_i=\Phi_i$. 
Under the early-time approximations described above,
the Hamiltonian constraint~\eqref{HamNmodel} then reduces to $6\Phi_i+3\delta_{r,i}\simeq0$, yielding
$\delta_{r,i}=-2\Phi_i$. 
Combining this result with the first relation
in Eq.~\eqref{AdiabaticICNewtonian}, we obtain
\be
\delta_{c,i}=\delta_{b,i}=-\frac{3}{2}\Phi_i\,,
\qquad
\delta_{r,i}=-2\Phi_i\,.
\label{AdiabaticICPhi}
\ee

As the first analytic step, we construct a transparent baseline for the
radiation--matter transition of the metric potential.  In this auxiliary
calculation only, we take $\kappa_k\to0$ and impose $\delta\phi=0$ at all
times.  The condition $\delta\phi=0$ is used only to close the transition
equation for $\Phi$ and is abandoned before the peak-response analysis; it
is not imposed in the CLASS results.  
Since the fluid continuity equations contain no explicit $\delta\phi$ terms, their super-Hubble 
limits derived below also apply to the full coupled system in which the scalar-field perturbation is
dynamically evolved.
During the transient braiding epoch, we use $\tilde{\alpha}_K\simeq6x_3$ and
$\alpha_B=-x_3/2$.  Equations~\eqref{contcNmodel}--\eqref{contrNmodel}
then give
\be
\delta_c'=3\Phi'\,,\qquad
\delta_b'=3\Phi'\,,\qquad
\delta_r'=4\Phi'\,,
\label{AdiabaticDeltaPhi}
\ee
where prime denotes derivative with respect to $N=\ln a$. 
Equivalently, $\delta_c-3\Phi$, $\delta_b-3\Phi$, and $\delta_r-4\Phi$
are conserved.  These continuity relations apply both to the auxiliary
baseline and to the restored coupled system in the strict
$\kappa_k\to0$ limit.

To obtain a closed equation for the evolution of $\Phi$ across the
radiation--matter transition, we first differentiate the Hamiltonian
constraint \eqref{HamNmodel} with respect to $N$. The fluid density
perturbations are then eliminated with Eq.~\eqref{AdiabaticDeltaPhi},
together with the adiabatic relation
$\delta_c=\delta_b=3\delta_r/4$, which is preserved by the continuity equations. We introduce $y\equiv\Omega_m/\Omega_r$, which obeys $y'=y$,
and adopt the approximate background constraint
$\Omega_m+\Omega_r+x_3\simeq1$. After the differentiation, we impose $\Psi=\Phi$ and retain the leading terms in the super-Hubble expansion.
This gives
\ba
& &
(2-x_3)y\Phi_{,yy}
+\left[4-5x_3-yx_{3,y}+\frac{3y+4}{1+y}\right]\Phi_{,y}
\nonumber\\
& &
-\left[4x_{3,y}+\frac{1}{(1+y)^2}\right]\Phi=0\,,
\label{PhiSecondYConstraint}
\ea
where $\Phi_{,y}\equiv{\rm d}\Phi/{\rm d}y$ and
$x_{3,y}\equiv{\rm d}x_3/{\rm d}y$.
In the limit $x_3\to0$, Eq.~\eqref{PhiSecondYConstraint} reduces to the
standard radiation--matter transition equation,
\be
\Phi_{,yy}
+\frac{21y^2+54y+32}{2y(1+y)(3y+4)}\Phi_{,y}
+\frac{1}{y(1+y)(3y+4)}\Phi=0\,.
\label{PhiGRTransitionEq}
\ee
After discarding the decaying mode, its regular solution is
\be
\Phi(y)=\Phi_{\rm rad}
\frac{16\sqrt{1+y}+9y^3+2y^2-8y-16}{10y^3}\,,
\label{PhiGRTransitionSol}
\ee
where $\Phi_{\rm rad}$ denotes the constant value of the potential deep
in the radiation era. This solution approaches
$\Phi_{\rm rad}$ as $y\to0$ and
$(9/10)\Phi_{\rm rad}$ as $y\to\infty$. 
The asymptotic factor $9/10$
is obtained in the idealized perfect-fluid treatment adopted here, in which the neutrino quadrupole and the associated anisotropic stress are neglected. 
The full Boltzmann evolution retains the free-streaming neutrino quadrupole,
which slightly changes the potential evolution and makes $\Psi$ and $\Phi$
not exactly equal in our sign convention.  Consequently, the $\Lambda$CDM
curve in Fig.~\ref{fig:fig9} need not reach precisely $0.9$ over the finite
interval shown~\cite{Dodelson:2020}.

Within this auxiliary baseline,
Eq.~\eqref{PhiSecondYConstraint} shows how the background braiding variable
$x_3$ and its localized variation modify the radiation--matter transition
when the scalar-field perturbation is omitted.  The $\delta\phi=0$
calculation ends at this point.  We now restore the scalar-field
perturbation and the CDM momentum-transfer terms and examine the localized
peak response of the coupled system.  To identify the corresponding
perturbative sources, we use the full super-Hubble momentum constraint
\eqref{MomNmodel} and introduce the dimensionless variables
\be
\pi\equiv \frac{H\delta\phi}{\dot\phi}\,,
\qquad
V_I\equiv H v_I\,.
\label{piVdefs}
\ee
In terms of these variables, the momentum constraint becomes
\ba
\Phi' &=& -(1+\alpha_B)\Phi+\alpha_B\pi'
- {\cal B}_\pi\pi
+\frac{3}{2}q_c\Omega_cV_c \nonumber\\
&&+\frac{3}{2}\Omega_bV_b+2\Omega_rV_r\,,
\label{PhiPrimeSuper}\\
{\cal B}_\pi&\equiv&
(1+\alpha_B)h+\frac{3}{2}(q_c\Omega_c+\Omega_b)
+2\Omega_r\,.
\label{BpiDef}
\ea
At the maximum of the $x_3$ peak, Eqs.~\eqref{OmegaPeak} and
\eqref{epshPeak} give ${\cal B}_\pi\simeq9\alpha_B/5$ for $q_c=1$.
Keeping the leading correction proportional to $q_c-1$ gives
\be
{\cal B}_\pi\simeq \frac{9}{5}\alpha_B
+\frac{3}{2}\Omega_c(q_c-1)\,.
\label{BpiPeakExpand}
\ee
For this source decomposition, a superscript $(0)$ denotes a quantity
evaluated in a reference solution of the coupled perturbation equations
with the source terms localized around the $x_3$ peak removed, while the
nonlocalized background evolution is retained. Unlike the auxiliary
baseline with $\delta\phi=0$, or equivalently $\pi=0$, introduced above,
this reference solution retains both the scalar-field and fluid
perturbations; hence, $\pi^{(0)}$ generally does not vanish. Subtracting
the momentum constraint for the reference solution from
Eq.~\eqref{PhiPrimeSuper} gives an inhomogeneous equation for the
peak-induced potential,
$\Phi_{\rm ind}\equiv\Phi-\Phi^{(0)}$. To leading order in the
peak-induced response, we evaluate the perturbations appearing in the
localized source terms on the reference solution. 
The resulting source is
\ba
{\cal S}_\Phi &\simeq&
-\alpha_B\Phi^{(0)}
+\alpha_B\left[(\pi^{(0)})'-\frac{9}{5}\pi^{(0)}\right] \nonumber \\
& &
+\frac{3}{2}\Omega_c(q_c-1)(V_c^{(0)}-\pi^{(0)})\,.
\label{SPhiFull}
\ea
The first term in Eq.~\eqref{SPhiFull},
$-\alpha_B\Phi^{(0)}=x_3\Phi^{(0)}/2$, represents the direct
metric--braiding channel evaluated on the reference solution. Although
useful for isolating one component of the source, it does not capture the
complete braiding-induced response. The second term,
$\alpha_B[(\pi^{(0)})'-9\pi^{(0)}/5]$, describes the corresponding
scalar-field channel and can oppose or even dominate the first term. The
last term represents the explicit contribution from CDM momentum transfer.
Because these terms act through the coupled 
evolution of the metric, scalar-field, and fluid perturbations, the sign and magnitude of the
peak-induced responses of $\Phi$ and $\delta_c$ cannot be inferred from $-\alpha_B\Phi^{(0)}$ alone. 
All these contributions are retained in the
full numerical calculation.

Let $N_{\rm p}$ denote the e-fold at which $x_3$ reaches its transient
maximum, and let $N_-$ and $N_+$ bracket the localized peak. We define the
net peak-induced change by
$\Delta\Phi_{\rm p}\equiv
\Phi_{\rm ind}(N_+)-\Phi_{\rm ind}(N_-)$. 
To estimate the direct
metric--braiding contribution, we retain only the source term proportional
to the reference metric potential,
${\cal S}_\Phi\simeq-\alpha_B\Phi^{(0)}
=x_3\Phi^{(0)}/2$. Approximating $\Phi^{(0)}$ as constant across the
narrow peak interval, integration gives
\be
\Delta\Phi_{\rm p}
\simeq \frac{1}{2}\Phi^{(0)}(N_{\rm p}){\cal A}_3\,,
\qquad
{\cal A}_3\equiv\int_{N_-}^{N_+}{\rm d}N\,x_3(N)\,.
\label{DeltaPhiPeak}
\ee
On super-Hubble scales, the velocity-divergence terms
$\kappa_k^2H^2v_c$ and $\kappa_k^2H^2v_b$ in
Eqs.~\eqref{contcNmodel} and \eqref{contbNmodel}, respectively, are
negligible for regular velocity potentials. Both the full and reference
solutions therefore satisfy
$\delta_c'=3\Phi'$ and $\delta_b'=3\Phi'$, as given in
Eq.~\eqref{AdiabaticDeltaPhi}. 
Subtracting the corresponding continuity
relations gives
$\delta_{c,{\rm ind}}'=3\Phi_{\rm ind}'$ and
$\delta_{b,{\rm ind}}'=3\Phi_{\rm ind}'$, where
$\delta_{I,{\rm ind}}\equiv\delta_I-\delta_I^{(0)}$. Hence, the peak-induced change
in the CDM density contrast is
\be
\Delta\delta_{c}^{\rm peak}
\simeq3\Delta\Phi_{\rm p}
\simeq \frac{3}{2}\Phi^{(0)}(N_{\rm p}){\cal A}_3\,.
\label{DeltadeltacPeak}
\ee
This estimate captures only the direct metric--braiding channel represented
by ${\cal S}_\Phi\simeq-\alpha_B\Phi^{(0)}$ in
Eq.~\eqref{SPhiFull}. Once the scalar-field and momentum-transfer terms in
Eq.~\eqref{SPhiFull}, together with the fully coupled perturbation dynamics,
are included, neither the effective 
coefficient relating
$\Delta\Phi_{\rm p}$ to $\Phi^{(0)}(N_{\rm p}){\cal A}_3$ nor the sign of
the total response relative to $\Lambda$CDM can be inferred from this term
alone. Both must instead be determined from the complete perturbation
system.

Figure~\ref{fig:fig9} presents the numerical CLASS evolution of the
normalized curvature potential.  The same normalization is used in
Figs.~\ref{fig:fig9} and \ref{fig:fig10}: each plotted quantity $Q$ is
divided by $Q_{z_i}\equiv Q(z_i)$.  Toward lower redshift, all three model
curves lie below the $\Lambda$CDM reference.  The departure increases in the
order (iii), (i), and (ii), which follows the ordering
$x_{3,{\rm (iii)}}^{\rm peak}\simeq2.286614\times10^{-3}$,
$x_{3,{\rm (i)}}^{\rm peak}\simeq 3.873008\times10^{-3}$, and
$x_{3,{\rm (ii)}}^{\rm peak}\simeq4.166489\times10^{-3}$.  This correspondence
suggests that the total braiding response controls the relative size of the
large-scale departure in these examples.  As emphasized by the source
decomposition in Eq.~\eqref{SPhiFull}, however, the result cannot be inferred
from the $-\alpha_B\Phi^{(0)}$ contribution alone; it follows from the
coupled metric, scalar-field, matter, and radiation evolution.

The velocity perturbation responds to the peak through an integral over the
source.  In terms of $V_c=Hv_c$, the CDM Euler equation is
\be
V_c'+(\epsilon_{q_c}-h)V_c
=\frac{\Phi}{q_c}
+\frac{q_c-1}{q_c}(\pi'-h\pi)
+\epsilon_{q_c}\pi\,,
\label{VcSuper}
\ee
where we have used $\Psi=\Phi$.  Its formal solution is
\be
V_c(N)=e^{-I_c(N)}\left[V_c(N_i)
+\int_{N_i}^{N}{\rm d}\widetilde N\,
e^{I_c(\widetilde N)}S_c(\widetilde N)
\right]\,,
\label{VcSolution}
\ee
where $N_i$ is the initial e-fold. 
The integrating factor $I_c$ and source $S_c$ are
\be
\begin{aligned}
I_c(N)&=\int_{N_i}^{N}{\rm d}\widetilde N\,
[\epsilon_{q_c}(\widetilde N)-h(\widetilde N)]\,,\\
S_c(N)&=\frac{\Phi}{q_c}
+\frac{q_c-1}{q_c}(\pi'-h\pi)
+\epsilon_{q_c}\pi\,.
\end{aligned}
\label{ScIcDef}
\ee
To estimate the velocity response induced by the localized peak, we use the
same interval $[N_-,N_+]$ introduced above and define
$\Delta N_{\rm p}\equiv N_+-N_-$.  For $q_c\simeq1$,
$\epsilon_{q_c}\simeq0$, and a narrow peak, the contribution generated by
$\Phi_{\rm ind}$ at the end of the peak is
\be
\begin{aligned}
\Delta V_{c,{\rm p}}
&\equiv V_c(N_+)-V_c^{(0)}(N_+) \\
&\simeq\int_{N_-}^{N_+}{\rm d}N\,
\exp\left[\int_N^{N_+}{\rm d}\widetilde N\,h(\widetilde N)\right]
\Phi_{\rm ind}(N)\\
&={\cal O}(\Delta N_{\rm p}\Delta\Phi_{\rm p})\,.
\end{aligned}
\label{DeltaVcPeak}
\ee
Thus, the velocity response is suppressed by the finite width of the peak,
whereas the density response is directly tied to $\Delta\Phi_{\rm p}$ through
the super-Hubble continuity equation.  When the momentum-transfer coupling
is non-negligible, the last two terms in Eq.~\eqref{VcSuper} provide
additional scalar-field sources for $V_c$ through $\pi'$ and $\pi$, while
the factor $1/q_c$ rescales the metric-force term $\Phi$.

Figure~\ref{fig:fig10} gives the numerical CLASS evolution of the
normalized CDM perturbations.  The density contrast in the upper panel grows
more rapidly than in $\Lambda$CDM, and its departure follows the same order
(iii), (i), and (ii) as the curvature response in Fig.~\ref{fig:fig9}.  The
normalized velocity potential in the lower panel is instead slightly smaller
than the $\Lambda$CDM result, with case (iii) closest to the reference curve.
The density and curvature trends are consistent with the conserved
super-Hubble combination fixed by the matter--radiation adiabatic initial
conditions.  
In the strict $\kappa_k\to0$ limit, Eq.~\eqref{AdiabaticDeltaPhi}
implies that $\delta_c-3\Phi$ is conserved. Using the initial condition
in Eq.~\eqref{AdiabaticICPhi}, we find
\be
\delta_c-3\Phi=\delta_{c,i}-3\Phi_i
=-\frac{9}{2}\Phi_i\,.
\label{AdiabaticConstExample}
\ee
Equivalently, this relation can be written as
\be
\frac{\delta_c}{\delta_{c,i}}
=3-2\frac{\Phi}{\Phi_i}\,.
\ee
Thus, a reduction of $\Phi/\Phi_i$ corresponds to an increase of
$\delta_c/\delta_{c,i}$ in this limit.  The finite-$k$ CLASS evolution also
retains the velocity-gradient terms, the dynamically evolved scalar-field
perturbation, and the full Boltzmann hierarchy.  Equations~\eqref{VcSuper} and \eqref{VcSolution} further show that
$V_c$ is a time-integrated response to the metric force $\Phi/q_c$ and
the $\pi$-dependent momentum-transfer terms, with their past contributions
weighted by the $q_c$-dependent integrating factor.

The density and velocity responses combine in the large-scale matter
spectrum through the gauge-invariant comoving total-matter density contrast,
\be
\Delta_m
=\frac{\Omega_c(\delta_c+3V_c)+\Omega_b(\delta_b+3V_b)}
{\Omega_c+\Omega_b}\,.
\label{DeltamLargeScale}
\ee
Its gauge invariance can be seen explicitly from an infinitesimal change
of time slicing, $\widetilde t=t+\xi^0$. For each pressureless matter
component $I=c,b$, the perturbations transform as
$\widetilde\delta_I=\delta_I+3H\xi^0$ and
$\widetilde V_I=V_I-H\xi^0$. The combination
$\delta_I+3V_I$ is therefore invariant, and so is the weighted total $\Delta_m$.

Let $a_{\rm p}\equiv a(N_{\rm p})$ and
$H_{\rm p}\equiv H(N_{\rm p})$ denote the scale factor and Hubble rate at
the transient peak. The usual super-Hubble suppression by $k^2$ follows
directly from the Hamiltonian and momentum constraints,
Eqs.~\eqref{HamNmodel} and \eqref{MomNmodel}. In the early-time GR limit
of the reference solution, the momentum constraint can be used to
eliminate the combination of the time derivative and lapse potential from
the Hamiltonian constraint, yielding
\be
2\kappa_k^2\Phi^{(0)}
+3\Omega_m\Delta_m^{(0)}
+3\Omega_r\Delta_r^{(0)}=0\,,
\ee
where
$\Delta_r^{(0)}\equiv\delta_r^{(0)}+4V_r^{(0)}$.
For the regular matter--radiation adiabatic mode, the relative velocity
vanishes at leading order in the gradient expansion, so that
$\Delta_r^{(0)}=4\Delta_m^{(0)}/3$. It follows that
\be
\Delta_m^{(0)}
=-\frac{2\kappa_k^2\Phi^{(0)}}
{3\Omega_m+4\Omega_r}\,.
\ee
This suppression reflects the cancellation, in the comoving density
contrast, of the leading ${\cal O}(k^0)$ density and velocity
contributions associated with a common local time shift. On the regular
adiabatic reference branch, the smooth scalar-field, braiding, and
momentum-transfer contributions can modify the coefficient but, provided
the constraint system remains regular, do not change the leading
$\kappa_k^2$ order of the gradient expansion~\cite{Dodelson:2020}.
Therefore, around the peak epoch,
\be
\Delta_m^{(0)}(k,N_{\rm p})
\simeq C_m{\cal R}_k
\left(\frac{k}{a_{\rm p}H_{\rm p}}\right)^2,
\qquad k\ll a_{\rm p}H_{\rm p},
\label{DeltamStandardScaling}
\ee
where ${\cal R}_k$ is the primordial comoving curvature perturbation and
$C_m$ is a dimensionless coefficient determined by the background evolution and matter--radiation composition of the reference solution.

The transient braiding peak generates an additional response through the
coupled metric, matter, velocity, and scalar-field perturbations.  The
estimates above show that this response is proportional to the peak area
${\cal A}_3$ defined in Eq.~\eqref{DeltaPhiPeak}.  Since the perturbation
equations are linear, it is also proportional to the primordial amplitude
${\cal R}_k$.  We therefore parametrize the peak-induced contribution to the
comoving matter contrast as
\be
\Delta_{m,{\rm ind}}(k,N_{\rm p})
\equiv \Delta_m-\Delta_m^{(0)}
\simeq C_3 {\cal A}_3 {\cal T}_{\Delta}(k,N_{\rm p}){\cal R}_k\,,
\label{DeltamPeakScaling}
\ee
where $C_3$ is a dimensionless response coefficient.  The factor
${\cal T}_{\Delta}(k,N_{\rm p})$ denotes the finite-wavelength response of
the comoving density contrast to the localized peak source, evaluated
around the peak epoch $N_{\rm p}$.  This notation is chosen in analogy
with the usual linear transfer function, since
${\cal T}_{\Delta}$ encodes the scale-dependent propagation of the source
through the coupled Einstein--Boltzmann system~\cite{Dodelson:2020}. 
Its normalization can be absorbed into $C_3$, whereas its $k$ dependence
is not fixed by the peak-area estimate.  In particular, the density and
velocity perturbations can separately acquire leading-order responses
that cancel in the gauge-invariant combination $\Delta_m$.  We therefore
leave the low-$k$ behavior of ${\cal T}_{\Delta}$ unspecified and determine
the finite-wavelength response from the full numerical evolution.

Combining Eqs.~\eqref{DeltamStandardScaling} and
\eqref{DeltamPeakScaling}, the relative peak-induced correction can be
written as
\be
\frac{\Delta_{m,{\rm ind}}}{\Delta_m^{(0)}}
\simeq \frac{C_3}{C_m}{\cal A}_3{\cal T}_{\Delta}(k,N_{\rm p})
\left(\frac{a_{\rm p}H_{\rm p}}{k}\right)^2.
\label{DeltamEnhanceRatio}
\ee
The corresponding power-spectrum ratio at the peak epoch is
\be
\begin{aligned}
\frac{P(k,N_{\rm p})}{P_{\Lambda}(k,N_{\rm p})}
&\simeq
\frac{P^{(0)}(k,N_{\rm p})}{P_{\Lambda}(k,N_{\rm p})}
\\
&\quad\times\biggl|1+\frac{C_3}{C_m}{\cal A}_3
{\cal T}_{\Delta}(k,N_{\rm p})
\left(\frac{a_{\rm p}H_{\rm p}}{k}\right)^2\biggr|^2\,.
\end{aligned}
\label{MatterEnhanceScaling}
\ee
Here $P^{(0)}$ denotes the power spectrum of $\Delta_m$ in the reference
solution with the source terms localized around the $x_3$ peak removed,
whereas $P_\Lambda$ is the $\Lambda$CDM spectrum evaluated with the same
standard cosmological parameters. The prefactor $P^{(0)}/P_\Lambda$
accounts for the fact that the reference solution can differ from
$\Lambda$CDM even in the absence of the localized peak contribution.
Equation~\eqref{MatterEnhanceScaling} estimates the response generated
around $N_{\rm p}$ only. The spectra at $z=0$ presented below include the
full subsequent evolution and are computed directly with CLASS.

The scale dependence of the relative response in
Eqs.~\eqref{DeltamEnhanceRatio} and \eqref{MatterEnhanceScaling} is
controlled by the combination
${\cal T}_{\Delta}(k,N_{\rm p})(a_{\rm p}H_{\rm p}/k)^2$. Since the
peak-area estimate does not determine the low-$k$ behavior of
${\cal T}_{\Delta}$, these equations do not imply a universal asymptotic
power law. In particular, if
${\cal T}_{\Delta}\propto k^2$ approximately over a finite wavenumber
range, it compensates the explicit $k^{-2}$ factor and leaves only a
residual scale dependence. 
For $k\ll a_{\rm p}H_{\rm p}$, the super-Hubble gradient expansion
provides a qualitative description of the peak-induced response, although
its precise scale dependence still requires the coupled perturbation
evolution. For $k\gtrsim a_{\rm p}H_{\rm p}$, spatial-gradient terms and
the higher radiation multipoles are no longer negligible, so the
super-Hubble approximation breaks down and the full Einstein--Boltzmann
hierarchy becomes essential. The $z=0$ spectra are therefore computed
with the full CLASS evolution for all wavenumbers. The distinct responses
of $\Phi$, $\delta_c$, and $V_c$ shown in
Figs.~\ref{fig:fig9} and \ref{fig:fig10} illustrate why both the amplitude
and scale dependence of the large-scale trend are determined by the
complete coupled system. The resulting matter spectra are discussed in
Sec.~\ref{sec:powerCMB}.

\section{Matter power spectrum and CMB power spectrum}
\label{sec:powerCMB}

We now turn from the long-wavelength diagnostic of
Sec.~\ref{sec:superHubble} to observable spectra computed with the modified
CLASS Boltzmann code.  This is the same full implementation used for
Figs.~\ref{fig:fig9} and \ref{fig:fig10}; in particular, $\delta\phi$ is
evolved dynamically together with all fluid and metric perturbations.  For all
spectra below, the density parameters in Eq.~\eqref{LCDMparams} are kept
fixed.  We also set the reduced Hubble constant to $h_{100}=0.67810$ and use
the primordial curvature spectrum
${\cal P}_{\cal R}(k)=A_s(k/k_*)^{n_s-1}$ with
$A_s=2.1\times10^{-9}$, $n_s=0.9649$, and the pivot wavenumber 
$k_*=0.05\,{\rm Mpc}^{-1}$.  
The Thomson optical depth associated with reionization is fixed to $\kappa_{\rm T,reio}=0.0544$.
We vary only the model parameters defining the
three stable background solutions in Eq.~\eqref{stableCasesTableBackground},
with $a_1=-1$.  As established in Secs.~\ref{subsec:stability_background}
and \ref{persec}, these solutions satisfy all linear stability conditions
throughout the numerical interval.

\subsection{Matter power spectrum}
\label{subsec:matterP}

The response parametrization in Sec.~\ref{sec:superHubble} shows that the
large-scale behavior depends on the undetermined transfer factor
${\cal T}_{\Delta}(k,N_{\rm p})$. Here we evaluate the complete
finite-wavelength response with CLASS, which reveals a finite enhancement
toward the lowest wavenumbers covered by the calculation. By contrast,
smaller-scale modes, which enter the Hubble radius earlier, are governed
mainly by quasi-static dynamics: for $\beta>0$, the CDM inertia factor
$q_c=1+4\beta x_1^2/\Omega_c$ exceeds unity, reducing the response of CDM
to the gravitational potentials and suppressing its growth.

\begin{figure}[t]
 \centering
 \includegraphics[width=\columnwidth]{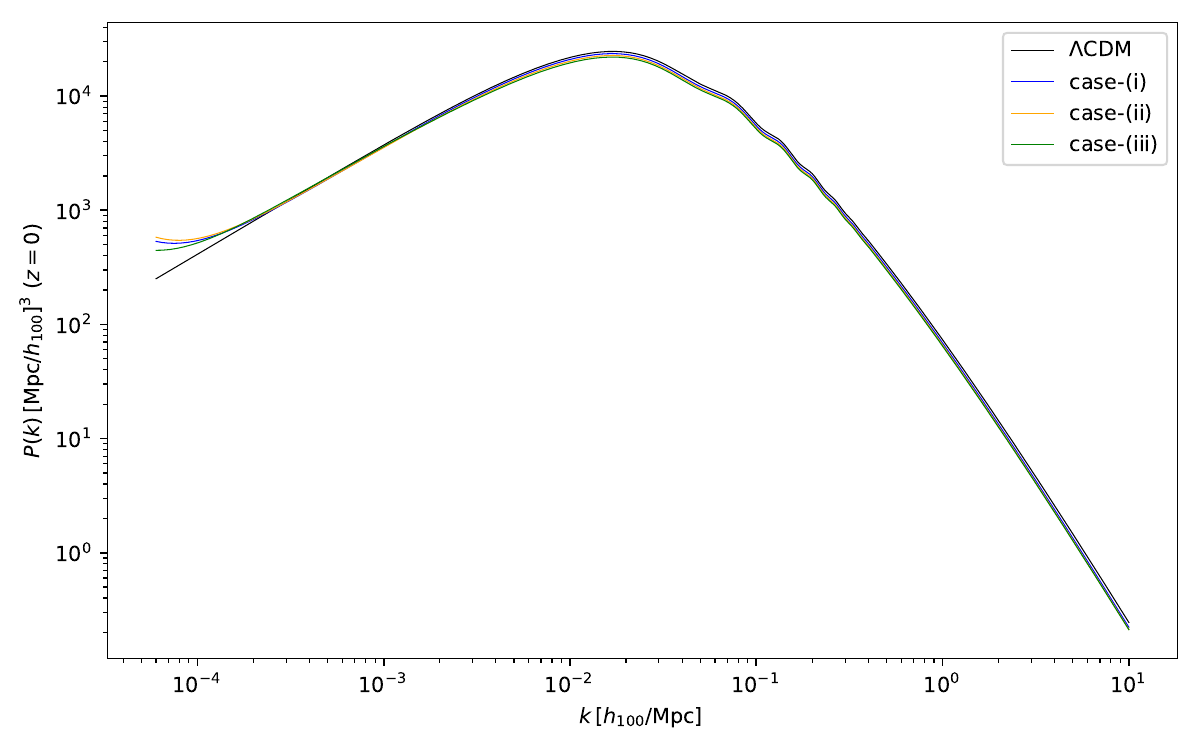}
 \caption{CLASS linear power spectra $P(k)$ of the comoving total-matter
 density contrast $\Delta_m$ at $z=0$ for the three stable parameter
 choices (i), (ii), and (iii) defined in Eq.~\eqref{stableCasesTableBackground}.
 The fixed reference cosmology uses the density parameters in
 Eq.~\eqref{LCDMparams}, together with $h_{100}=0.67810$,
 $A_s=2.1\times10^{-9}$, $n_s=0.9649$, and
 $\kappa_{\rm T, reio}=0.0544$.
 The black curve is the corresponding $\Lambda$CDM spectrum, while the blue,
 orange, and green curves show the three model cases.  The horizontal and
 vertical axes use $k$ in $h_{100}\,{\rm Mpc}^{-1}$ and $P(k)$ in $({\rm Mpc}/h_{100})^3$, respectively.
At the lowest wavenumber shown,
$k\simeq10^{-4}\,h_{100}\,{\rm Mpc}^{-1}$, the enhancement relative to
$\Lambda$CDM reflects the fully coupled large-scale perturbation response
generated around the transient braiding peak. By contrast, the suppression
on smaller scales is governed primarily by the momentum-transfer coupling
$\beta$.}
 \label{fig:fig11}
\end{figure}

Figure~\ref{fig:fig11} shows the matter power spectra computed with CLASS over
the wide wavenumber range
$10^{-4}\lesssim k/(h_{100}\,{\rm Mpc}^{-1})\lesssim10$. The enhancement
toward the low-$k$ edge is the finite-wavelength response parametrized in
Eq.~\eqref{MatterEnhanceScaling}. Its scale dependence is determined by
${\cal T}_{\Delta}(k,N_{\rm p})$ through the full CLASS evolution and
should not be interpreted as a universal asymptotic power law. For modes with
$k\gtrsim a_{\rm p}H_{\rm p}$, which are close to or inside the Hubble radius
during the transient peak, the super-Hubble approximation breaks down and the
enhancement is moderated by the full coupled dynamics of radiation, the
scalar field, and the metric perturbations. 
For modes deep inside the Hubble radius during the transient peak,
$k\gg a_{\rm p}H_{\rm p}$, the increased CDM inertia instead dominates
the response, leading to a suppression relative to $\Lambda$CDM.

\begin{figure}[t]
 \centering
 \includegraphics[width=\columnwidth]{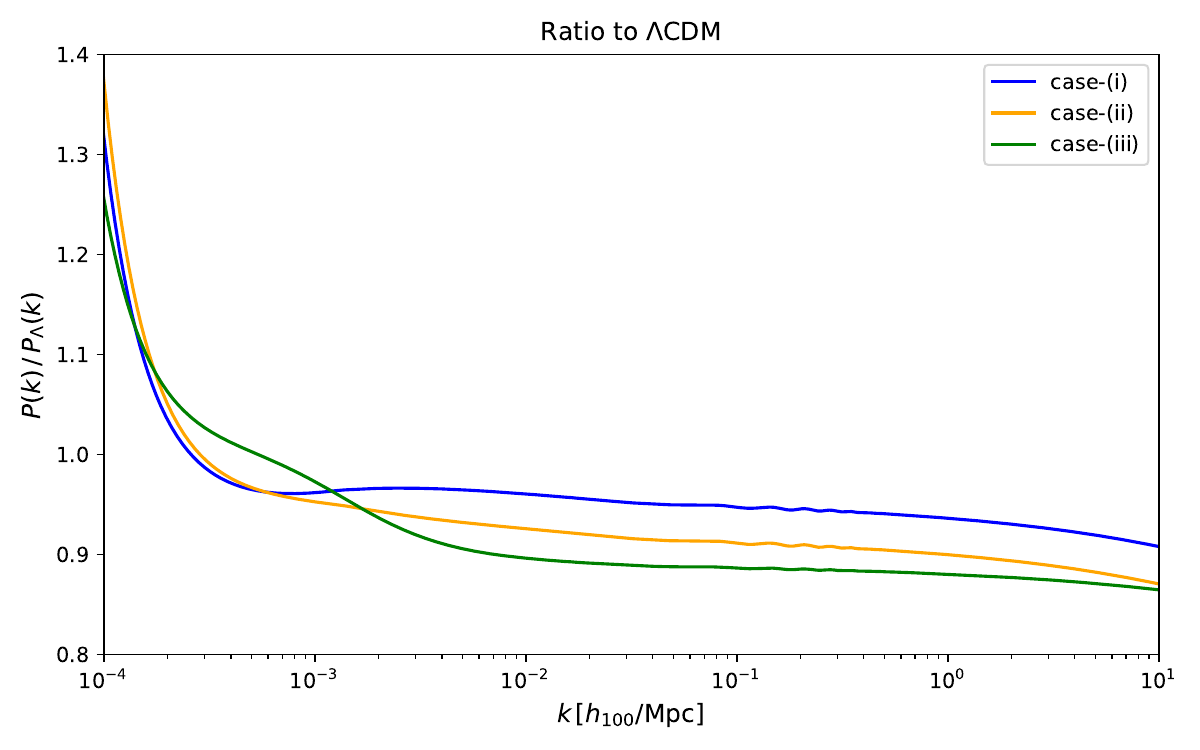}
\caption{Ratio of the CLASS linear matter power spectra for cases (i),
(ii), and (iii) to the corresponding $\Lambda$CDM spectrum,
$P(k)/P_{\Lambda}(k)$, for the same three stable parameter choices as in
Fig.~\ref{fig:fig11}.  The enhancement near the lowest wavenumber shown,
$k\simeq10^{-4}\,h_{100}\,{\rm Mpc}^{-1}$, is associated with the coupled
response of the scalar-field, metric, and matter perturbations during the
transient $x_3$ peak.  On smaller scales, the momentum-transfer coupling
suppresses the power.  The suppression follows the hierarchy of $\beta$ and
is strongest in case (iii), which has the largest $\beta$.}
 \label{fig:fig12}
\end{figure}

The ratio plot in Fig.~\ref{fig:fig12} makes the two competing effects
more transparent. At
$k\simeq10^{-4}\,h_{100}\,{\rm Mpc}^{-1}$, the ratios are approximately
$1.29$, $1.35$, and $1.24$ for cases (i), (ii), and (iii), respectively.
This ordering broadly tracks the transient-braiding amplitude, since
$x_3^{\rm peak}$ increases in the order (iii), (i), and (ii).

The examples shown in Figs.~\ref{fig:fig11} and \ref{fig:fig12} have
$x_3^{\rm peak}={\cal O}(10^{-3})$, for which the enhancement becomes
apparent around
$k\simeq10^{-4}\,h_{100}\,{\rm Mpc}^{-1}$. Additional CLASS calculations
with larger values of $x_3^{\rm peak}$ confirm that the enhanced region
extends toward larger wavenumbers and that, at a fixed $k$ in the
large-scale regime, $P(k)/P_\Lambda(k)$ generally increases with
$x_3^{\rm peak}$. These trends should not be interpreted as an exact
one-parameter prediction, because the finite width of the peak, the
scalar-field and radiation perturbations, and the metric response all
enter the full Boltzmann evolution.

By contrast, on smaller scales, the dominant trend is the
$\beta$-dependent suppression of CDM growth. Around
$k\simeq0.1\,h_{100}\,{\rm Mpc}^{-1}$, one finds
$P(k)/P_{\Lambda}(k)\simeq0.95$, $0.91$, and $0.89$ for cases (i),
(ii), and (iii), respectively. The suppression is strongest for case
(iii), which has the largest $\beta$. Thus, the small-scale suppression
is ordered primarily by the momentum-transfer hierarchy, whereas the
enhancement toward the lowest wavenumbers is governed mainly by the
fully coupled perturbation response during the transient braiding epoch.

\subsection{CMB power spectrum}
\label{subsec:CMBP}

The CMB temperature power spectrum probes the same metric perturbations
through a different combination of source terms. On large angular scales,
the anisotropies receive the ordinary Sachs--Wolfe contribution at last
scattering and the integrated Sachs--Wolfe (ISW) contribution accumulated
along the line of sight. By contrast, the acoustic-scale structure is governed
primarily by photon--baryon oscillations before and at last scattering. We
therefore discuss the large-angle and acoustic-scale regimes separately.

We first consider the large-angle range $2\leq\ell\leq30$, 
where $\ell$ denotes the CMB angular multipole and smaller values of $\ell$ correspond to larger angular scales.  
Let $\tau$ be conformal time, defined by
${\rm d}\tau={\rm d}t/a$, and let $\tau_0$ denote 
its present value. 
We write the observed dimensionless temperature
anisotropy as
\be
\begin{aligned}
\frac{\Delta T(\hat{\bm n})}{T_0}
&=\int \frac{{\rm d}^3 k}{(2\pi)^3}\,{\cal R}({\bm k})
\sum_{\ell=0}^{\infty}(-i)^\ell(2\ell+1)
\\
&\quad\times
\Theta_\ell(k,\tau_0)P_\ell(\hat{\bm k}\cdot\hat{\bm n})\,.
\end{aligned}
\label{ThetaMultipoleDef}
\ee
Here $T_0$ is the present-day mean CMB temperature, 
$\hat{\bm n}$ is the line-of-sight unit vector, 
${\bm k}$ is the comoving wavevector with $\hat{\bm k}\equiv{\bm k}/k$, and $P_\ell$ is the Legendre polynomial of
multipole order $\ell$. 
The quantity ${\cal R}({\bm k})$ is the primordial comoving curvature perturbation, and $\Theta_\ell(k,\tau_0)$ is the corresponding linear
temperature transfer multipole evaluated at the present conformal time. Equivalently, with
$\Delta T(\hat{\bm n})=\sum_{\ell m}a_{\ell m}^{\rm T}Y_{\ell m}(\hat{\bm n})$,
statistical isotropy gives
\be
\left\langle a_{\ell m}^{\rm T}a_{\ell' m'}^{{\rm T}\,*}\right\rangle
=C_\ell^{\rm TT}\delta_{\ell\ell'}\delta_{mm'}\,.
\label{ClTTdef}
\ee
Here $Y_{\ell m}$ are spherical harmonics, $a_{\ell m}^{\rm T}$ are their
temperature-anisotropy coefficients, and $\langle\cdots\rangle$ denotes an
ensemble average.  For the primordial spectrum
$\langle {\cal R}({\bm k}){\cal R}^{*}({\bm k}')\rangle
=(2\pi)^3\delta^{(3)}({\bm k}-{\bm k}')
(2\pi^2/k^3){\cal P}_{\cal R}(k)$, the temperature spectrum is
\be
C_\ell^{\rm TT}=4\pi T_0^2
\int {\rm d} \ln k\,{\cal P}_{\cal R}(k)\,
\Theta_\ell^2(k,\tau_0)\,.
\label{ClTTTransfer}
\ee
Here ${\cal P}_{\cal R}(k)$ is the dimensionless primordial curvature power
spectrum. For a blackbody photon perturbation, the temperature monopole is the
angular average of the photon temperature fluctuation and satisfies
$\Theta_0=\delta_\gamma/4$, where $\delta_\gamma$ is the photon density
contrast. Neglecting the Doppler and polarization sources and adopting the
instantaneous-last-scattering approximation, the line-of-sight solution for
the temperature multipoles can be written schematically
as~\cite{Hu:1994uz,Seljak:1996is}
\be
\begin{aligned}
\Theta_\ell(k,\tau_0)&\simeq{}
\left[\Theta_0(k,\tau_*)+\Psi(k,\tau_*)\right]
j_\ell\!\left[k(\tau_0-\tau_*)\right]
\\
&+\int_{\tau_*}^{\tau_0}{\rm d}\tau\,
e^{-\kappa_{\rm T}(\tau)}
\left(\Phi'+\Psi'\right)
j_\ell\!\left[k(\tau_0-\tau)\right]
+\cdots\,,
\end{aligned}
\label{ThetaLOSsource}
\ee
where $\tau_*$ is the conformal time of last scattering, $j_\ell$ is
the spherical Bessel function, and a prime in this subsection denotes
${\rm d}/{\rm d}\tau$. The quantity $\kappa_{\rm T}(\tau)$ is the Thomson
optical depth accumulated between conformal time $\tau$ and today,
with $\kappa_{\rm T}(\tau_0)=0$. The first term represents the last-scattering
contribution obtained after integrating the visibility function in the
instantaneous-last-scattering approximation, whereas the second term is
the ISW contribution. Since the latter is generated continuously along
the photon trajectory, its integrand is weighted by
$e^{-\kappa_{\rm T}(\tau)}$, the probability that a photon propagates from
conformal time $\tau$ to the observer without rescattering.

The part accumulated shortly after last scattering is conventionally
called the early ISW contribution, whereas the part generated during
DE domination is the late-time ISW contribution~\cite{Dodelson:2020}.
The full visibility function, including
reionization effects, is retained in the numerical CLASS calculation.

For adiabatic perturbations on super-Hubble scales in the idealized
matter-dominated limit, anisotropic stress is negligible and
$\Psi=\Phi$ in the metric convention of Eq.~\eqref{Newtonianmetric}.
The ordinary Sachs--Wolfe source 
at last scattering then satisfies
\be
\Theta_0(k,\tau_*)+\Psi(k,\tau_*)
\simeq \frac{1}{3}\Phi(k,\tau_*)\,.
\label{ThetaPsiPhiThird}
\ee

For the modes contributing mainly to $2\leq\ell\leq30$, transient
braiding around radiation--matter equality changes both the metric potential
at last scattering and its subsequent approach to the matter-era solution.
Figure~\ref{fig:fig9} illustrates the corresponding large-scale behavior for
a representative mode: the normalized curvature potential lies below the
$\Lambda$CDM result in all three cases.  This modifies both the ordinary
Sachs--Wolfe source in Eq.~\eqref{ThetaPsiPhiThird} and the early-ISW
integral in Eq.~\eqref{ThetaLOSsource}.  Because these contributions are
coherent and correlated, their net change cannot be inferred from the
potential amplitude alone; the full CLASS result retains their sum.

The metric potential evolves again after the onset of DE domination, producing the late-time ISW contribution.  Momentum transfer affects this
part through the $\beta$-dependent CDM inertia and the resulting late-time
evolution of the metric potentials.  The low-$\ell$ spectrum therefore
reflects the coherent sum and interference of the ordinary Sachs--Wolfe,
early ISW, and late-time ISW sources rather than any one contribution in
isolation.

\begin{figure}[t]
\centering
\includegraphics[width=\columnwidth]{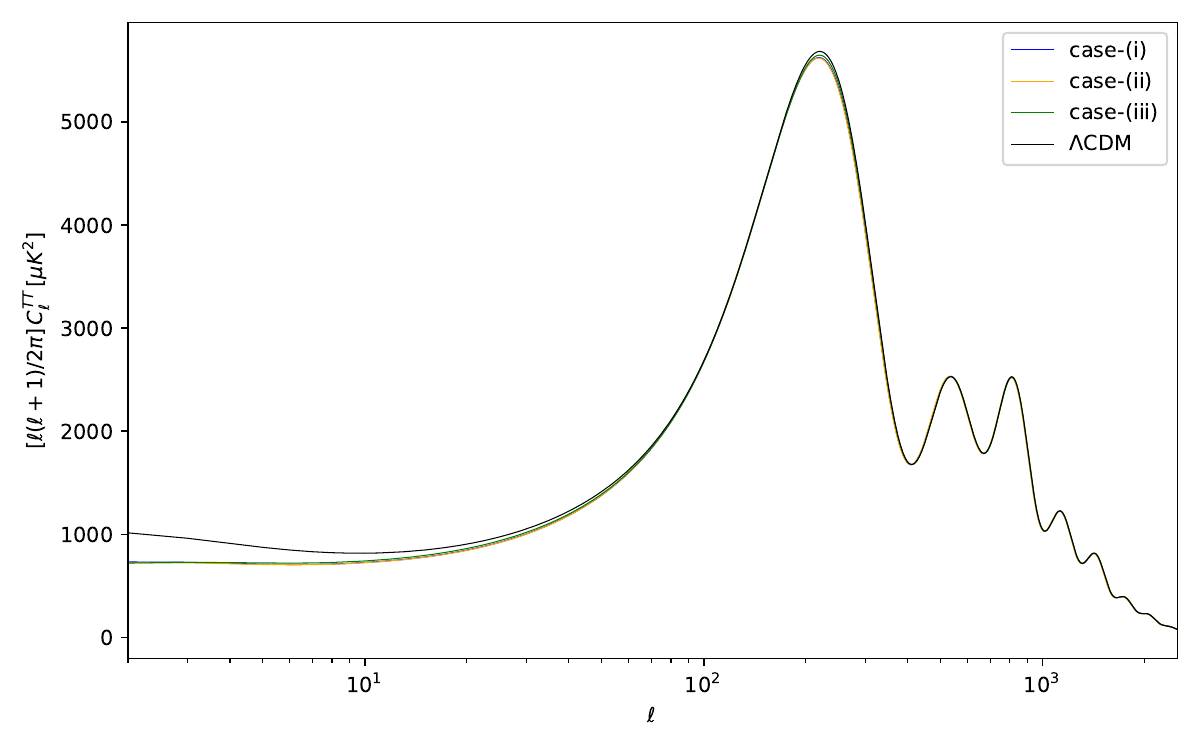}
\caption{CMB temperature power spectra computed with CLASS for the same three model
cases as in Figs.~\ref{fig:fig11} and \ref{fig:fig12}, together with the $\Lambda$CDM spectrum.  The scalar-field perturbation is included in the
full Boltzmann evolution.  The vertical axis shows
$\ell(\ell+1)C_\ell^{\rm TT}/(2\pi)$ in units of $\mu{\rm K}^2$.
All three model cases have lower power than $\Lambda$CDM over
$2\leq\ell\leq30$.  In the acoustic range, the spectra remain close to the
$\Lambda$CDM result, with a slight shift of the first acoustic peak toward
smaller $\ell$ and a modest reduction in its height.}
 \label{fig:fig13}
\end{figure}

\begin{figure}[t]
 \centering
 \includegraphics[width=\columnwidth]{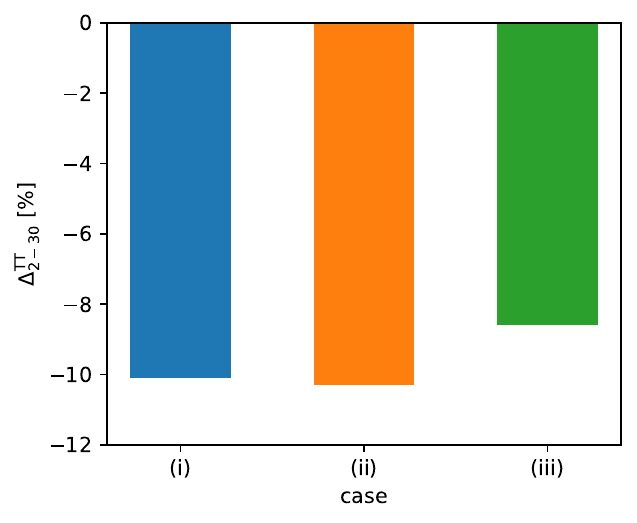}
 \caption{Fractional change of the large-angle CMB temperature power over
 $2\leq\ell\leq30$, defined by Eq.~\eqref{lowellTTchange}, for cases (i),
 (ii), and (iii).  The blue, orange, and green bars are computed from the
 CLASS spectra in Fig.~\ref{fig:fig13}.  All three changes are negative.}
 \label{fig:fig14}
\end{figure}

Figure~\ref{fig:fig13} shows the resulting suppression at large angular
scales.  This does not contradict the enhancement of the matter power
spectrum toward the lowest wavenumbers shown in Fig.~\ref{fig:fig12},
because the latter is governed by the comoving density response,
whereas the CMB temperature anisotropy depends directly on the metric
potentials and their line-of-sight evolution.  
To quantify the percentage change in the large-angle CMB temperature power relative to $\Lambda$CDM over $2\leq\ell\leq30$, we define
\be
\begin{aligned}
\Delta_{2-30}^{\rm TT}
&\equiv
100\left[
\frac{\sum_{\ell=2}^{30}D_\ell^{\rm model}}
 {\sum_{\ell=2}^{30}D_\ell^{\Lambda{\rm CDM}}}-1\right],
\\
D_\ell&\equiv \frac{\ell(\ell+1)C_\ell^{\rm TT}}{2\pi}\,.
\end{aligned}
\label{lowellTTchange}
\ee
Figure~\ref{fig:fig14} shows the values extracted from the spectra in
Fig.~\ref{fig:fig13}.  They are approximately $-10.1\%$, $-10.3\%$, and
$-8.6\%$ for cases (i), (ii), and (iii), respectively.
The magnitude of the departure increases in the order (iii), (i), and (ii), which
matches the ordering of $x_3^{\rm peak}$ rather than the monotonic hierarchy
in $\beta$.  This correspondence suggests that transient early-time
braiding is the main origin of the case-to-case ordering of the low-$\ell$
suppression.  The late-time, $\beta$-dependent ISW contribution changes the
precise amplitudes and can reinforce or partly cancel the early contribution,
so the percentages must be obtained from the complete line-of-sight
calculation.

The large-angle suppression found here is qualitatively similar to that obtained 
in the Galileon ghost condensate model, which follows from the present action 
by setting $V(\phi)=0$ and $\beta=0$~\cite{Peirone:2019aua}. In that analysis,
the reduced low-$\ell$ power improves the CMB fit and contributes to the statistical 
preference over $\Lambda$CDM. Since the cubic Galileon interaction is common 
to the two models, this similarity suggests that kinetic braiding plays an important role 
in producing the suppression. In the present model, however, the scalar potential and 
momentum transfer modify its precise amplitude through the late-time evolution of the metric potentials. Determining whether the suppression also improves the global fit in the present model 
requires a dedicated likelihood analysis.

A complementary probe of the same late-time metric evolution is provided
by ISW--galaxy cross-correlations.  Scalar Galileon DE models can predict a
negative cross-correlation, in tension with observational indications of a
positive signal~\cite{Kimura:2011td,Barreira:2012kk,Renk:2017rzu,Kable:2021yws}.
In the present model, both the scalar potential and the
momentum-transfer-induced weakening of gravity in the CDM sector modify
the evolution of the metric potentials and matter perturbations.  They
could therefore alter the sign and amplitude of the ISW--galaxy
cross-correlation relative to those in scalar Galileon models.  A dedicated
calculation over the relevant redshift and multipole ranges is needed to
establish whether the tension is alleviated, 
and we leave this analysis for future work.

We next consider the primary acoustic range
$30<\ell\lesssim10^3$.  The direct momentum-transfer correction is
proportional to $q_c-1=4\beta x_1^2/\Omega_c$ and is very small before
recombination, as follows from the early-time hierarchy in
Eq.~\eqref{earlyhierarchy}.  The pronounced $\beta$-dependent suppression
of CDM growth and the associated change of the metric potentials occur mainly
at low redshifts, after the primary acoustic source has formed.  They therefore
have little influence on the primary peak phase and amplitude, entering the
CMB mainly through late-time secondary contributions such as the ISW effect
and lensing.  By contrast, the transient $x_3$ peak occurs around
radiation--matter equality and directly modifies the pre-recombination
expansion and gravitational driving. 
It therefore primarily controls the reduction 
of the sound horizon and the phase-dependent 
pre-recombination driving, whereas the geometrical
acoustic scale also reflects the integrated 
expansion history from last
scattering to the present epoch.

The geometric part of the peak positions is summarized by the CMB shift
parameters
\be
\begin{aligned}
{\cal R}_{\rm sh}&\equiv \sqrt{\Omega_{m0}}H_0D_M(z_*)\,,\\
\ell_A&\equiv \frac{\pi D_M(z_*)}{r_s(z_*)}
=\frac{\pi {\cal R}_{\rm sh}}
{\sqrt{\Omega_{m0}}H_0 r_s(z_*)}\,.
\end{aligned}
\label{CMBshiftParams}
\ee
Here $\Omega_{m0}=\Omega_{c0}+\Omega_{b0}$ is the present total density
parameter of nonrelativistic matter, $z_*$ is the photon last-scattering
redshift, and
$D_M(z_*)=\int_0^{z_*}{\rm d}z/H(z)$ is the comoving angular-diameter
distance in a spatially flat background. 
The comoving sound horizon at
last scattering is
\be
r_s(z_*)=\int_0^{\tau_*}c_\gamma(\tau)\,{\rm d}\tau
=\int_{z_*}^{\infty}\frac{c_\gamma(z)}{H(z)}\,{\rm d}z\,,
\label{rsdefCMB}
\ee
where $\tau_*$ is the conformal time corresponding to $z_*$ and
$c_\gamma$ is the photon--baryon sound speed, satisfying
$c_\gamma^2=1/[3(1+R_b)]$. Here
$R_b\equiv3\rho_b/(4\rho_\gamma)$ is the baryon-to-photon inertia ratio,
and $\rho_\gamma$ is the photon energy density. 
Since $\Omega_{m0}$ and
$H_0$ are fixed,
\be
\frac{\Delta\ell_A}{\ell_A}
\simeq
\frac{\Delta {\cal R}_{\rm sh}}{{\cal R}_{\rm sh}}
-
\frac{\Delta r_s}{r_s}
\label{ellARshRelation}
\ee
at linear order.  Here and below, each fractional shift is defined relative
to $\Lambda$CDM; for example,
$\Delta\ell_A/\ell_A\equiv
(\ell_A^{\rm model}-\ell_A^{\Lambda{\rm CDM}})/
\ell_A^{\Lambda{\rm CDM}}$,
and analogously for ${\cal R}_{\rm sh}$ and $r_s$.
For $\Lambda$CDM and all three cases of the interacting model, the photon-decoupling redshift is
$z_*=1088.28$.  
The corresponding $\Lambda$CDM reference values are
\be
\begin{aligned}
r_s^{\Lambda{\rm CDM}}(z_*)&=143.031376~{\rm Mpc},\\
D_M^{\Lambda{\rm CDM}}(z_*)&=13694.663097~{\rm Mpc},\\
{\cal R}_{\rm sh}^{\Lambda{\rm CDM}}&=1.752256,\\
\ell_A^{\Lambda{\rm CDM}}&=300.7945.
\end{aligned}
\label{LCDMshiftReference}
\ee
These values are obtained from the CLASS output for the chosen parameter values.
For the three stable cases (i), (ii), and (iii), the transient braiding
variable reaches its maximum at approximately the same redshift,
$z_{\rm peak}\simeq5275.5$, whereas its peak amplitude differs among the cases. The peak properties, CMB shift parameters, sound horizons, and comoving
angular-diameter distances obtained from the same CLASS runs used for
Fig.~\ref{fig:fig13} are listed below, 
together with their fractional
changes relative to $\Lambda$CDM:
\be
\resizebox{0.98\columnwidth}{!}{$
\begin{gathered}
\begin{array}{c|cccc}
\text{Model}
 & x_3^{\rm peak}
 & z_{\rm peak}
 & {\cal R}_{\rm sh}
 & \ell_A
\\
\hline
\Lambda{\rm CDM}
 & -
 & -
 & 1.752256
 & 300.7945
\\
{\rm (i)}
 & 3.873008\times10^{-3}
 & 5275.5
 & 1.748379
 & 300.6263
\\
{\rm (ii)}
 & 4.166489\times10^{-3}
 & 5275.5
 & 1.745106
 & 300.1013
\\
{\rm (iii)}
 & 2.286614\times10^{-3}
 & 5275.5
 & 1.753524
 & 301.3063
\end{array}
\\[1mm]
\begin{array}{c|ccccc}
\text{Model}
 & \substack{r_s(z_*)\\ {\rm [Mpc]}}
 & \substack{D_M(z_*)\\ {\rm [Mpc]}}
 & \Delta{\cal R}_{\rm sh}/{\cal R}_{\rm sh}
 & \Delta r_s/r_s
 & \Delta\ell_A/\ell_A
\\
\hline
\Lambda{\rm CDM}
 & 143.031376
 & 13694.663097
 & -
 & -
 & -
\\
{\rm (i)}
 & 142.794773
 & 13664.362437
 & -0.2213\%
 & -0.1654\%
 & -0.0559\%
\\
{\rm (ii)}
 & 142.776816
 & 13638.785056
 & -0.4080\%
 & -0.1780\%
 & -0.2305\%
\\
{\rm (iii)}
 & 142.891769
 & 13704.574254
 & +0.0724\%
 & -0.0976\%
 & +0.1701\%
\end{array}
\end{gathered}
$}
\label{shiftParamTable}
\ee
Here all fractional differences are defined relative to the
$\Lambda$CDM reference model.  Since $\Omega_{m0}$ and $H_0$ are held
fixed, Eq.~\eqref{CMBshiftParams} implies
\be
\frac{\Delta{\cal R}_{\rm sh}}{{\cal R}_{\rm sh}}
=
\frac{\Delta D_M}{D_M}.
\label{RshDMrelation}
\ee
The reduction of the sound horizon is largest in case (ii), followed
by cases (i) and (iii).  This ordering,
\be
\left|\frac{\Delta r_s}{r_s}\right|_{\rm (iii)}
<
\left|\frac{\Delta r_s}{r_s}\right|_{\rm (i)}
<
\left|\frac{\Delta r_s}{r_s}\right|_{\rm (ii)},
\ee
agrees with the ordering of the transient braiding amplitudes,
\be
x_{3,{\rm (iii)}}^{\rm peak}
<
x_{3,{\rm (i)}}^{\rm peak}
<
x_{3,{\rm (ii)}}^{\rm peak}.
\ee
This correspondence indicates that the modification of the
pre-recombination sound horizon is controlled predominantly by the
transient early-time braiding.

The change in the acoustic scale is determined by the competition
between the distance and sound-horizon shifts.  The exact relation is
\be
1+\frac{\Delta\ell_A}{\ell_A}
=
\frac{1+\Delta D_M/D_M}
     {1+\Delta r_s/r_s},
\label{ellAExactShift}
\ee
which, at linear order, reduces to Eq.~\eqref{ellARshRelation}.
The fractional values reported in
Eq.~\eqref{shiftParamTable} are calculated directly from the numerical
CLASS outputs, rather than from the linearized relation.

In case (i), both $D_M(z_*)$ and $r_s(z_*)$ are smaller than their
$\Lambda$CDM values.
The fractional reduction in $D_M(z_*)$, $0.2213\%$, is larger than the
$0.1654\%$ reduction in $r_s(z_*)$, yielding
$\Delta\ell_A/\ell_A\simeq-0.0559\%$. 
The geometrical acoustic scale is therefore shifted slightly 
toward smaller multipoles.

The same competition is more pronounced in case (ii). Relative to their
$\Lambda$CDM values, $D_M(z_*)$ and $r_s(z_*)$ are reduced by
$0.4080\%$ and $0.1780\%$, respectively. Consequently,
$\Delta\ell_A/\ell_A\simeq-0.2305\%$, which is the most negative
acoustic-scale shift among the three cases. 

Case (iii) behaves differently. Relative to $\Lambda$CDM,
$r_s(z_*)$ decreases by $0.0976\%$, whereas $D_M(z_*)$ increases by
$0.0724\%$. Both changes increase
$\ell_A=\pi D_M(z_*)/r_s(z_*)$, producing the positive shift
$\Delta\ell_A/\ell_A\simeq+0.1701\%$. Thus, although the reduction in
$r_s(z_*)$ follows the ordering of $x_3^{\rm peak}$, the resulting
acoustic-scale shift does not and can even change sign. This is because
$r_s(z_*)$ is governed mainly by the pre-recombination expansion,
including the transient braiding epoch, whereas $D_M(z_*)$ depends on
the integrated post-recombination expansion from last scattering to the
present.

The common braiding peak redshift is also consistent with the analytic
estimate derived in Eq.~\eqref{OmegaPeak}. Since the interaction does not
transfer energy at the background level, radiation and nonrelativistic
matter scale as $\rho_r\propto a^{-4}$ and $\rho_m\propto a^{-3}$,
respectively. Moreover, $\Omega_r/\Omega_m=\rho_r/\rho_m$, and
$\rho_r=\rho_m$ at $z=z_{\rm eq}$. Using the equality redshift obtained
from the corresponding CLASS background, $z_{\rm eq}\simeq3517$, together
with $z_{\rm peak}\simeq5275.5$, we obtain
\be
\frac{\Omega_r^{\rm peak}}{\Omega_m^{\rm peak}}
=
\frac{1+z_{\rm peak}}{1+z_{\rm eq}}
=
\frac{5276.5}{3518}
\simeq1.50.
\ee
This agrees with the analytic result
$\Omega_r^{\rm peak}/\Omega_m^{\rm peak}\simeq3/2$ in
Eq.~\eqref{OmegarOmeagmPeakRatio}. The redshift of the transient braiding
peak is therefore set mainly by the radiation--matter transition, whereas
its amplitude depends on the model parameters.

The shift parameters in Eq.~\eqref{shiftParamTable} encode the geometrical contribution to the acoustic-peak positions through $D_M(z_*)$ and
$r_s(z_*)$. They do not by themselves determine the precise locations or
heights of the CMB temperature peaks, which also depend on the phase shifts
and gravitational driving of the photon--baryon oscillations. To describe these effects, we define
\be
S_\gamma(k,\tau)\equiv\Theta_0(k,\tau)+\Psi(k,\tau)\,,
\label{SgammaDef}
\ee
whose evolution in the tight-coupling regime takes the schematic form
\be
S_\gamma''+\frac{R_b'}{1+R_b}S_\gamma'
+k^2c_\gamma^2S_\gamma={\cal F}_{\Phi\Psi}\,,
\label{AcousticOscillatorSchematic}
\ee
where ${\cal F}_{\Phi\Psi}$ encodes the driving by the metric potentials. 
After integration by parts, the driven part of the
last-scattering source contains terms of the schematic form
\ba
\hspace{-0.5cm}
\Delta S_\gamma(k,\tau_*)
&\sim& \int^{\tau_*} \! {\rm d}\tau\,
(\Phi'+\Psi')
\nonumber \\
\hspace{-0.5cm}
& &\times
\cos\!\left[k\{r_s(\tau_*)-r_s(\tau)\}\right]+\cdots,
\label{AcousticDrivingIntegral}
\ea
where $r_s(\tau)=\int_0^\tau c_\gamma(\tilde{\tau})
{\rm d}\tilde{\tau}$.  Equation~\eqref{AcousticDrivingIntegral} describes
pre-recombination gravitational driving, not the post-recombination ISW
integral in Eq.~\eqref{ThetaLOSsource}.

Schematically, the position of the first temperature peak can be written as
$\ell_1\simeq\ell_A(1-\varphi_1)$, where $\varphi_1$ is the effective
phase shift generated by the pre-recombination evolution of the metric
potentials and radiation perturbations~\cite{Dodelson:2020}. In this
convention, a larger $\varphi_1$ shifts the peak toward smaller $\ell$.

For cases (i) and (ii), the negative values of $\Delta\ell_A$ move the
geometrical acoustic scale toward lower multipoles, whereas the positive
$\Delta\ell_A$ in case (iii) moves it toward higher multipoles. The precise
location of the first peak is determined by the combined changes in
$\ell_A$ and the phase shift $\varphi_1$. Therefore, the net displacement
toward lower multipoles seen in Fig.~\ref{fig:fig13} for case (iii),
despite its positive $\Delta\ell_A$, requires an increase in $\varphi_1$
relative to $\Lambda$CDM large enough to overcome the geometrical shift.

The first peak is also slightly lower than its $\Lambda$CDM counterpart.
Since the standard baryon density, primordial spectrum, and reionization
optical depth are held fixed, the peak-height differences arise mainly
from the modification of the pre-recombination metric driving induced by
the transient $x_3$ peak around radiation--matter equality. The direct
$\beta$-dependent momentum-transfer effect remains subdominant at that
epoch. The full CLASS calculation retains the Doppler terms,
diffusion damping, reionization, and the finite width of the visibility
function.

\section{Conclusions}
\label{sec:conclude}

We have investigated a potential-extended cubic-Galileon DE model with
elastic momentum exchange between the scalar field and CDM.  The model
combines two complementary mechanisms: the scalar potential explicitly
breaks shift symmetry and allows an upward crossing from
$w_{\rm DE}<-1$ to $w_{\rm DE}>-1$, while the interaction $\beta Z^2$
increases the dynamical inertia of CDM without modifying its background
dilution law.  We derived the background and Newtonian-gauge perturbation
equations, obtained analytic approximations in the relevant cosmological
regimes, and evolved the complete linear perturbation system using a
modified implementation of CLASS.

The background analysis identified a viable branch satisfying
$A=a_1+2\beta<0$, $x_2>0$, $x_3>0$, $\lambda x_1>0$, and $x_4>0$.
At high redshift, the cubic Galileon term gives the dominant contribution
to the subdominant DE density. The system subsequently enters a stable
phantom regime at intermediate redshift. At lower redshift, the scalar
potential becomes dynamically important and changes the relative evolution
of the background contributions. The decomposition
${\cal C}={\cal C}_0+{\cal B}x_4$ shows that the competition among these
contributions, rather than any single term, produces the second crossing
back to $w_{\rm DE}>-1$. All three representative
solutions satisfy $q_s>0$, $q_c>0$, and $c_s^2>0$ throughout the numerical
interval and asymptotically approach the stable Galileon de Sitter branch.

For perturbations deep inside the Hubble radius, where the quasi-static
approximation is valid, the $\beta$-dependent momentum transfer affects
CDM growth through $q_c=1+4\beta x_1^2/\Omega_c$ and through its
associated mixing with Galileon braiding. These effects drive $G_c$
below $G$ at low redshifts, whereas $G_b$ remains enhanced above $G$
by Galileon braiding. As $\beta$ increases, the resulting suppression
of CDM clustering becomes more pronounced, leading to a lower
present-day $f\sigma_8$ and reduced matter power on small scales.
Since the displayed growth histories share the same present-day
$\sigma_8$ normalization, however, their $f\sigma_8$ curves need not
remain below the $\Lambda$CDM prediction at intermediate redshifts.

The transient $x_3$ peak near radiation--matter equality generates a
distinctive perturbation response on very large scales. The analytic
source decomposition identifies a contribution proportional to the peak
area ${\cal A}_3$, although the sign and magnitude of the full response
are determined by the coupled evolution of the metric, scalar-field,
matter, and radiation perturbations. For the illustrative Fourier mode
$k=H_0$ considered in Sec.~\ref{sec:superHubble}, the curvature potential
and CDM velocity potential, normalized at the initial epoch, are suppressed
relative to their $\Lambda$CDM counterparts, whereas the similarly
normalized CDM density contrast is enhanced.

At the lowest wavenumbers probed by the CLASS calculation, the combined
density and velocity response enhances the amplitude of the comoving
total-matter density contrast and hence its linear power spectrum. The
analytic parametrization shows that the scale dependence of this
enhancement is governed by ${\cal T}_{\Delta}(k,N_{\rm p})$, whose strict
$k\to0$ behavior is not determined by the peak-area estimate. The
numerical results therefore establish a finite enhancement toward the
largest scales probed, rather than a universal asymptotic power law. On
smaller spatial scales, by contrast, the matter power spectrum is
suppressed, with the suppression becoming stronger as the
momentum-transfer coupling $\beta$ increases.

The CMB temperature spectrum provides a complementary probe of this
large-scale perturbation response. Over the angular scales corresponding
to the low multipoles $2\leq\ell\leq30$, the temperature power is
suppressed relative to $\Lambda$CDM in all three cases. This suppression
arises from the coherent combination of the ordinary Sachs--Wolfe and
early- and late-time ISW contributions and is quantified by
$\Delta_{2-30}^{\rm TT}\simeq-10.1\%$, $-10.3\%$, and $-8.6\%$ for
cases (i), (ii), and (iii), respectively. The case-to-case ordering of
the suppression magnitude follows $x_3^{\rm peak}$ rather than $\beta$,
suggesting that transient braiding near radiation--matter equality largely
controls the differences over this multipole range, while momentum
transfer modifies the late-time ISW contribution.

Across the acoustic-peak region, $30<\ell\lesssim10^3$, the direct
$\beta$-dependent modification of the perturbation dynamics remains
small before recombination. Relative to $\Lambda$CDM, the sound horizon
$r_s(z_*)$ is reduced in all three cases, with the magnitude of its
reduction increasing in the order (iii), (i), and (ii), consistent with
the ordering of $x_3^{\rm peak}$. Since
$\ell_A=\pi D_M(z_*)/r_s(z_*)$, however, the acoustic-scale shift also
depends on the change in $D_M(z_*)$. The resulting fractional shifts are
$\Delta\ell_A/\ell_A\simeq-0.0559\%$, $-0.2305\%$, and $+0.1701\%$
for cases (i), (ii), and (iii), respectively. The shifts are negative
in cases (i) and (ii) because the fractional reduction of $D_M(z_*)$
exceeds that of $r_s(z_*)$. In case (iii), by contrast, $D_M(z_*)$
increases while $r_s(z_*)$ decreases, with both changes increasing
$\ell_A$.

The precise location of the first temperature peak is determined by the
interplay between the geometrical acoustic-scale shift and the effective
phase shift $\varphi_1$ induced by the pre-recombination evolution of the
metric potentials and radiation perturbations. Relative to $\Lambda$CDM,
the peak is shifted toward lower multipoles in all three cases. In case
(iii), this occurs despite $\Delta\ell_A>0$, indicating that the change
in $\varphi_1$ overcomes the geometrical shift. The associated modification
of the gravitational driving of the photon--baryon oscillations slightly
suppresses the first-peak height in all three cases. The reduction of
$r_s(z_*)$ and the changes in $\varphi_1$ and gravitational driving are
therefore mainly controlled by transient braiding near radiation--matter
equality, whereas the change in $D_M(z_*)$ also reflects the integrated
background expansion from last scattering to the present epoch.

The representative solutions studied here were chosen to demonstrate
viable cosmological evolution rather than obtained from a fit to
observational data. A joint Markov chain Monte Carlo analysis incorporating
BAO, Type Ia supernova, CMB, redshift-space-distortion, and weak-lensing
data is therefore an important next step. Such an analysis will establish 
whether the combined signatures of phantom-divide crossing, 
suppressed small-scale growth, and reduced low-$\ell$ CMB power 
improve the fit relative to $\Lambda$CDM.
Further directions include nonlinear structure formation, ISW--galaxy
and lensing cross-correlations, and generalizations to broader classes 
of scalar potentials and
momentum-transfer interactions.

\acknowledgments
We thank Jose Beltr\'an Jim\'enez and Antonio De Felice for useful 
discussions. M.~C.~P. acknowledges support from JSPS KAKENHI Grant No.~26KF0150.
S.~T. acknowledges support from JSPS KAKENHI Grant Nos.~26K07090
and 26H00847, as well as from the Waseda University Grant for
Special Research Projects (Project No.~2026C-486).

\bibliographystyle{mybibstyle}
\bibliography{bib}

@article{Hu:1994uz,
    author = "Hu, Wayne and Sugiyama, Naoshi",
    title = "{Anisotropies in the cosmic microwave background: An Analytic approach}",
    eprint = "astro-ph/9407093",
    archivePrefix = "arXiv",
    doi = "10.1086/175624",
    journal = "Astrophys. J.",
    volume = "444",
    pages = "489--506",
    year = "1995"
}

@article{Seljak:1996is,
    author = "Seljak, Uros and Zaldarriaga, Matias",
    title = "{A Line of sight integration approach to cosmic microwave background anisotropies}",
    eprint = "astro-ph/9603033",
    archivePrefix = "arXiv",
    doi = "10.1086/177793",
    journal = "Astrophys. J.",
    volume = "469",
    pages = "437--444",
    year = "1996"
}

@article{Wolf:2025acj,
    author = "Wolf, William J. and Ferreira, Pedro G. and Garc{\'\i}a-Garc{\'\i}a, Carlos",
    title = "{Cosmological constraints on Galileon dark energy with broken shift symmetry}",
    eprint = "2509.17586",
    archivePrefix = "arXiv",
    primaryClass = "astro-ph.CO",
    doi = "10.1103/bxvj-bsv1",
    journal = "Phys. Rev. D",
    volume = "113",
    number = "2",
    pages = "023551",
    year = "2026"
}

@article{Tsujikawa:2026xqm,
    author = "Tsujikawa, Shinji",
    title = "{Realizing the phantom-divide crossing with vector and scalar fields}",
    eprint = "2601.21274",
    archivePrefix = "arXiv",
    primaryClass = "astro-ph.CO",
    reportNumber = "WUCG-26-01",
    doi = "10.1088/1475-7516/2026/06/009",
    journal = "JCAP",
    volume = "06",
    pages = "009",
    year = "2026"
}

@article{Calderon:2026hbr,
    author = "Calderon, Rodrigo and Linder, Eric V.",
    title = "{Charging Across the Phantom Divide with Modified Gravity}",
    eprint = "2605.26259",
    archivePrefix = "arXiv",
    primaryClass = "gr-qc",
    month = "5",
    year = "2026"
}

@article{Naidoo:2026umv,
    author = "Naidoo, Krishna and Hallam, James and Baker, Tessa and Sirera, Sergi",
    title = "{Constraints on Horndeski Gravity with Phantom Crossing}",
    eprint = "2606.20794",
    archivePrefix = "arXiv",
    primaryClass = "astro-ph.CO",
    month = "6",
    year = "2026"
}

@article{Garcia-Garcia:2026nzy,
    author = "Garc{\'\i}a-Garc{\'\i}a, Carlos and Ferreira, Pedro G. and Wolf, William J.",
    title = "{The Status of Single Scalar Field Dark Energy}",
    eprint = "2607.07777",
    archivePrefix = "arXiv",
    primaryClass = "astro-ph.CO",
    month = "7",
    year = "2026"
}

@article{Hallam:2026qsk,
    author = "Hallam, James and Naidoo, Krishna and Sirera, Sergi and Baker, Tessa",
    title = "{Rolling Galileons: Evolving Braiding Strength for Viable Dark Energy}",
    eprint = "2607.16395",
    archivePrefix = "arXiv",
    primaryClass = "astro-ph.CO",
    month = "7",
    year = "2026"
}

@book{Dodelson:2020,
  author    = {Dodelson, Scott and Schmidt, Fabian},
  title     = {Modern Cosmology},
  edition   = {2},
  publisher = {Academic Press},
  year      = {2021},
  isbn      = {978-0-12-815948-4}
}

@article{Boisseau:2000pr,
    author = "Boisseau, B. and Esposito-Farese, Gilles and Polarski, D. and Starobinsky, Alexei A.",
    title = "{Reconstruction of a scalar tensor theory of gravity in an accelerating universe}",
    eprint = "gr-qc/0001066",
    archivePrefix = "arXiv",
    reportNumber = "CPT-99-P-3917",
    doi = "10.1103/PhysRevLett.85.2236",
    journal = "Phys. Rev. Lett.",
    volume = "85",
    pages = "2236",
    year = "2000"
}

@article{Tsujikawa:2007gd,
    author = "Tsujikawa, Shinji",
    title = "{Matter density perturbations and effective gravitational constant in modified gravity models of dark energy}",
    eprint = "0705.1032",
    archivePrefix = "arXiv",
    primaryClass = "astro-ph",
    doi = "10.1103/PhysRevD.76.023514",
    journal = "Phys. Rev. D",
    volume = "76",
    pages = "023514",
    year = "2007"
}

@article{Lesgourgues:2011re,
    author = "Lesgourgues, Julien",
    title = "{The Cosmic Linear Anisotropy Solving System (CLASS) I: Overview}",
    eprint = "1104.2932",
    archivePrefix = "arXiv",
    primaryClass = "astro-ph.IM",
    month = "4",
    year = "2011"
}

@article{Blas:2011rf,
    author = "Blas, Diego and Lesgourgues, Julien and Tram, Thomas",
    title = "{The Cosmic Linear Anisotropy Solving System (CLASS) II: Approximation schemes}",
    eprint = "1104.2933",
    archivePrefix = "arXiv",
    primaryClass = "astro-ph.CO",
    reportNumber = "CERN-PH-TH-2011-082, LAPTH-010-11",
    doi = "10.1088/1475-7516/2011/07/034",
    journal = "JCAP",
    volume = "07",
    pages = "034",
    year = "2011"
}

@article{BeltranJimenez:2020qdu,
    author = "Beltr{\'a}n Jim{\'e}nez, Jose and Bettoni, Dario and Figueruelo, David and Teppa Pannia, Florencia A. and Tsujikawa, Shinji",
    title = "{Velocity-dependent interacting dark energy and dark matter with a Lagrangian description of perfect fluids}",
    eprint = "2012.12204",
    archivePrefix = "arXiv",
    primaryClass = "astro-ph.CO",
    doi = "10.1088/1475-7516/2021/03/085",
    journal = "JCAP",
    volume = "03",
    pages = "085",
    year = "2021"
}

@article{BeltranJimenez:2021wbq,
    author = "Beltr{\'a}n Jim{\'e}nez, Jose and Bettoni, Dario and Figueruelo, David and Teppa Pannia, Florencia Anabella and Tsujikawa, Shinji",
    title = "{Probing elastic interactions in the dark sector and the role of S8}",
    eprint = "2106.11222",
    archivePrefix = "arXiv",
    primaryClass = "astro-ph.CO",
    doi = "10.1103/PhysRevD.104.103503",
    journal = "Phys. Rev. D",
    volume = "104",
    number = "10",
    pages = "103503",
    year = "2021"
}

@article{Khoury:2003aq,
    author = "Khoury, Justin and Weltman, Amanda",
    title = "{Chameleon fields: Awaiting surprises for tests of gravity in space}",
    eprint = "astro-ph/0309300",
    archivePrefix = "arXiv",
    doi = "10.1103/PhysRevLett.93.171104",
    journal = "Phys. Rev. Lett.",
    volume = "93",
    pages = "171104",
    year = "2004"
}

@article{Amendola:2006we,
    author = "Amendola, Luca and Gannouji, Radouane and Polarski, David and Tsujikawa, Shinji",
    title = "{Conditions for the cosmological viability of f(R) dark energy models}",
    eprint = "gr-qc/0612180",
    archivePrefix = "arXiv",
    doi = "10.1103/PhysRevD.75.083504",
    journal = "Phys. Rev. D",
    volume = "75",
    pages = "083504",
    year = "2007"
}

@article{Peebles:1982ff,
    author  = "Peebles, P. J. E.",
    title   = "{Large-scale background temperature and mass fluctuations due to scale-invariant primeval perturbations}",
    journal = "Astrophys. J. Lett.",
    volume  = "263",
    pages   = "L1--L5",
    year    = "1982",
    doi     = "10.1086/183911"
}

@article{Peebles:1984ge,
    author  = "Peebles, P. J. E.",
    title   = "{Tests of Cosmological Models Constrained by Inflation}",
    journal = "Astrophys. J.",
    volume  = "284",
    pages   = "439--444",
    year    = "1984",
    doi     = "10.1086/162425"
}

@article{Turner:1984nf,
    author  = "Turner, Michael S. and Steigman, Gary and Krauss, Lawrence M.",
    title   = "{Flatness of the Universe: Reconciling Theoretical Prejudices with Observational Data}",
    journal = "Phys. Rev. Lett.",
    volume  = "52",
    pages   = "2090--2093",
    year    = "1984",
    doi     = "10.1103/PhysRevLett.52.2090"
}

@article{Efstathiou:1990xe,
    author  = "Efstathiou, G. and Sutherland, W. J. and Maddox, S. J.",
    title   = "{The Cosmological Constant and Cold Dark Matter}",
    journal = "Nature",
    volume  = "348",
    pages   = "705--707",
    year    = "1990",
    doi     = "10.1038/348705a0"
}

@article{Ostriker:1995su,
    author  = "Ostriker, J. P. and Steinhardt, Paul J.",
    title   = "{The Observational Case for a Low-Density Universe with a Nonzero Cosmological Constant}",
    journal = "Nature",
    volume  = "377",
    pages   = "600--602",
    year    = "1995",
    doi     = "10.1038/377600a0"
}

@article{Li:2023tui,
    author = "Li, Xiangchong and others",
    title = "{Hyper Suprime-Cam Year 3 results: Cosmology from cosmic shear two-point correlation functions}",
    eprint = "2304.00702",
    archivePrefix = "arXiv",
    primaryClass = "astro-ph.CO",
    doi = "10.1103/PhysRevD.108.123518",
    journal = "Phys. Rev. D",
    volume = "108",
    number = "12",
    pages = "123518",
    year = "2023"
}

@article{Creminelli:2017sry,
    author = "Creminelli, Paolo and Vernizzi, Filippo",
    title = "{Dark Energy after GW170817 and GRB170817A}",
    eprint = "1710.05877",
    archivePrefix = "arXiv",
    primaryClass = "astro-ph.CO",
    doi = "10.1103/PhysRevLett.119.251302",
    journal = "Phys. Rev. Lett.",
    volume = "119",
    number = "25",
    pages = "251302",
    year = "2017"
}

@article{Ezquiaga:2017ekz,
    author = "Ezquiaga, Jose Mar{\'\i}a and Zumalac{\'a}rregui, Miguel",
    title = "{Dark Energy After GW170817: Dead Ends and the Road Ahead}",
    eprint = "1710.05901",
    archivePrefix = "arXiv",
    primaryClass = "astro-ph.CO",
    reportNumber = "IFT-UAM-CSIC-17-096, NORDITA-2017-109",
    doi = "10.1103/PhysRevLett.119.251304",
    journal = "Phys. Rev. Lett.",
    volume = "119",
    number = "25",
    pages = "251304",
    year = "2017"
}

@article{Sakstein:2017xjx,
    author = "Sakstein, Jeremy and Jain, Bhuvnesh",
    title = "{Implications of the Neutron Star Merger GW170817 for Cosmological Scalar-Tensor Theories}",
    eprint = "1710.05893",
    archivePrefix = "arXiv",
    primaryClass = "astro-ph.CO",
    doi = "10.1103/PhysRevLett.119.251303",
    journal = "Phys. Rev. Lett.",
    volume = "119",
    number = "25",
    pages = "251303",
    year = "2017"
}

@article{Baker:2017hug,
    author = "Baker, T. and Bellini, E. and Ferreira, P. G. and Lagos, M. and Noller, J. and Sawicki, I.",
    title = "{Strong constraints on cosmological gravity from GW170817 and GRB 170817A}",
    eprint = "1710.06394",
    archivePrefix = "arXiv",
    primaryClass = "astro-ph.CO",
    doi = "10.1103/PhysRevLett.119.251301",
    journal = "Phys. Rev. Lett.",
    volume = "119",
    number = "25",
    pages = "251301",
    year = "2017"
}

@article{Kase:2019mox,
    author = "Kase, Ryotaro and Tsujikawa, Shinji",
    title = "{Weak cosmic growth in coupled dark energy with a Lagrangian formulation}",
    eprint = "1911.02179",
    archivePrefix = "arXiv",
    primaryClass = "gr-qc",
    doi = "10.1016/j.physletb.2020.135400",
    journal = "Phys. Lett. B",
    volume = "804",
    pages = "135400",
    year = "2020"
}

@article{Aoki:2025bmj,
    author = "Aoki, Katsuki and Beltr{\'a}n Jim{\'e}nez, Jose and Pookkillath, 
    Masroor C. and Tsujikawa, Shinji",
    title = "{Effective field theory of coupled dark energy and dark matter}",
    eprint = "2504.17293",
    archivePrefix = "arXiv",
    primaryClass = "astro-ph.CO",
    reportNumber = "YITP-25-56, WUCG-25-04",
    doi = "10.1103/5wsb-1vk6",
    journal = "Phys. Rev. D",
    volume = "113",
    number = "4",
    pages = "044053",
    year = "2026"
}

@article{Goldstein:2017mmi,
    author = "Goldstein, A. and others",
    title = "{An Ordinary Short Gamma-Ray Burst with 
    Extraordinary Implications: Fermi-GBM Detection of GRB 170817A}",
    eprint = "1710.05446",
    archivePrefix = "arXiv",
    primaryClass = "astro-ph.HE",
    doi = "10.3847/2041-8213/aa8f41",
    journal = "Astrophys. J. Lett.",
    volume = "848",
    number = "2",
    pages = "L14",
    year = "2017"
}

@article{Nesseris:2010pc,
    author = "Nesseris, Savvas and De Felice, Antonio and Tsujikawa, Shinji",
    title = "{Observational constraints on Galileon cosmology}",
    eprint = "1010.0407",
    archivePrefix = "arXiv",
    primaryClass = "astro-ph.CO",
    doi = "10.1103/PhysRevD.82.124054",
    journal = "Phys. Rev. D",
    volume = "82",
    pages = "124054",
    year = "2010"
}

@article{Shlivko:2025fgv,
    author = "Shlivko, David and Steinhardt, Paul J. and Steinhardt, Charles L.",
    title = "{Optimal parameterizations for observational constraints 
    on thawing dark energy}",
    eprint = "2504.02028",
    archivePrefix = "arXiv",
    primaryClass = "astro-ph.CO",
    doi = "10.1088/1475-7516/2025/06/054",
    month = "4",
    year = "2025"
}

@article{Akrami:2025zlb,
    author = "Akrami, Yashar and Alestas, George and Nesseris, Savvas",
    title = "{Has DESI detected exponential quintessence?}",
    eprint = "2504.04226",
    archivePrefix = "arXiv",
    primaryClass = "astro-ph.CO",
    reportNumber = "IFT-UAM/CSIC-25-36",
    month = "4",
    year = "2025"
}

@article{Bayat:2025xfr,
    author = "Bayat, Zahra and Hertzberg, Mark P.",
    title = "{Examining quintessence models with DESI data}",
    eprint = "2505.18937",
    archivePrefix = "arXiv",
    primaryClass = "astro-ph.CO",
    doi = "10.1088/1475-7516/2025/08/065",
    journal = "JCAP",
    volume = "08",
    pages = "065",
    year = "2025"
}

@article{Cline:2025sbt,
    author = "Cline, James M. and Muralidharan, Varun",
    title = "{Simple quintessence models in light of DESI-BAO observations}",
    eprint = "2506.13047",
    archivePrefix = "arXiv",
    primaryClass = "astro-ph.CO",
    doi = "10.1103/8z2m-nbv6",
    journal = "Phys. Rev. D",
    volume = "112",
    number = "6",
    pages = "063539",
    year = "2025"
}

@article{Gialamas:2025pwv,
    author = {Gialamas, Ioannis D. and H{\"u}tsi, Gert and Raidal, Martti and Urrutia, Juan and Vasar, Martin and Veerm{\"a}e, Hardi},
    title = "{Quintessence and phantoms in light of DESI 2025}",
    eprint = "2506.21542",
    archivePrefix = "arXiv",
    primaryClass = "astro-ph.CO",
    doi = "10.1103/kdqc-y37v",
    journal = "Phys. Rev. D",
    volume = "112",
    number = "6",
    pages = "063551",
    year = "2025"
}

@article{Alestas:2025syk,
    author = "Alestas, George and Caldarola, Marienza and Ocampo, Indira and Nesseris, Savvas and Tsujikawa, Shinji",
    title = "{DESI constraints on two-field quintessence with exponential potentials}",
    eprint = "2510.21627",
    archivePrefix = "arXiv",
    primaryClass = "astro-ph.CO",
    reportNumber = "IFT-UAM/CSIC-25-77, WUCG-25-12",
    doi = "10.1103/ckf5-jkvv",
    journal = "Phys. Rev. D",
    volume = "114",
    number = "2",
    pages = "023532",
    year = "2026"
}

@article{Shlivko:2025krk,
    author = "Shlivko, David",
    title = "{Thawing Quintessence: Priors, evidence, and likely trajectories}",
    eprint = "2512.20832",
    archivePrefix = "arXiv",
    primaryClass = "astro-ph.CO",
    month = "12",
    year = "2025"
}

@article{Caldwell:2003vq,
    author = "Caldwell, Robert R. and Kamionkowski, Marc and Weinberg, Nevin N.",
    title = "{Phantom energy and cosmic doomsday}",
    eprint = "astro-ph/0302506",
    archivePrefix = "arXiv",
    doi = "10.1103/PhysRevLett.91.071301",
    journal = "Phys. Rev. Lett.",
    volume = "91",
    pages = "071301",
    year = "2003"
}

@article{Singh:2003vx,
    author = "Singh, Parampreet and Sami, M. and Dadhich, Naresh",
    title = "{Cosmological dynamics of phantom field}",
    eprint = "hep-th/0305110",
    archivePrefix = "arXiv",
    doi = "10.1103/PhysRevD.68.023522",
    journal = "Phys. Rev. D",
    volume = "68",
    pages = "023522",
    year = "2003"
}

@article{Vainshtein:1972sx,
    author = "Vainshtein, A. I.",
    title = "{To the problem of nonvanishing gravitation mass}",
    doi = "10.1016/0370-2693(72)90147-5",
    journal = "Phys. Lett. B",
    volume = "39",
    pages = "393--394",
    year = "1972"
}

@article{Faulkner:2006ub,
    author = "Faulkner, Thomas and Tegmark, Max and Bunn, Emory F. and Mao, Yi",
    title = "{Constraining f(R) Gravity as a Scalar Tensor Theory}",
    eprint = "astro-ph/0612569",
    archivePrefix = "arXiv",
    doi = "10.1103/PhysRevD.76.063505",
    journal = "Phys. Rev. D",
    volume = "76",
    pages = "063505",
    year = "2007"
}

@article{Capozziello:2007eu,
    author = "Capozziello, Salvatore and Tsujikawa, Shinji",
    title = "{Solar system and equivalence principle constraints on f(R) gravity by chameleon approach}",
    eprint = "0712.2268",
    archivePrefix = "arXiv",
    primaryClass = "gr-qc",
    doi = "10.1103/PhysRevD.77.107501",
    journal = "Phys. Rev. D",
    volume = "77",
    pages = "107501",
    year = "2008"
}

@article{Tsujikawa:2013fta,
    author = "Tsujikawa, Shinji",
    title = "{Quintessence: A Review}",
    eprint = "1304.1961",
    archivePrefix = "arXiv",
    primaryClass = "gr-qc",
    doi = "10.1088/0264-9381/30/21/214003",
    journal = "Class. Quant. Grav.",
    volume = "30",
    pages = "214003",
    year = "2013"
}

@article{Barreira:2012kk,
    author = "Barreira, Alexandre and Li, Baojiu and Baugh, Carlton M. 
    and Pascoli, Silvia",
    title = "{Linear perturbations in Galileon gravity models}",
    eprint = "1208.0600",
    archivePrefix = "arXiv",
    primaryClass = "astro-ph.CO",
    doi = "10.1103/PhysRevD.86.124016",
    journal = "Phys. Rev. D",
    volume = "86",
    pages = "124016",
    year = "2012"
}

@article{Renk:2017rzu,
    author = "Renk, Janina and Zumalac{\'a}rregui, Miguel and Montanari, Francesco 
    and Barreira, Alexandre",
    title = "{Galileon gravity in light of ISW, CMB, BAO and H$_0$ data}",
    eprint = "1707.02263",
    archivePrefix = "arXiv",
    primaryClass = "astro-ph.CO",
    reportNumber = "NORDITA-2017-068, HIP-2017-17-TH",
    doi = "10.1088/1475-7516/2017/10/020",
    journal = "JCAP",
    volume = "10",
    pages = "020",
    year = "2017"
}

@article{DeFelice:2010as,
    author = "De Felice, Antonio and Kase, Ryotaro and Tsujikawa, Shinji",
    title = "{Matter perturbations in Galileon cosmology}",
    eprint = "1011.6132",
    archivePrefix = "arXiv",
    primaryClass = "astro-ph.CO",
    doi = "10.1103/PhysRevD.83.043515",
    journal = "Phys. Rev. D",
    volume = "83",
    pages = "043515",
    year = "2011"
}

@article{Kimura:2011td,
    author = "Kimura, Rampei and Kobayashi, Tsutomu and Yamamoto, Kazuhiro",
    title = "{Observational Constraints on Kinetic Gravity Braiding from the Integrated Sachs-Wolfe Effect}",
    eprint = "1110.3598",
    archivePrefix = "arXiv",
    primaryClass = "astro-ph.CO",
    doi = "10.1103/PhysRevD.85.123503",
    journal = "Phys. Rev. D",
    volume = "85",
    pages = "123503",
    year = "2012"
}

@article{Kable:2021yws,
    author = "Kable, Joshua A. and Benevento, Giampaolo and Frusciante, 
    Noemi and De Felice, Antonio and Tsujikawa, Shinji",
    title = "{Probing modified gravity with integrated Sachs-Wolfe CMB 
    and galaxy cross-correlations}",
    eprint = "2111.10432",
    archivePrefix = "arXiv",
    primaryClass = "astro-ph.CO",
    doi = "10.1088/1475-7516/2022/09/002",
    journal = "JCAP",
    volume = "09",
    pages = "002",
    year = "2022"
}

@article{DeFelice:2011hq,
    author = "De Felice, Antonio and Kobayashi, Tsutomu and Tsujikawa, Shinji",
    title = "{Effective gravitational couplings for cosmological perturbations 
    in the most general scalar-tensor theories with second-order field equations}",
    eprint = "1108.4242",
    archivePrefix = "arXiv",
    primaryClass = "gr-qc",
    doi = "10.1016/j.physletb.2011.11.028",
    journal = "Phys. Lett. B",
    volume = "706",
    pages = "123--133",
    year = "2011"
}

@article{LIGOScientific:2017zic,
    author = "Abbott, B. P. and others",
    collaboration = "LIGO Scientific, Virgo, Fermi-GBM, INTEGRAL",
    title = "{Gravitational Waves and Gamma-rays from a Binary Neutron 
    Star Merger: GW170817 and GRB 170817A}",
    eprint = "1710.05834",
    archivePrefix = "arXiv",
    primaryClass = "astro-ph.HE",
    reportNumber = "LIGO-P1700308",
    doi = "10.3847/2041-8213/aa920c",
    journal = "Astrophys. J. Lett.",
    volume = "848",
    number = "2",
    pages = "L13",
    year = "2017"
}

@article{Copeland:1997et,
    author = "Copeland, Edmund J. and Liddle, Andrew R and Wands, David",
    title = "{Exponential potentials and cosmological scaling solutions}",
    eprint = "gr-qc/9711068",
    archivePrefix = "arXiv",
    reportNumber = "SUSX-TH-97-022, SUSSEX-AST-97-11-1, PU-RCG-97-20",
    doi = "10.1103/PhysRevD.57.4686",
    journal = "Phys. Rev. D",
    volume = "57",
    pages = "4686--4690",
    year = "1998"
}

@article{Fujii:1982ms,
    author = "Fujii, Yasunori",
    title = "{Origin of the Gravitational Constant and Particle Masses in Scale Invariant Scalar - Tensor Theory}",
    reportNumber = "UT-KOMABA-82-8",
    doi = "10.1103/PhysRevD.26.2580",
    journal = "Phys. Rev. D",
    volume = "26",
    pages = "2580",
    year = "1982"
}

@article{Ratra:1987rm,
    author = "Ratra, Bharat and Peebles, P. J. E.",
    title = "{Cosmological Consequences of a Rolling Homogeneous Scalar Field}",
    reportNumber = "PUPT-1072",
    doi = "10.1103/PhysRevD.37.3406",
    journal = "Phys. Rev. D",
    volume = "37",
    pages = "3406",
    year = "1988"
}

@article{Wetterich:1987fm,
    author = "Wetterich, C.",
    title = "{Cosmology and the Fate of Dilatation Symmetry}",
    eprint = "1711.03844",
    archivePrefix = "arXiv",
    primaryClass = "hep-th",
    reportNumber = "PRINT-87-0756, DESY-87-123",
    doi = "10.1016/0550-3213(88)90193-9",
    journal = "Nucl. Phys. B",
    volume = "302",
    pages = "668--696",
    year = "1988"
}

@article{Chiba:1997ej,
    author = "Chiba, Takeshi and Sugiyama, Naoshi and Nakamura, Takashi",
    title = "{Cosmology with x matter}",
    eprint = "astro-ph/9704199",
    archivePrefix = "arXiv",
    reportNumber = "YITP-97-18, KUNS-1440",
    doi = "10.1093/mnras/289.2.L5",
    journal = "Mon. Not. Roy. Astron. Soc.",
    volume = "289",
    pages = "L5--L9",
    year = "1997"
}

@article{Ferreira:1997au,
    author = "Ferreira, Pedro G. and Joyce, Michael",
    title = "{Structure formation with a selftuning scalar field}",
    eprint = "astro-ph/9707286",
    archivePrefix = "arXiv",
    reportNumber = "CFPA-97-TH-07",
    doi = "10.1103/PhysRevLett.79.4740",
    journal = "Phys. Rev. Lett.",
    volume = "79",
    pages = "4740--4743",
    year = "1997"
}

@article{Caldwell:1997ii,
    author = "Caldwell, R. R. and Dave, Rahul and Steinhardt, Paul J.",
    title = "{Cosmological imprint of an energy component with general equation of state}",
    eprint = "astro-ph/9708069",
    archivePrefix = "arXiv",
    doi = "10.1103/PhysRevLett.80.1582",
    journal = "Phys. Rev. Lett.",
    volume = "80",
    pages = "1582--1585",
    year = "1998"
}

@article{Armendariz-Picon:1999hyi,
    author = "Armendariz-Picon, C. and Damour, T. and Mukhanov, Viatcheslav F.",
    title = "{k - inflation}",
    eprint = "hep-th/9904075",
    archivePrefix = "arXiv",
    doi = "10.1016/S0370-2693(99)00603-6",
    journal = "Phys. Lett. B",
    volume = "458",
    pages = "209--218",
    year = "1999"
}

@article{Chiba:1999ka,
    author = "Chiba, Takeshi and Okabe, Takahiro and Yamaguchi, Masahide",
    title = "{Kinetically driven quintessence}",
    eprint = "astro-ph/9912463",
    archivePrefix = "arXiv",
    reportNumber = "UTAP-352",
    doi = "10.1103/PhysRevD.62.023511",
    journal = "Phys. Rev. D",
    volume = "62",
    pages = "023511",
    year = "2000"
}

@article{Armendariz-Picon:2000nqq,
    author = "Armendariz-Picon, C. and Mukhanov, Viatcheslav F. and Steinhardt, Paul J.",
    title = "{A Dynamical solution to the problem of a small cosmological constant 
    and late time cosmic acceleration}",
    eprint = "astro-ph/0004134",
    archivePrefix = "arXiv",
    doi = "10.1103/PhysRevLett.85.4438",
    journal = "Phys. Rev. Lett.",
    volume = "85",
    pages = "4438--4441",
    year = "2000"
}

@article{Copeland:2006wr,
    author = "Copeland, Edmund J. and Sami, M. and Tsujikawa, Shinji",
    title = "{Dynamics of dark energy}",
    eprint = "hep-th/0603057",
    archivePrefix = "arXiv",
    doi = "10.1142/S021827180600942X",
    journal = "Int. J. Mod. Phys. D",
    volume = "15",
    pages = "1753--1936",
    year = "2006"
}

@article{Silvestri:2009hh,
    author = "Silvestri, Alessandra and Trodden, Mark",
    title = "{Approaches to Understanding Cosmic Acceleration}",
    eprint = "0904.0024",
    archivePrefix = "arXiv",
    primaryClass = "astro-ph.CO",
    doi = "10.1088/0034-4885/72/9/096901",
    journal = "Rept. Prog. Phys.",
    volume = "72",
    pages = "096901",
    year = "2009"
}

@article{Clifton:2011jh,
    author = "Clifton, Timothy and Ferreira, Pedro G. and Padilla, 
    Antonio and Skordis, Constantinos",
    title = "{Modified Gravity and Cosmology}",
    eprint = "1106.2476",
    archivePrefix = "arXiv",
    primaryClass = "astro-ph.CO",
    doi = "10.1016/j.physrep.2012.01.001",
    journal = "Phys. Rept.",
    volume = "513",
    pages = "1--189",
    year = "2012"
}

@article{Joyce:2014kja,
    author = "Joyce, Austin and Jain, Bhuvnesh and Khoury, Justin and Trodden, Mark",
    title = "{Beyond the Cosmological Standard Model}",
    eprint = "1407.0059",
    archivePrefix = "arXiv",
    primaryClass = "astro-ph.CO",
    doi = "10.1016/j.physrep.2014.12.002",
    journal = "Phys. Rept.",
    volume = "568",
    pages = "1--98",
    year = "2015"
}

@article{Koyama:2015vza,
    author = "Koyama, Kazuya",
    title = "{Cosmological Tests of Modified Gravity}",
    eprint = "1504.04623",
    archivePrefix = "arXiv",
    primaryClass = "astro-ph.CO",
    doi = "10.1088/0034-4885/79/4/046902",
    journal = "Rept. Prog. Phys.",
    volume = "79",
    number = "4",
    pages = "046902",
    year = "2016"
}

@article{Kase:2018aps,
    author = "Kase, Ryotaro and Tsujikawa, Shinji",
    title = "{Dark energy in  theories after GW170817: A review}",
    eprint = "1809.08735",
    archivePrefix = "arXiv",
    primaryClass = "gr-qc",
    doi = "10.1142/S0218271819420057",
    journal = "Int. J. Mod. Phys. D",
    volume = "28",
    number = "05",
    pages = "1942005",
    year = "2019"
}

@article{Aoki:2024ktc,
    author = "Aoki, Katsuki and Gorji, Mohammad Ali and Hiramatsu, Takashi and Mukohyama, Shinji and Pookkillath, Masroor C. and Takahashi, Kazufumi",
    title = "{CMB spectrum in unified EFT of dark energy: scalar-tensor and vector-tensor theories}",
    eprint = "2405.04265",
    archivePrefix = "arXiv",
    primaryClass = "astro-ph.CO",
    reportNumber = "YITP-24-56, RUP-24-8, IPMU24-0016",
    doi = "10.1088/1475-7516/2024/07/056",
    journal = "JCAP",
    volume = "07",
    pages = "056",
    year = "2024"
}

@article{Tsujikawa:2025wca,
    author = "Tsujikawa, Shinji",
    title = "{Crossing the phantom divide in scalar-tensor and vector-tensor theories}",
    eprint = "2508.17231",
    archivePrefix = "arXiv",
    primaryClass = "astro-ph.CO",
    reportNumber = "WUCG-25-09",
    doi = "10.1103/y858-4swl",
    journal = "Phys. Rev. D",
    volume = "113",
    number = "4",
    pages = "L041301",
    year = "2026"
}

@article{SupernovaSearchTeam:1998fmf,
    author = "Riess, Adam G. and others",
    collaboration = "Supernova Search Team",
    title = "{Observational evidence from supernovae for an accelerating universe 
    and a cosmological constant}",
    eprint = "astro-ph/9805201",
    archivePrefix = "arXiv",
    doi = "10.1086/300499",
    journal = "Astron. J.",
    volume = "116",
    pages = "1009--1038",
    year = "1998"
}

@article{SupernovaCosmologyProject:1998vns,
    author = "Perlmutter, S. and others",
    collaboration = "Supernova Cosmology Project",
    title = "{Measurements of $\Omega$ and $\Lambda$ from 42 High Redshift Supernovae}",
    eprint = "astro-ph/9812133",
    archivePrefix = "arXiv",
    reportNumber = "LBNL-41801, LBL-41801",
    doi = "10.1086/307221",
    journal = "Astrophys. J.",
    volume = "517",
    pages = "565--586",
    year = "1999"
}

@article{WMAP:2003elm,
    author = "Spergel, D. N. and others",
    collaboration = "WMAP",
    title = "{First year Wilkinson Microwave Anisotropy Probe (WMAP) 
    observations: Determination of cosmological parameters}",
    eprint = "astro-ph/0302209",
    archivePrefix = "arXiv",
    doi = "10.1086/377226",
    journal = "Astrophys. J. Suppl.",
    volume = "148",
    pages = "175--194",
    year = "2003"
}

@article{SDSS:2005xqv,
    author = "Eisenstein, Daniel J. and others",
    collaboration = "SDSS",
    title = "{Detection of the Baryon Acoustic Peak in the Large-Scale 
    Correlation Function of SDSS Luminous Red Galaxies}",
    eprint = "astro-ph/0501171",
    archivePrefix = "arXiv",
    reportNumber = "FERMILAB-PUB-05-057-A-CD",
    doi = "10.1086/466512",
    journal = "Astrophys. J.",
    volume = "633",
    pages = "560--574",
    year = "2005"
}

@article{LIGOScientific:2017vwq,
    author = "Abbott, B. P. and others",
    collaboration = "LIGO Scientific, Virgo",
    title = "{GW170817: Observation of Gravitational Waves from a Binary Neutron Star Inspiral}",
    eprint = "1710.05832",
    archivePrefix = "arXiv",
    primaryClass = "gr-qc",
    reportNumber = "LIGO-P170817",
    doi = "10.1103/PhysRevLett.119.161101",
    journal = "Phys. Rev. Lett.",
    volume = "119",
    number = "16",
    pages = "161101",
    year = "2017"
}

@article{Chakraborty:2025syu,
    author = "Chakraborty, Amlan and Chanda, Prolay K. and Das, Subinoy and Dutta, Koushik",
    title = "{DESI results: hint towards coupled dark matter and dark energy}",
    eprint = "2503.10806",
    archivePrefix = "arXiv",
    primaryClass = "astro-ph.CO",
    doi = "10.1088/1475-7516/2025/11/047",
    journal = "JCAP",
    volume = "11",
    pages = "047",
    year = "2025"
}

@article{Ye:2024ywg,
    author = "Ye, Gen and Martinelli, Matteo and Hu, Bin 
    and Silvestri, Alessandra",
    title = "{Hints of Nonminimally Coupled Gravity in DESI 2024 Baryon Acoustic Oscillation Measurements}",
    eprint = "2407.15832",
    archivePrefix = "arXiv",
    primaryClass = "astro-ph.CO",
    doi = "10.1103/PhysRevLett.134.181002",
    journal = "Phys. Rev. Lett.",
    volume = "134",
    number = "18",
    pages = "181002",
    year = "2025"
}

@article{Wolf:2024stt,
    author = "Wolf, William J. and Ferreira, Pedro G. and Garc{\'\i}a-Garc{\'\i}a, Carlos",
    title = "{Matching current observational constraints with nonminimally coupled dark energy}",
    eprint = "2409.17019",
    archivePrefix = "arXiv",
    primaryClass = "astro-ph.CO",
    doi = "10.1103/PhysRevD.111.L041303",
    journal = "Phys. Rev. D",
    volume = "111",
    number = "4",
    pages = "L041303",
    year = "2025"
}

@article{Pan:2025psn,
    author = "Pan, Jiaming and Ye, Gen",
    title = "{Non-minimally coupled gravity constraints from DESI DR2 data}",
    eprint = "2503.19898",
    archivePrefix = "arXiv",
    primaryClass = "astro-ph.CO",
    month = "3",
    year = "2025"
}

@article{Wang:2025znm,
    author = "Wang, Jia-Qi and Cai, Rong-Gen and Guo,, Zong-Kuan and Wang, Shao-Jiang",
    title = "{Resolving the Planck-DESI tension by non-minimally coupled quintessence}",
    eprint = "2508.01759",
    archivePrefix = "arXiv",
    primaryClass = "astro-ph.CO",
    month = "8",
    year = "2025"
}

@article{Adam:2025kve,
    author = "Adam, Husam and Hertzberg, Mark P. and Jim{\'e}nez-Aguilar, Daniel and Khan, Iman",
    title = "{Comparing Minimal and Non-Minimal Quintessence Models to 2025 DESI Data}",
    eprint = "2509.13302",
    archivePrefix = "arXiv",
    primaryClass = "astro-ph.CO",
    month = "9",
    year = "2025"
}

@article{SanchezLopez:2025uzw,
    author = "S{\'a}nchez L{\'o}pez, Samuel and Karam, Alexandros and Hazra, Dhiraj Kumar",
    title = "{Non-Minimally Coupled Quintessence in Light of DESI}",
    eprint = "2510.14941",
    archivePrefix = "arXiv",
    primaryClass = "astro-ph.CO",
    month = "10",
    year = "2025"
}

@article{Amendola:2016saw,
    author = "Amendola, Luca and others",
    title = "{Cosmology and fundamental physics with the Euclid satellite}",
    eprint = "1606.00180",
    archivePrefix = "arXiv",
    primaryClass = "astro-ph.CO",
    doi = "10.1007/s41114-017-0010-3",
    journal = "Living Rev. Rel.",
    volume = "21",
    number = "1",
    pages = "2",
    year = "2018"
}

@article{Babichev:2011iz,
    author = "Babichev, Eugeny and Deffayet, Cedric and Esposito-Farese, Gilles",
    title = "{Constraints on Shift-Symmetric Scalar-Tensor Theories with a Vainshtein Mechanism from Bounds on the Time Variation of G}",
    eprint = "1107.1569",
    archivePrefix = "arXiv",
    primaryClass = "gr-qc",
    doi = "10.1103/PhysRevLett.107.251102",
    journal = "Phys. Rev. Lett.",
    volume = "107",
    pages = "251102",
    year = "2011"
}

@article{Kimura:2011dc,
    author = "Kimura, Rampei and Kobayashi, Tsutomu and Yamamoto, Kazuhiro",
    title = "{Vainshtein screening in a cosmological background in the most general second-order scalar-tensor theory}",
    eprint = "1111.6749",
    archivePrefix = "arXiv",
    primaryClass = "astro-ph.CO",
    doi = "10.1103/PhysRevD.85.024023",
    journal = "Phys. Rev. D",
    volume = "85",
    pages = "024023",
    year = "2012"
}

@article{Hofmann:2018myc,
    author = {Hofmann, F. and M{\"u}ller, J.},
    title = "{Relativistic tests with lunar laser ranging}",
    doi = "10.1088/1361-6382/aa8f7a",
    journal = "Class. Quant. Grav.",
    volume = "35",
    number = "3",
    pages = "035015",
    year = "2018"
}

@article{Tsujikawa:2019pih,
    author = "Tsujikawa, Shinji",
    title = "{Lunar Laser Ranging constraints on nonminimally coupled 
    dark energy and standard sirens}",
    eprint = "1903.07092",
    archivePrefix = "arXiv",
    primaryClass = "gr-qc",
    doi = "10.1103/PhysRevD.100.043510",
    journal = "Phys. Rev. D",
    volume = "100",
    number = "4",
    pages = "043510",
    year = "2019"
}

@article{Perivolaropoulos:2005yv,
    author = "Perivolaropoulos, Leandros",
    title = "{Crossing the phantom divide barrier with scalar tensor theories}",
    eprint = "astro-ph/0504582",
    archivePrefix = "arXiv",
    doi = "10.1088/1475-7516/2005/10/001",
    journal = "JCAP",
    volume = "10",
    pages = "001",
    year = "2005"
}

@article{Amendola:2007nt,
    author = "Amendola, Luca and Tsujikawa, Shinji",
    title = "{Phantom crossing, equation-of-state singularities, and local gravity constraints in f(R) models}",
    eprint = "0705.0396",
    archivePrefix = "arXiv",
    primaryClass = "astro-ph",
    doi = "10.1016/j.physletb.2007.12.041",
    journal = "Phys. Lett. B",
    volume = "660",
    pages = "125--132",
    year = "2008"
}

@article{Motohashi:2010tb,
    author = "Motohashi, Hayato and Starobinsky, Alexei A. and Yokoyama, Jun'ichi",
    title = "{Phantom boundary crossing and anomalous growth index of fluctuations in viable f(R) models of cosmic acceleration}",
    eprint = "1002.1141",
    archivePrefix = "arXiv",
    primaryClass = "astro-ph.CO",
    doi = "10.1143/PTP.123.887",
    journal = "Prog. Theor. Phys.",
    volume = "123",
    pages = "887--902",
    year = "2010"
}

@article{Planck:2018vyg,
    author = "Aghanim, N. and others",
    collaboration = "Planck",
    title = "{Planck 2018 results. VI. Cosmological parameters}",
    eprint = "1807.06209",
    archivePrefix = "arXiv",
    primaryClass = "astro-ph.CO",
    doi = "10.1051/0004-6361/201833910",
    journal = "Astron. Astrophys.",
    volume = "641",
    pages = "A6",
    year = "2020",
    note = "[Erratum: Astron.Astrophys. 652, C4 (2021)]"
}

@article{Nicolis:2008in,
    author = "Nicolis, Alberto and Rattazzi, Riccardo and Trincherini, Enrico",
    title = "{The Galileon as a local modification of gravity}",
    eprint = "0811.2197",
    archivePrefix = "arXiv",
    primaryClass = "hep-th",
    doi = "10.1103/PhysRevD.79.064036",
    journal = "Phys. Rev. D",
    volume = "79",
    pages = "064036",
    year = "2009"
}

@article{Brax:2008hh,
    author = "Brax, Philippe and van de Bruck, Carsten and Davis, Anne-Christine 
    and Shaw, Douglas J.",
    title = "{f(R) Gravity and Chameleon Theories}",
    eprint = "0806.3415",
    archivePrefix = "arXiv",
    primaryClass = "astro-ph",
    doi = "10.1103/PhysRevD.78.104021",
    journal = "Phys. Rev. D",
    volume = "78",
    pages = "104021",
    year = "2008"
}

@article{DESI:2024mwx,
    author = "Adame, A. G. and others",
    collaboration = "DESI",
    title = "{DESI 2024 VI: cosmological constraints from the measurements 
    of baryon acoustic oscillations}",
    eprint = "2404.03002",
    archivePrefix = "arXiv",
    primaryClass = "astro-ph.CO",
    reportNumber = "FERMILAB-PUB-24-0154-PPD",
    doi = "10.1088/1475-7516/2025/02/021",
    journal = "JCAP",
    volume = "02",
    pages = "021",
    year = "2025"
}

@article{DESI:2024aqx,
    author = "Calderon, R. and others",
    collaboration = "DESI",
    title = "{DESI 2024: reconstructing dark energy using crossing statistics with DESI DR1 BAO data}",
    eprint = "2405.04216",
    archivePrefix = "arXiv",
    primaryClass = "astro-ph.CO",
    doi = "10.1088/1475-7516/2024/10/048",
    journal = "JCAP",
    volume = "10",
    pages = "048",
    year = "2024"
}

@article{DESI:2025zgx,
    author = "Abdul Karim, M. and others",
    collaboration = "DESI",
    title = "{DESI DR2 Results II: Measurements of 
    Baryon Acoustic Oscillations and Cosmological Constraints}",
    eprint = "2503.14738",
    archivePrefix = "arXiv",
    primaryClass = "astro-ph.CO",
    reportNumber = "FERMILAB-PUB-25-0169-PPD",
    month = "3",
    year = "2025"
}

@article{DESI:2025fii,
    author = "Lodha, K. and others",
    collaboration = "DESI",
    title = "{Extended Dark Energy analysis using DESI DR2 BAO measurements}",
    eprint = "2503.14743",
    archivePrefix = "arXiv",
    primaryClass = "astro-ph.CO",
    reportNumber = "FERMILAB-PUB-25-0164-PPD",
    month = "3",
    year = "2025"
}

@article{Chevallier:2000qy,
    author = "Chevallier, Michel and Polarski, David",
    title = "{Accelerating universes with scaling dark matter}",
    eprint = "gr-qc/0009008",
    archivePrefix = "arXiv",
    doi = "10.1142/S0218271801000822",
    journal = "Int. J. Mod. Phys. D",
    volume = "10",
    pages = "213--224",
    year = "2001"
}

@article{Linder:2002et,
    author = "Linder, Eric V.",
    title = "{Exploring the expansion history of the universe}",
    eprint = "astro-ph/0208512",
    archivePrefix = "arXiv",
    doi = "10.1103/PhysRevLett.90.091301",
    journal = "Phys. Rev. Lett.",
    volume = "90",
    pages = "091301",
    year = "2003"
}

@article{Caldwell:1999ew,
    author = "Caldwell, R. R.",
    title = "{A Phantom menace?}",
    eprint = "astro-ph/9908168",
    archivePrefix = "arXiv",
    doi = "10.1016/S0370-2693(02)02589-3",
    journal = "Phys. Lett. B",
    volume = "545",
    pages = "23--29",
    year = "2002"
}

@article{Feng:2004ad,
    author = "Feng, Bo and Wang, Xiu-Lian and Zhang, Xin-Min",
    title = "{Dark energy constraints from the cosmic age and supernova}",
    eprint = "astro-ph/0404224",
    archivePrefix = "arXiv",
    doi = "10.1016/j.physletb.2004.12.071",
    journal = "Phys. Lett. B",
    volume = "607",
    pages = "35--41",
    year = "2005"
}

@article{Guo:2004fq,
    author = "Guo, Zong-Kuan and Piao, Yun-Song and Zhang, Xin-Min 
    and Zhang, Yuan-Zhong",
    title = "{Cosmological evolution of a quintom model of dark energy}",
    eprint = "astro-ph/0410654",
    archivePrefix = "arXiv",
    doi = "10.1016/j.physletb.2005.01.017",
    journal = "Phys. Lett. B",
    volume = "608",
    pages = "177--182",
    year = "2005"
}

@article{Carroll:2003st,
    author = "Carroll, Sean M. and Hoffman, Mark and Trodden, Mark",
    title = "{Can the dark energy equation-of-state parameter $w$ 
    be less than $−1$?}",
    eprint = "astro-ph/0301273",
    archivePrefix = "arXiv",
    reportNumber = "EFI-2003-01, SU-GP-03-1-1",
    doi = "10.1103/PhysRevD.68.023509",
    journal = "Phys. Rev. D",
    volume = "68",
    pages = "023509",
    year = "2003"
}

@article{Cline:2003gs,
    author = "Cline, James M. and Jeon, Sangyong and Moore, Guy D.",
    title = "{The Phantom menaced: Constraints on low-energy effective ghosts}",
    eprint = "hep-ph/0311312",
    archivePrefix = "arXiv",
    reportNumber = "MCGILL-03-25",
    doi = "10.1103/PhysRevD.70.043543",
    journal = "Phys. Rev. D",
    volume = "70",
    pages = "043543",
    year = "2004"
}

@article{Deffayet:2009wt,
    author = "Deffayet, C. and Esposito-Farese, Gilles and Vikman, A.",
    title = "{Covariant Galileon}",
    eprint = "0901.1314",
    archivePrefix = "arXiv",
    primaryClass = "hep-th",
    doi = "10.1103/PhysRevD.79.084003",
    journal = "Phys. Rev. D",
    volume = "79",
    pages = "084003",
    year = "2009"
}

@article{DeFelice:2010pv,
    author = "De Felice, Antonio and Tsujikawa, Shinji",
    title = "{Cosmology of a covariant Galileon field}",
    eprint = "1007.2700",
    archivePrefix = "arXiv",
    primaryClass = "astro-ph.CO",
    doi = "10.1103/PhysRevLett.105.111301",
    journal = "Phys. Rev. Lett.",
    volume = "105",
    pages = "111301",
    year = "2010"
}

@article{DeFelice:2010nf,
    author = "De Felice, Antonio and Tsujikawa, Shinji",
    title = "{Generalized Galileon cosmology}",
    eprint = "1008.4236",
    archivePrefix = "arXiv",
    primaryClass = "hep-th",
    doi = "10.1103/PhysRevD.84.124029",
    journal = "Phys. Rev. D",
    volume = "84",
    pages = "124029",
    year = "2011"
}

@article{Peirone:2019aua,
    author = "Peirone, Simone and Benevento, Giampaolo and Frusciante, Noemi and Tsujikawa, Shinji",
    title = "{Cosmological data favor Galileon ghost condensate over $\Lambda$CDM}",
    eprint = "1905.05166",
    archivePrefix = "arXiv",
    primaryClass = "astro-ph.CO",
    doi = "10.1103/PhysRevD.100.063540",
    journal = "Phys. Rev. D",
    volume = "100",
    number = "6",
    pages = "063540",
    year = "2019"
}

@article{DeFelice:2010aj,
    author = "De Felice, Antonio and Tsujikawa, Shinji",
    title = "{f(R) theories}",
    eprint = "1002.4928",
    archivePrefix = "arXiv",
    primaryClass = "gr-qc",
    doi = "10.12942/lrr-2010-3",
    journal = "Living Rev. Rel.",
    volume = "13",
    pages = "3",
    year = "2010"
}

@article{Hu:2007nk,
    author = "Hu, Wayne and Sawicki, Ignacy",
    title = "{Models of f(R) Cosmic Acceleration that Evade Solar-System Tests}",
    eprint = "0705.1158",
    archivePrefix = "arXiv",
    primaryClass = "astro-ph",
    doi = "10.1103/PhysRevD.76.064004",
    journal = "Phys. Rev. D",
    volume = "76",
    pages = "064004",
    year = "2007"
}

@article{Starobinsky:2007hu,
    author = "Starobinsky, Alexei A.",
    title = "{Disappearing cosmological constant in f(R) gravity}",
    eprint = "0706.2041",
    archivePrefix = "arXiv",
    primaryClass = "astro-ph",
    doi = "10.1134/S0021364007150027",
    journal = "JETP Lett.",
    volume = "86",
    pages = "157--163",
    year = "2007"
}

@article{Appleby:2007vb,
    author = "Appleby, Stephen A. and Battye, Richard A.",
    title = "{Do consistent $F(R)$ models mimic General Relativity 
    plus $\Lambda$?}",
    eprint = "0705.3199",
    archivePrefix = "arXiv",
    primaryClass = "astro-ph",
    doi = "10.1016/j.physletb.2007.08.037",
    journal = "Phys. Lett. B",
    volume = "654",
    pages = "7--12",
    year = "2007"
}

@article{Tsujikawa:2007xu,
    author = "Tsujikawa, Shinji",
    title = "{Observational signatures of $f(R)$ dark energy models 
    that satisfy cosmological and local gravity constraints}",
    eprint = "0709.1391",
    archivePrefix = "arXiv",
    primaryClass = "astro-ph",
    doi = "10.1103/PhysRevD.77.023507",
    journal = "Phys. Rev. D",
    volume = "77",
    pages = "023507",
    year = "2008"
}

@article{Horndeski:1974wa,
    author = "Horndeski, Gregory Walter",
    title = "{Second-order scalar-tensor field equations in a four-dimensional space}",
    doi = "10.1007/BF01807638",
    journal = "Int. J. Theor. Phys.",
    volume = "10",
    pages = "363--384",
    year = "1974"
}

@article{Deffayet:2011gz,
    author = "Deffayet, C. and Gao, Xian and Steer, D. A. and Zahariade, G.",
    title = "{From k-essence to generalised Galileons}",
    eprint = "1103.3260",
    archivePrefix = "arXiv",
    primaryClass = "hep-th",
    doi = "10.1103/PhysRevD.84.064039",
    journal = "Phys. Rev. D",
    volume = "84",
    pages = "064039",
    year = "2011"
}

@article{Kobayashi:2011nu,
    author = "Kobayashi, Tsutomu and Yamaguchi, Masahide and 
    Yokoyama, Jun'ichi",
    title = "{Generalized G-inflation: Inflation with the most general 
    second-order field equations}",
    eprint = "1105.5723",
    archivePrefix = "arXiv",
    primaryClass = "hep-th",
    reportNumber = "KUNS-2339, RESCEU-9-11",
    doi = "10.1143/PTP.126.511",
    journal = "Prog. Theor. Phys.",
    volume = "126",
    pages = "511--529",
    year = "2011"
}

@article{Charmousis:2011bf,
    author = "Charmousis, Christos and Copeland, Edmund J. and 
    Padilla, Antonio and Saffin, Paul M.",
    title = "{General second order scalar-tensor theory, self tuning, 
    and the Fab Four}",
    eprint = "1106.2000",
    archivePrefix = "arXiv",
    primaryClass = "hep-th",
    doi = "10.1103/PhysRevLett.108.051101",
    journal = "Phys. Rev. Lett.",
    volume = "108",
    pages = "051101",
    year = "2012"
}

@article{Kase:2020hst,
    author = "Kase, Ryotaro and Tsujikawa, Shinji",
    title = "{General formulation of cosmological perturbations in scalar-tensor dark energy coupled to dark matter}",
    eprint = "2005.13809",
    archivePrefix = "arXiv",
    primaryClass = "gr-qc",
    doi = "10.1088/1475-7516/2020/11/032",
    journal = "JCAP",
    volume = "11",
    pages = "032",
    year = "2020"
}

@article{BeltranJimenez:2026ymd,
    author = "Beltr{\'a}n Jim{\'e}nez, Jose and Ichiki, Kiyotomo and Liu, Xiaolin and Teppa Pannia, Florencia Anabella and Tsujikawa, Shinji",
    title = "{Revisiting observational constraints on coupled exponential quintessence with energy and momentum transfers: degeneracy with massive neutrinos}",
    eprint = "2603.15805",
    archivePrefix = "arXiv",
    primaryClass = "astro-ph.CO",
    year = "2026"
}

@article{Simpson:2010vh,
    author = "Simpson, Fergus",
    title = "{Scattering of Dark Matter and Dark Energy}",
    eprint = "1007.1034",
    archivePrefix = "arXiv",
    primaryClass = "astro-ph.CO",
    doi = "10.1103/PhysRevD.82.083505",
    journal = "Phys. Rev. D",
    volume = "82",
    pages = "083505",
    year = "2010"
}

@article{Asghari:2019qld,
    author = "Asghari, Mahnaz and Beltr{\'a}n Jim{\'e}nez, Jose and Khosravi, Shahram and Mota, David F.",
    title = "{On structure formation from a small-scales-interacting dark sector}",
    eprint = "1902.05532",
    archivePrefix = "arXiv",
    primaryClass = "astro-ph.CO",
    doi = "10.1088/1475-7516/2019/04/042",
    journal = "JCAP",
    volume = "04",
    pages = "042",
    year = "2019"
}

@article{BeltranJimenez:2020tme,
    author = "Beltr{\'a}n Jim{\'e}nez, Jose and Bettoni, Dario and Figueruelo, David and Teppa Pannia, Florencia A.",
    title = "{On cosmological signatures of baryons-dark energy elastic couplings}",
    eprint = "2004.14661",
    archivePrefix = "arXiv",
    primaryClass = "astro-ph.CO",
    doi = "10.1088/1475-7516/2020/08/020",
    journal = "JCAP",
    volume = "08",
    pages = "020",
    year = "2020"
}

@article{Cardona:2022lcz,
    author = "Cardona, Wilmar and Figueruelo, David",
    title = "{Momentum transfer in the dark sector and lensing convergence in upcoming galaxy surveys}",
    eprint = "2209.12583",
    archivePrefix = "arXiv",
    primaryClass = "astro-ph.CO",
    doi = "10.1088/1475-7516/2022/12/010",
    journal = "JCAP",
    volume = "12",
    pages = "010",
    year = "2022"
}

@article{BeltranJimenez:2022ixm,
    author = "Beltr{\'a}n Jim{\'e}nez, Jose and Di Dio, Enea and Figueruelo, David",
    title = "{A smoking gun from the power spectrum dipole for elastic interactions in the dark sector}",
    eprint = "2212.08617",
    archivePrefix = "arXiv",
    primaryClass = "astro-ph.CO",
    doi = "10.1088/1475-7516/2023/11/088",
    journal = "JCAP",
    volume = "11",
    pages = "088",
    year = "2023"
}

@article{BeltranJimenez:2024nbz,
    author = "Beltr{\'a}n Jim{\'e}nez, Jose and Figueruelo, David and Teppa Pannia, Florencia A.",
    title = "{Nondegeneracy of massive neutrinos and elastic interactions in the dark sector}",
    eprint = "2403.03216",
    archivePrefix = "arXiv",
    primaryClass = "astro-ph.CO",
    doi = "10.1103/PhysRevD.110.023527",
    journal = "Phys. Rev. D",
    volume = "110",
    number = "2",
    pages = "023527",
    year = "2024"
}

@article{BeltranJimenez:2024rkg,
    author = "Beltr{\'a}n Jim{\'e}nez, Jose and Bettoni, Dario and Figueruelo, David and Teppa Pannia, Florencia A.",
    title = "{On Evidence for Elastic Interactions in the Dark Sector}",
    eprint = "2410.18645",
    archivePrefix = "arXiv",
    primaryClass = "astro-ph.CO",
    doi = "10.1016/j.dark.2024.101761",
    journal = "Phys. Dark Univ.",
    volume = "47",
    pages = "101761",
    year = "2025"
}

@article{Poulin:2022sgp,
    author = "Poulin, Vivian and Bernal, Jos{\'e} Luis and Kovetz, Ely D. and Kamionkowski, Marc",
    title = "{The Sigma-8 Tension is a Drag}",
    eprint = "2209.06217",
    archivePrefix = "arXiv",
    primaryClass = "astro-ph.CO",
    doi = "10.1103/PhysRevD.107.123538",
    journal = "Phys. Rev. D",
    volume = "107",
    number = "12",
    pages = "123538",
    year = "2023"
}

@article{Cruickshank:2025wjy,
    author = "Cruickshank, Nathan and Crittenden, Robert and Koyama, Kazuya and Bruni, Marco",
    title = "{Forecasts for interacting dark energy with time-dependent momentum exchange}",
    eprint = "2504.03555",
    archivePrefix = "arXiv",
    primaryClass = "astro-ph.CO",
    doi = "10.1088/1475-7516/2025/10/052",
    journal = "JCAP",
    volume = "10",
    pages = "052",
    year = "2025"
}

@article{Cruickshank:2025rqa,
    author = "Cruickshank, Nathan and Crittenden, Robert and Koyama, Kazuya and Bruni, Marco",
    title = "{Dark sector interactions in the $w\\rightarrow-1$ limit: velocity locking in pure momentum exchange models}",
    eprint = "2512.11639",
    archivePrefix = "arXiv",
    primaryClass = "astro-ph.CO",
    doi = "10.1088/1475-7516/2026/07/019",
    journal = "JCAP",
    volume = "07",
    pages = "019",
    year = "2026"
}

@article{BeltranJimenez:2025nls,
    author = "Beltr{\'a}n Jim{\'e}nez, Jose and Figueruelo, David and Mota, David F. and Winther, Hans A.",
    title = "{Non-linear structure formation with elastic interactions in the dark sector}",
    eprint = "2510.12551",
    archivePrefix = "arXiv",
    primaryClass = "astro-ph.CO",
    doi = "10.1051/0004-6361/202557845",
    journal = "Astron. Astrophys.",
    volume = "707",
    pages = "A269",
    year = "2026"
}

@article{Pourtsidou:2013nha,
    author = "Pourtsidou, Alkistis and Skordis, Constantinos and Copeland, Edmund J.",
    title = "{Models of coupled dark matter to dark energy}",
    eprint = "1307.0458",
    archivePrefix = "arXiv",
    primaryClass = "astro-ph.CO",
    doi = "10.1103/PhysRevD.88.083505",
    journal = "Phys. Rev. D",
    volume = "88",
    number = "8",
    pages = "083505",
    year = "2013"
}

@article{Boehmer:2015sha,
    author = "Boehmer, Christian G. and Tamanini, Nicola and Wright, Matthew",
    title = "{Interacting quintessence from a variational approach Part II: derivative couplings}",
    eprint = "1502.04030",
    archivePrefix = "arXiv",
    primaryClass = "gr-qc",
    doi = "10.1103/PhysRevD.91.123003",
    journal = "Phys. Rev. D",
    volume = "91",
    number = "12",
    pages = "123003",
    year = "2015"
}

@article{Skordis:2015yra,
    author = "Skordis, Constantinos and Pourtsidou, Alkistis and Copeland, Edmund J.",
    title = "{The Parameterized Post-Friedmannian Framework for Interacting Dark Energy Theories}",
    eprint = "1502.07297",
    archivePrefix = "arXiv",
    primaryClass = "astro-ph.CO",
    doi = "10.1103/PhysRevD.91.083537",
    journal = "Phys. Rev. D",
    volume = "91",
    number = "8",
    pages = "083537",
    year = "2015"
}

@article{Koivisto:2015qua,
    author = "Koivisto, Tomi S. and Saridakis, Emmanuel N. and Tamanini, Nicola",
    title = "{Scalar-Fluid theories: cosmological perturbations and large-scale structure}",
    eprint = "1505.07556",
    archivePrefix = "arXiv",
    primaryClass = "astro-ph.CO",
    reportNumber = "NORDITA-2015-56",
    doi = "10.1088/1475-7516/2015/09/047",
    journal = "JCAP",
    volume = "09",
    pages = "047",
    year = "2015"
}

@article{Pourtsidou:2016ico,
    author = "Pourtsidou, Alkistis and Tram, Thomas",
    title = "{Reconciling CMB and structure growth measurements with dark energy interactions}",
    eprint = "1604.04222",
    archivePrefix = "arXiv",
    primaryClass = "astro-ph.CO",
    doi = "10.1103/PhysRevD.94.043518",
    journal = "Phys. Rev. D",
    volume = "94",
    number = "4",
    pages = "043518",
    year = "2016"
}

@article{Dutta:2017fjw,
    author = "Dutta, Jibitesh and Khyllep, Wompherdeiki and Tamanini, Nicola",
    title = "{Scalar-Fluid interacting dark energy: cosmological dynamics beyond the exponential potential}",
    eprint = "1701.00744",
    archivePrefix = "arXiv",
    primaryClass = "gr-qc",
    doi = "10.1103/PhysRevD.95.023515",
    journal = "Phys. Rev. D",
    volume = "95",
    number = "2",
    pages = "023515",
    year = "2017"
}

@article{Linton:2017cep,
    author = "Linton, Mark S. and Pourtsidou, Alkistis and Crittenden, Robert and Maartens, Roy",
    title = "{Variable sound speed in interacting dark energy models}",
    eprint = "1711.05196",
    archivePrefix = "arXiv",
    primaryClass = "astro-ph.CO",
    doi = "10.1088/1475-7516/2018/04/043",
    journal = "JCAP",
    volume = "04",
    pages = "043",
    year = "2018"
}

@article{Kase:2019veo,
    author = "Kase, Ryotaro and Tsujikawa, Shinji",
    title = "{Scalar-field dark energy nonminimally and kinetically coupled to dark matter}",
    eprint = "1910.02699",
    archivePrefix = "arXiv",
    primaryClass = "gr-qc",
    doi = "10.1103/PhysRevD.101.063511",
    journal = "Phys. Rev. D",
    volume = "101",
    number = "6",
    pages = "063511",
    year = "2020"
}

@article{Chamings:2019qak,
    author = "Chamings, Finlay Noble and Avgoustidis, Anastasios and Copeland, Edmund J. and Green, Anne M. and Pourtsidou, Alkistis",
    title = "{Understanding the suppression of structure formation from dark matter-dark energy momentum coupling}",
    eprint = "1912.09858",
    archivePrefix = "arXiv",
    primaryClass = "astro-ph.CO",
    doi = "10.1103/PhysRevD.101.043531",
    journal = "Phys. Rev. D",
    volume = "101",
    number = "4",
    pages = "043531",
    year = "2020"
}

@article{Amendola:2020lnd,
    author = "Amendola, Luca and Tsujikawa, Shinji",
    title = "{Scaling solutions and weak gravity in dark energy with energy and momentum couplings}",
    eprint = "2003.02686",
    archivePrefix = "arXiv",
    primaryClass = "gr-qc",
    doi = "10.1088/1475-7516/2020/06/020",
    journal = "JCAP",
    volume = "06",
    pages = "020",
    year = "2020"
}

@article{DeFelice:2020cpt,
    author = "De Felice, Antonio and Nakamura, Shintaro and Tsujikawa, Shinji",
    title = "{Suppressed cosmic growth in coupled vector-tensor theories}",
    eprint = "2004.09384",
    archivePrefix = "arXiv",
    primaryClass = "gr-qc",
    reportNumber = "WUCG-20-02",
    doi = "10.1103/PhysRevD.102.063531",
    journal = "Phys. Rev. D",
    volume = "102",
    number = "6",
    pages = "063531",
    year = "2020"
}

@article{Linton:2021gqk,
    author = "Linton, Mark S. and Crittenden, Robert and Pourtsidou, Alkistis",
    title = "{Momentum transfer models of interacting dark energy}",
    eprint = "2107.03235",
    archivePrefix = "arXiv",
    primaryClass = "astro-ph.CO",
    doi = "10.1088/1475-7516/2022/08/075",
    journal = "JCAP",
    volume = "08",
    pages = "075",
    year = "2022"
}

@article{Liu:2023yaw,
    author = "Liu, Xiaolin and Tsujikawa, Shinji and Ichiki, Kiyotomo",
    title = "{Observational constraints on interactions between dark energy and dark matter with momentum and energy transfers}",
    eprint = "2309.13946",
    archivePrefix = "arXiv",
    primaryClass = "astro-ph.CO",
    doi = "10.1103/PhysRevD.109.043533",
    journal = "Phys. Rev. D",
    volume = "109",
    number = "4",
    pages = "043533",
    year = "2024"
}

@article{KiDS:2016kqt,
    author = "Hildebrandt, H. and others",
    collaboration = "KiDS",
    title = "{KiDS-450: Cosmological parameter constraints from tomographic weak gravitational lensing}",
    eprint = "1606.05338",
    archivePrefix = "arXiv",
    primaryClass = "astro-ph.CO",
    doi = "10.1093/mnras/stw2805",
    journal = "Mon. Not. Roy. Astron. Soc.",
    volume = "465",
    number = "2",
    pages = "1454--1498",
    year = "2017"
}

@article{DES:2017myr,
    author = "Abbott, T. M. C. and others",
    collaboration = "DES",
    title = "{Dark Energy Survey Year 1 Results: Cosmological Constraints from Galaxy Clustering and Weak Lensing}",
    eprint = "1708.01530",
    archivePrefix = "arXiv",
    primaryClass = "astro-ph.CO",
    reportNumber = "FERMILAB-PUB-17-294-PPD",
    doi = "10.1103/PhysRevD.98.043526",
    journal = "Phys. Rev. D",
    volume = "98",
    number = "4",
    pages = "043526",
    year = "2018"
}

@article{DES:2021bvc,
    author = "Abbott, T. M. C. and others",
    collaboration = "DES",
    title = "{Dark Energy Survey Year 3 Results: Cosmological Constraints from Galaxy Clustering and Weak Lensing}",
    eprint = "2105.13549",
    archivePrefix = "arXiv",
    primaryClass = "astro-ph.CO",
    reportNumber = "FERMILAB-PUB-21-239-AE-E-LD-PPD-SCD",
    doi = "10.1103/PhysRevD.105.023520",
    journal = "Phys. Rev. D",
    volume = "105",
    number = "2",
    pages = "023520",
    year = "2022"
}

@article{Asgari:2020wuj,
    author = "Asgari, Marika and others",
    title = "{KiDS-1000 Cosmology: Cosmic shear constraints and comparison between two point statistics}",
    eprint = "2007.15633",
    archivePrefix = "arXiv",
    primaryClass = "astro-ph.CO",
    doi = "10.1051/0004-6361/202039070",
    journal = "Astron. Astrophys.",
    volume = "645",
    pages = "A104",
    year = "2021"
}

@article{Heymans:2020gsg,
    author = "Heymans, Catherine and others",
    title = "{KiDS-1000 Cosmology: Multi-probe weak gravitational lensing and spectroscopic galaxy clustering constraints}",
    eprint = "2007.15632",
    archivePrefix = "arXiv",
    primaryClass = "astro-ph.CO",
    doi = "10.1051/0004-6361/202039063",
    journal = "Astron. Astrophys.",
    volume = "646",
    pages = "A140",
    year = "2021"
}

@article{Ma:1995ey,
  author        = {Ma, Chung-Pei and Bertschinger, Edmund},
  title         = {Cosmological perturbation theory in the synchronous and conformal Newtonian gauges},
  journal       = {Astrophys. J.},
  volume        = {455},
  pages         = {7--25},
  year          = {1995},
  eprint        = {astro-ph/9506072},
  archivePrefix = {arXiv},
  doi           = {10.1086/176550}
}

\end{document}